\documentclass{IEEEtran}
\usepackage{cite}
\usepackage{amsmath,amssymb,amsfonts}
\allowdisplaybreaks
\usepackage{graphicx}
\usepackage{textcomp,nicefrac}
\makeatletter
\let\@OrigAlph\@Alph
\renewcommand{\@Alph}[1]{%
  \ifnum#1>26
    A\@OrigAlph{\numexpr#1-26\relax}%
  \else
    \@OrigAlph{#1}%
  \fi
}
\makeatother
\def\BibTeX{{\rm B\kern-.05em{\sc i\kern-.025em b}\kern-.08em
T\kern-.1667em\lower.7ex\hbox{E}\kern-.125emX}}
\begin{document}
\title{A Unified Analytical Framework for LYSO-SiPM Scintillation Pulse Dynamics}

\author{Ao Qiu, and Qingguo Xie
    \thanks{This work was supported in part by the National Natural Science Foundation of China, under Grant 625B2079, 61927801, 62250002, and 62050288. (\textit{Corresponding author: Qingguo Xie})}
    \thanks{Ao Qiu is with the Department of Biomedical Engineering, Huazhong University of Science and Technology, Wuhan, 430074 China (e-mail: aqiu@hust.edu.cn).}
    \thanks{Qingguo Xie is with the Department of Biomedical Engineering, Huazhong University of Science and Technology, Wuhan, 430074 China; Wuhan National Laboratory for Optoelectronics, Wuhan, 430074 China; and the Department of Electronic Engineering and Information Science, University of Science and Technology of China, Hefei, 230026 China (e-mail: qgxie@hust.edu.cn).}}

\maketitle

\begin{abstract}
    Existing scintillation-detector models typically treat scintillation kinetics, optical transport, silicon photomultiplier (SiPM) response, and timing statistics separately, limiting end-to-end prediction of waveform formation and detector performance. We present a unified analytical framework for lutetium-yttrium oxyorthosilicate (LYSO)-SiPM scintillation detectors that links these processes within a single forward model. The framework incorporates finite thermalization, depth-dependent optical transit-time spread, and microcell occupancy dynamics to provide a physics-based description of macroscopic pulse formation. It yields closed-form exponentially modified Gaussian pulses in the linear regime, state-dependent integral solutions in saturation, and recovers the conventional bi-exponential pulse model---ubiquitously used yet hitherto only empirically justified in scintillation pulse fitting and sparse-sampling reconstruction---as a controlled reduction of the full optoelectronic cascade. Experimental validation on 10,000 directly digitized Na-22 pulses shows that the dynamic saturation model captures amplitude-dependent waveform distortion and is favored by the Akaike information criterion (AIC) over a matched bi-exponential baseline in 100/100 high-amplitude pulses and 98/100 medium-amplitude pulses. By coupling the dynamic triggering rate to compound Poisson statistics, the framework also predicts current-variance envelopes and Fisher-information-based timing limits, including an intrinsic coincidence timing resolution lower bound of about 100 ps full width at half maximum (FWHM) for a reference 511-keV LYSO-SiPM configuration. These results deepen the physical understanding of scintillation-detector waveform formation and timing limits by clarifying how scintillation kinetics, optical transport, and SiPM microcell dynamics jointly shape the observed response.
\end{abstract}


\section{Introduction}
\label{sec:introduction}
\IEEEPARstart{A}{ccurate} end-to-end models of scintillation waveform formation and detector timing are central to applications ranging from Time-of-Flight Positron Emission Tomography (ToF-PET) and high-energy physics calorimetry to nuclear safeguards and gamma-ray astronomy~\cite{Surti2015Update, Schaart2021Time}. Lutetium-Yttrium Oxyorthosilicate (LYSO) crystals coupled to Silicon Photomultipliers (SiPMs) have become a standard detector platform in these settings because they combine high light yield, fast timing, compact size, and high gain~\cite{Lecoq2017Pushing}. Yet most existing models still treat scintillation kinetics, optical transport, SiPM response, and timing statistics as separate layers, making it difficult to explain how they jointly determine the observed waveform and its performance limits.

Previous analytical studies have clarified individual parts of this chain rather than the full cascade. Birowosuto \textit{et al.}~\cite{birowosuto_novel_2009} characterized scintillation properties of Ce$^{3+}$-doped halides under assumptions that effectively collapse the excitation stage to instantaneous deposition. Marano \textit{et al.}~\cite{marano_accurate_2014} developed analytical models for the transient single-photoelectron response of SiPMs, and Xie \textit{et al.}~\cite{xie_characterization_2006} introduced a static SiPM saturation model without intra-pulse recovery. What remains missing is a unified analytical framework that couples finite thermalization, optical transport, microcell occupancy and recovery, and single-cell current response, and then carries that description forward to noise and timing limits.

In this work, we develop such a framework for LYSO-SiPM detectors. The model links scintillation kinetics, depth-of-interaction (DOI)-dependent optical transport, SiPM microcell occupancy dynamics, and timing statistics within a single forward description. Because DOI enters the transport kernel explicitly, the same formulation also suggests a route to waveform-based DOI inference. Our main contributions are threefold:

1) \textbf{Finite kinetics and optical transport:} We replace the instantaneous-excitation approximation with finite thermalization and couple the resulting emission profile to a DOI-dependent optical transit-time spread (TTS) model. This yields a physically bounded photon-arrival description and clarifies how early pulse formation depends jointly on crystal kinetics and optical dispersion.

2) \textbf{Coupled SiPM response and saturation:} We connect this photon-arrival process to SiPM microcell activation, recovery, and multi-exponential single-cell response. The resulting formulation yields closed-form exponentially modified Gaussian (EMG) pulses in the linear regime and state-dependent integral solutions in saturation. As a corollary, the conventional bi-exponential pulse model---which is almost universally adopted for scintillation pulse fitting and sparse-sampling (e.g., Multi-Voltage Threshold) reconstruction, yet whose physical origin within the LYSO-SiPM detection chain has hitherto been left empirical---emerges as a controlled asymptotic reduction of this framework. To our knowledge, this is the first time the model is obtained as a limiting case of the full optoelectronic cascade.

3) \textbf{Variance and timing limits:} We couple the dynamic triggering-rate formulation to generalized non-stationary compound Poisson statistics, including dark counts and crosstalk-induced variance scaling. This extends the framework from deterministic pulse prediction to current-variance envelopes and Cram\'er-Rao timing bounds, clarifying how the same physical cascade governs both waveform formation and timing performance.

\section{Theory and Methods}

To facilitate a structured mathematical exposition, a consolidated notation summary is provided at the end of the main text.

\subsection{Continuous Scintillation Light-Generation Kinetics}
The physical sequence in LYSO involves a complex multistep kinetic cascade, ranging from sub-picosecond hot-carrier interactions to slower picosecond exciton migration and subsequent trap-center capture~\cite{Vasiliev2008Microscopic}. If the intermediate transitions are fast relative to a dominant rate-limiting step, this cascade can be approximated as a single-exponential relaxation process. The formal rate-limiting-step reduction is detailed in Appendix~\ref{app:rate_limiting}. At the macroscopic level, the cascaded states are therefore lumped into a single bottleneck time constant $\tau_r$. This reduction preserves Linear Time-Invariant (LTI) mathematical tractability while retaining the finite rise-time mechanism. For an initial population of $N_0$ excitation centers (primary electron-hole pairs) and a bottleneck thermalization time constant $\tau_r$, the generation rate is given by $R(t) = \frac{N_0}{\tau_r} e^{-t/\tau_r}$ for $t \ge 0$. Its temporal integral satisfies $\int_0^\infty R(t) dt = N_0$, thereby conserving the initial excitation population.

Within the crystal, photon emission is modulated by recursive self-absorption and re-emission~\cite{Pichon2023Ce}. While actual LYSO crystals exhibit multiple luminescent decay components, we utilize a single effective intrinsic decay time for analytical clarity. Let $\tau$ be the primary intrinsic decay time, $a$ the self-absorption probability, $\eta$ the quantum yield of re-emission, and $\epsilon$ the probability that an emitted photon falls into the self-absorption overlap band. The model assumes $0 \le \epsilon, a, \eta \le 1$ and, more specifically, $0 \le \epsilon a \eta < 1$ so that the recursive generation series converges and the effective decay time remains finite. This parameterization implicitly assumes that re-emitted photons undergo an identical intrinsic decay mechanism with the same primary time constant $\tau$, serving as a macroscopic mean-field simplification for the potentially varying local Ce$^{3+}$ emitting environments. Following the self-absorption cascade logic discussed by Birowosuto \textit{et al.}~\cite{birowosuto_novel_2009}, we construct the escaping-light response as a sum over successive emission generations. The zeroth-generation term corresponds to photons that escape on their first emission attempt and contributes the fraction $(1-\epsilon a)$ of the intrinsic decay kernel $\tau^{-1} e^{-t/\tau}$, while each additional re-absorption/re-emission cycle contributes an extra factor $\epsilon a \eta$ and one further self-convolution of that intrinsic kernel. Under the spatially uniform mean-field assumption that every cycle sees the same overlap probability and the same intrinsic decay kernel, this cascade sums in closed form to a single exponential envelope with effective time constant $\tau_{eff} = \frac{\tau}{1 - \epsilon a \eta}$, as shown in Appendix~\ref{app:self_absorption}.

The macroscopic intrinsic impulse response therefore becomes:
\begin{equation}
    h(t) = \frac{1-\epsilon a}{\tau} e^{-t/\tau_{eff}} u(t)
    \label{eq:ht}
\end{equation}
Here and throughout, $u(t)$ denotes the Heaviside step function enforcing causality.
Physically, $h(t)$ represents the macroscopic intrinsic photon emission rate per initial excitation center. The prefactor $(1-\epsilon a)/\tau$ is set by the first-pass escape fraction and the intrinsic emission attempt rate, whereas the envelope $e^{-t/\tau_{eff}}$ represents the closed-form sum of all higher-order self-absorption generations. The infinite temporal integral physically represents the net macroscopic photon escape probability fraction. As detailed in Appendix~\ref{app:ht_integral}, integrating the rate exactly evaluates to:
\begin{equation}
    \int_0^\infty h(t) dt = \frac{1-\epsilon a}{1-\epsilon a \eta}
    \label{eq:ht_integral}
\end{equation}
This gives the net macroscopic photon escape probability fraction, accounting for recursive self-absorption trapping, so that the expected total detectable internal photon yield scales as $Y_{total} = N_0 \frac{1-\epsilon a}{1-\epsilon a \eta}$.

The modified intrinsic photon emission rate $Y_{mod}(t)$ is obtained by computing the temporal convolution of the intrinsic impulse response $h(t)$ with the continuous single-exponential thermalization cascade $R(t)$. The resulting closed form, derived in Appendix~\ref{app:convolution_ymod}, is a causal continuous bi-exponential light-generation profile. Defining the global generation rate constant $C_{gen} = \frac{N_0(1-\epsilon a)\tau_{eff}}{\tau(\tau_{eff} - \tau_r)}$ (with the dimension of $[\mathrm{photons} \cdot \mathrm{time}^{-1}]$; the verification that photon number is conserved through this convolution, $\int_0^\infty Y_{mod}(t)\,dt = Y_{total}$, is given in Appendix~\ref{app:photon_conservation}):
\begin{equation}
    Y_{mod}(t) = C_{gen} \left( e^{-\frac{t}{\tau_{eff}}} - e^{-\frac{t}{\tau_r}} \right) u(t)
    \label{eq:ymod}
\end{equation}

In the degenerate case where the thermalization time constant matches the effective decay time constant ($\tau_r \to \tau_{eff}$), direct substitution yields an indeterminate $0/0$ form. Applying L'H\^opital's rule by evaluating the limit with respect to $\tau_r$ resolves the singularity into a bounded critically damped response (see Appendix~\ref{app:convolution_ymod}):
\begin{equation}
    \lim_{\tau_r \to \tau_{eff}} Y_{mod}(t) = \frac{N_0(1-\epsilon a)}{\tau \tau_{eff}} t e^{-\frac{t}{\tau_{eff}}} u(t)
    \label{eq:ymod_limit}
\end{equation}

\subsection{Optical Transport and Depth-Dependent TTS}
In practical scintillation detectors employing elongated crystals, varying photon reflection trajectories introduce an Optical Transit Time Spread (TTS). TTS is a Depth-of-Interaction (DOI)-dependent phenomenon.

For an elongated crystal, assuming $z$ denotes the longitudinal distance from the interaction vertex to the photodetector, the mean transit time $\mu_{TTS}(z)$ and the transit spread variance $\sigma_{TTS}^2(z)$ can be parameterized via first-order representations:
\begin{align}
    \mu_{TTS}(z)      & \approx \mu_0 + \frac{z}{v_{eff}} \label{eq:mu_tts}  \\
    \sigma_{TTS}^2(z) & \approx \sigma_0^2 + k_{disp} z \label{eq:sigma_tts}
\end{align}
where $v_{eff}$ acts as an effective macroscopic longitudinal optical velocity. Notably, $v_{eff} < c/n$ (where $n \approx 1.82$ for LYSO, yielding $c/n \approx 1.65 \times 10^8$~m/s)~\cite{Mao2008Optical}. This velocity reduction occurs because the macroscopic longitudinal propagation velocity must be geometrically projected over the internal scattering angles. Assuming isotropic internal photon emission uniformly distributed across the forward hemisphere, the expected longitudinal velocity component evaluates to $\langle \cos \theta \rangle = 0.5$ via angular integration (Appendix~\ref{app:geom_velocity}).

However, in real LYSO crystals, the dipole orientation preferences of $\mathrm{Ce}^{3+}$ transitions, restricted grazing angles, multi-path scattering typical of elongated geometries, and modifications to the angular acceptance cone via total internal reflection cutoffs at the crystal-SiPM index boundary deviate the transport from this ideal uniform expectation. These effects motivate using an operational effective-velocity range $v_{eff} \approx 0.5$--$0.8 \times c/n$. The lower end $0.5 \times c/n$ follows from the isotropic forward-hemisphere average in Appendix~\ref{app:geom_velocity}, whereas the upper end is consistent with preferential transmission of near-normal photons through the TIR acceptance cone in Appendix~\ref{app:tir_correction}. In practice, $v_{eff}$ is treated as an empirical effective parameter that also absorbs multiple reflections, wrapping geometry, and residual angular-selection effects. The parameter $k_{disp}$ characterizes the spatial variance accumulation rate due to sequential reflections, while $\mu_0$ and $\sigma_0$ parameterize the near-surface mean delay and standard deviation, so that $\sigma_0^2$ appears as the variance intercept in \eqref{eq:sigma_tts}.

Motivated by the Central Limit Theorem (CLT), multiple randomized reflections cumulatively smear the arrival times. If the total transport delay is approximated as the sum of many approximately independent reflection-induced increments with finite variance and finite third absolute moment, the Berry-Esseen theorem bounds the deviation from the Gaussian cumulative distribution by a constant times the normalized third absolute moment divided by the square root of the number of contributing increments~\cite{Berry1941Accuracy}. We therefore adopt a Gaussian approximation for intermediate-to-long crystals ($10$--$30$~mm) coupled with diffuse surfaces or rough wrapping (i.e., length-to-width aspect ratio $L/d \gtrsim 5$), where numerous randomized multipath events dominate the transport and suppress the skewness of individual trajectories. Thus, the optical transit spread is approximated as:
\begin{align}
    f_{TTS}(t; z) & = \frac{1}{\sqrt{2\pi}\sigma_{TTS}(z)} \nonumber                                \\
                  & \quad \times \exp\left( - \frac{(t-\mu_{TTS}(z))^2}{2\sigma_{TTS}^2(z)} \right)
    \label{eq:ftts}
\end{align}
However, for highly polished short crystals coupled without diffuse reflectors (e.g., $L/d \lesssim 2$), direct unscattered Euclidean pathways and low-order specular reflections dominate~\cite{Vinke2014Time}. The arrival-time distribution can then remain multi-modal and positively skewed, so the Gaussian approximation is no longer well controlled. In such configurations, the analytical Exponentially Modified Gaussian (EMG) superpositions derived later in \eqref{eq:ilin_emg} may lose applicability, necessitating a fallback to numerical integration.

Assuming a bulk optical transfer efficiency $k_{trans}$, the macroscopic incident photon rate reaching the SiPM surface $r_{ph}(t; z)$ is the convolution of the intrinsic generation and the spatial geometric kernel. Because $Y_{mod}(s)$ is causal, i.e. $Y_{mod}(s)=0$ for $s<0$, the physical emission history vanishes before excitation. When $\mu_{TTS}(z)$ exceeds $\sigma_{TTS}(z)$ by several standard deviations, the Gaussian probability mass leaking into the non-physical negative temporal domain is negligible. For near-surface interactions ($z \approx 0$), however, the interface delays typically yield $\mu_0 \sim 50$~ps and $\sigma_0 \sim 30$~ps, resulting in a ratio of $\mu_0 / \sigma_0 \approx 1.67$. This ratio corresponds to a negative-time tail probability of approximately $\Phi(-1.67) \approx 4.75\%$, where $\Phi$ is the standard normal cumulative distribution function. We therefore treat the extension of the transport convolution lower boundary to $s \to -\infty$ as a controlled analytical approximation rather than a physically exact statement. It preserves closed-form Linear Time-Invariant (LTI) convolution properties, but it slightly overestimates the ultra-early pre-arrival slope near $t=0$; for polished short crystals or analyses focused on sub-100-ps leading-edge structure, truncated or numerically integrated transport kernels are preferable. Utilizing a distinct integration variable $s$:
\begin{equation}
    r_{ph}(t; z) = k_{trans} \int_{-\infty}^{\infty} Y_{mod}(s) f_{TTS}(t-s; z) ds
    \label{eq:rph_tts}
\end{equation}
Consequently, its cumulative integral is defined over the same extended time domain:
\begin{equation}
    n_{ph}(t; z) = \int_{-\infty}^{t} r_{ph}(t'; z) dt'
    \label{eq:nph_tts}
\end{equation}

\subsection{SiPM Microcell Activation and Recovery Behavior}
Following the static SiPM occupancy model introduced by Xie \textit{et al.}~\cite{xie_characterization_2006}, we model the expected cumulative occupancy by treating the cumulative incident photon count $n_{ph}(t; z)$ as a sequence of independent triggering trials randomly distributed over the $M$ microcells, with global Photon Detection Efficiency (PDE) $q$. The underlying combinatorial argument is exact for an integer number of incident photons; in the present macroscopic formulation we analytically continue it to the continuous expectation variable $n_{ph}(t; z)$ as a mean-field closure for the expected occupancy. For any specific microcell, the probability that one photon does \emph{not} trigger that cell is $1-q/M$, so after $n_{ph}(t; z)$ photons the probability that it remains unfired is $\left(1-\frac{q}{M}\right)^{n_{ph}(t; z)}$. The complementary probability is therefore the firing probability of that cell, and multiplying by $M$ gives the expected cumulative occupancy:
\begin{equation}
    N_{fired}(t; z) = M \left[ 1 - \left(1 - \frac{q}{M}\right)^{n_{ph}(t; z)} \right]
    \label{eq:nfired}
\end{equation}
This closure therefore assumes independent photon-trigger attempts, uniform random allocation over microcells, and no explicit prompt optical-crosstalk occupancy term inside $N_{fired}(t; z)$; prompt crosstalk is introduced later only through the mean current gain and variance model, not as a separate depletion channel in the occupancy state.

By defining the effective triggering parameter $\alpha = -\ln\left(1 - \frac{q}{M}\right)$ (whose combinatorial derivation and dilute-limit expansion are given in Appendix~\ref{app:alpha_derivation}), we rewrite the static exponential relationship as $N_{fired}(t; z) = M \left[1 - e^{-\alpha n_{ph}(t; z)}\right]$. By analytically differentiating this static occupancy state via the chain rule and applying algebraic substitution over the cumulative photon states (detailed in Appendix~\ref{app:lambda_derive}), the macroscopic static activation rate $\lambda(t; z) = \frac{d}{dt} N_{fired}(t; z)$ resolves to an exact residual availability form:
\begin{equation}
    \lambda(t; z) = \alpha r_{ph}(t; z) \left[ M - N_{fired}(t; z) \right]
    \label{eq:lambda}
\end{equation}
This form explicitly mirrors the structure of the dynamic recovery model introduced below.

Continuous intra-pulse recovery relies on a state-dependent dynamic model tracking the transient recovering microcell population $N_{busy}(t; z)$ coupled to a microcell RC-recharging recovery time constant $\tau_{rec}$~\cite{Acerbi2019Understanding}. In this phenomenological framework, $\tau_{rec}$ and the macroscopic multi-exponential avalanche current decay poles $\tau_x$ are treated as independently parameterized constants, although physically they share underlying circuit elements (such as the diode capacitance and quench resistance). Treating $z$ as a fixed parameter, the time evolution is governed by:
\begin{equation}
    \frac{d}{dt} N_{busy}(t; z) = \lambda_{dyn}(t; z) - \frac{N_{busy}(t; z)}{\tau_{rec}}
    \label{eq:dynamic_rec}
\end{equation}

The dynamic activation rate is $\lambda_{dyn}(t; z) = \alpha r_{ph}(t; z) [M - N_{busy}(t; z)]$. In adopting this binary recovery closure, we make two explicit approximations. First, substituting the static triggering parameter $\alpha = -\ln(1-q/M)$ directly into the dynamic ODE assumes that recovering microcells regain their full PDE instantaneously upon leaving the completely depleted state, so microcells are tracked only as fully active or completely depleted. Second, prompt optical-crosstalk avalanches are not modeled here as an additional busy-cell depletion channel in $N_{busy}(t; z)$; their mean effect is introduced later through $I_{cell}^{macro}$ and the variance model rather than through the occupancy ODE itself. The resulting dynamic ODE should therefore be interpreted as a first-order mean-field saturation model for primary photon-driven occupancy, and it tends to overestimate the instantaneous recovery capacity when partial recharge physics is important. In physical SiPM arrays, recovering microcells continuously restore their overvoltage as $\Delta V(t') \propto 1 - e^{-t'/\tau_{rec}}$, where $t'$ is the time elapsed since its previous discharge~\cite{Acerbi2019Understanding}. Assuming that the triggering probability and avalanche gain both scale approximately linearly with restored overvoltage, the expected macroscopic charge yield from a partial avalanche scales quadratically as $Y_{phys}(t') = (1 - e^{-t'/\tau_{rec}})^2$. While this idealized quadratic scaling slightly deviates near the avalanche breakdown threshold, it provides a useful envelope for estimating the absolute macroscopic saturation dead-time loss. Conversely, the macroscopic binary ODE ensemble aggregates these sub-states into a linear proportional recovery expectation $Y_{ODE}(t') = 1 - e^{-t'/\tau_{rec}}$.

To map this discrepancy to macroscopic saturation error, we compute the effective integrated dead-time per avalanche. By evaluating the integrals of the complementary lost detection probabilities over an infinite recovery cycle (expanded in Appendix~\ref{app:dead_time_integrals}), the physical effective dead-time evaluates to $\tau_{dead, phys} = 1.5\tau_{rec}$. However, the binary ODE model perceives a systematically shorter effective dead-time of $\tau_{dead, ODE} = 1.0\tau_{rec}$.

In normalized units $\xi = t'/\tau_{rec}$, the arithmetic integral difference is $0.5$, corresponding to $0.5 \tau_{rec}$ in physical time. Physically, this means that by linearly aggregating partial recoveries, the binary ODE ensemble overestimates the integrated recovery capacity by $0.5 \tau_{rec}$ per recovering microcell cycle. Because this overestimation expects more integrated current than is physically generated, the binary model underestimates the absolute saturation dead-time loss. Because the binary ODE overestimates the per-cell recovery yield at every instant ($Y_{phys}(t') \le Y_{ODE}(t')$ for all $t' \ge 0$), the aggregate macroscopic current is expected to be biased high under dense pileup; correspondingly, downstream timing-variance predictions should be interpreted as having an optimistic tendency rather than as a rigorously proven non-conservative bound. Grouping the $N_{busy}(t; z)$ terms, we form the first-order ODE:
\begin{align}
    \frac{d}{dt} N_{busy}(t; z) & + \left( \alpha r_{ph}(t; z) + \frac{1}{\tau_{rec}} \right) N_{busy}(t; z) \nonumber \\
                                & = \alpha M r_{ph}(t; z)
    \label{eq:dynamic_ode}
\end{align}
subject to the initial condition $N_{busy}(-\infty; z) = 0$, representing the physical requirement that all microcells are quiescent prior to the scintillation event.

To solve this ODE analytically, we multiply both sides by an integrating factor $\mu(t; z) = \exp \left( \alpha n_{ph}(t; z) + \frac{t}{\tau_{rec}} \right)$~\cite{Boyce2017Elementary}. Recognizing that the left side collapses via the product rule into an exact differential, followed by applying integration by parts across the domain (with the complete formulation provided in Appendix~\ref{app:ibp_ode} and expanded intermediate steps in Appendix~\ref{app:integrating_factor_detail}), the tracking problem resolves into a state-dependent integral representation:
\begin{equation}
    \begin{split}
        N_{busy}(t; z)
            &= M \\
            &\quad - \frac{M}{\tau_{rec}}
                \int_{-\infty}^t
                \exp\!\Biggl(
                    -\alpha [n_{ph}(t; z) - n_{ph}(t'; z)] \\
            &\qquad\qquad
                    - \frac{t - t'}{\tau_{rec}}
                \Biggr) dt'
    \end{split}
    \label{eq:analytical_dynamic}
\end{equation}

To demonstrate the physical consistency of this integral form, we examine its asymptotic reduction for an infinitely slow recovering SiPM ($\tau_{rec} \to \infty$). Before integration by parts, we rewrite the exact differential as a Riemann-Stieltjes integral with respect to the state measure generated by $e^{\alpha n_{ph}(t'; z)}$. The limit can then be passed inside the integral via the Lebesgue Dominated Convergence Theorem (DCT). Evaluating the converged boundary conditions, as detailed in Appendix~\ref{app:asymptotic_static_limit}, yields the exact reduction:
\begin{equation}
    \lim_{\tau_{rec} \to \infty} N_{busy}(t; z) = M \left[ 1 - \left(1 - \frac{q}{M}\right)^{n_{ph}(t; z)} \right]
    \label{eq:static_limit}
\end{equation}
This naturally reduces to the static binomial occupancy model formulated in \eqref{eq:nfired}, bypassing cumbersome secondary integrations.

\subsection{Extraction of Macroscopic Current}
\label{subsec:extraction_current}
In this macroscopic framework, the expectation of the total macroscopic current scales with the mean secondary cascade gain $\langle G \rangle = 1/(1-P_{ct})$ introduced by optical crosstalk, where $G$ denotes the discrete total avalanche multiplicity generated by one primary trigger. To maintain algebraic clarity while explicitly distinguishing between the deterministic continuous expectation current and the subsequent stochastic discrete variance bounds, we absorb this mean gain scalar into an equivalent macroscopic single-cell amplitude $I_{cell}^{macro} = \langle G \rangle I_{cell}^{single}$, where $I_{cell}^{single}$ corresponds to the pure unmultiplied physical single-avalanche peak current. Following the analytical single-photoelectron response modeling of Marano \textit{et al.}~\cite{marano_accurate_2014}, we represent the expected causal single-cell macroscopic current response for elapsed delays $t \ge 0$ as a multi-exponential decay:
\begin{equation}
    i_{ser}(t) = I_{cell}^{macro} \sum_{x \in \{d, p1, p2\}} A_x e^{-t/\tau_x}
    \label{eq:iser}
\end{equation}
and set $i_{ser}(t)=0$ for $t<0$ by causality. The weights $A_x$ are taken to be dimensionless, non-negative, and normalized, $A_x \ge 0$ with $\sum_{x \in \{d, p1, p2\}} A_x = 1$, so the kernel remains non-negative on the convolution domain and satisfies the boundary condition $i_{ser}(0) = I_{cell}^{macro}$. At the deterministic mean-current level, this representation assumes that prompt optical crosstalk rescales only the single-cell amplitude through $I_{cell}^{macro}$ and does not alter the pole set $\{\tau_x, A_x\}$ itself; higher-order cascade fluctuations are deferred to the variance model in Section~\ref{sec:poisson_noise_modeling}. This functional form models an instantaneous avalanche rise, neglecting the intrinsic sub-nanosecond avalanche plasma build-up kinetics (typically $50$--$150$~ps). This approximation remains valid for macroscopic system readout bandwidths constrained below $\sim 1$~GHz where intrinsic diffusion limitations are low-pass filtered. For analog front-ends exceeding $2$~GHz bandwidths, omitting this kinetic build-up stage establishes an analytical simplification boundary that may overestimate the macroscopic pulse's initial nanosecond slope. This normalization also enables the algebraic simplification used during integration by parts.

In the permanently depleted saturation regime, the static macroscopic current is given by the convolution $I_{stat}(t; z) = \int_{-\infty}^t \lambda(t'; z) i_{ser}(t-t') dt'$. By recognizing that $\lambda(t; z) = \frac{d}{dt} N_{fired}(t; z)$ and applying integration by parts over the convolution domain (see Appendix~\ref{app:ibp_macroscopic} for the full expansion and boundary limits matching; boundary condition convergence is validated in Appendix~\ref{app:boundary_validation}), this bypasses the numerical convolution of the transient incident rate function:
\begin{align}
    I_{stat}(t; z) & = I_{cell}^{macro} N_{fired}(t; z) \nonumber                                                                        \\
                   & \quad - I_{cell}^{macro} \int_{-\infty}^t N_{fired}(t'; z) \sum_{x} \frac{A_x}{\tau_x} e^{-\frac{t-t'}{\tau_x}} dt'
    \label{eq:i_stat}
\end{align}

Incorporating the dynamic intra-pulse recovery model, the macroscopic output current $I_{dyn}(t; z)$ is given by the convolution:
\begin{equation}
    I_{dyn}(t; z) = \int_{-\infty}^t \lambda_{dyn}(t'; z) i_{ser}(t-t') dt'
    \label{eq:i_dyn}
\end{equation}
By substituting the state relationship $\lambda_{dyn}(t; z) = \frac{\partial}{\partial t} N_{busy}(t; z) + \frac{N_{busy}(t; z)}{\tau_{rec}}$ from \eqref{eq:dynamic_rec}, we symmetrically apply integration by parts to the explicit derivative term. Following identical boundary cancellations and isolating the internal derivative $\frac{\partial}{\partial t'} i_{ser}(t-t')$ (as detailed in Appendix~\ref{app:ibp_macroscopic}), the representation condenses into a robust state-dependent integral solution:
\begin{align}
    I_{dyn}(t; z) & = I_{cell}^{macro} N_{busy}(t; z) \nonumber                                                                                                      \\
                  & \quad - I_{cell}^{macro} \int_{-\infty}^t N_{busy}(t'; z) \nonumber                                                                              \\
                  & \quad \times \sum_{x} A_x \left( \frac{1}{\tau_x} - \frac{1}{\tau_{rec}} \right) e^{-\frac{t-t'}{\tau_x}} dt'
    \label{eq:i_dyn_explicit}
\end{align}

The factored coefficient $\left( \frac{1}{\tau_x} - \frac{1}{\tau_{rec}} \right)$ is positive whenever $\tau_x < \tau_{rec}$ and negative whenever $\tau_x > \tau_{rec}$. In the present parameterization, all current-response poles satisfy $\tau_x < \tau_{rec}$, so this coefficient is strictly positive for every pole $x$. The general expression \eqref{eq:i_dyn_explicit} nonetheless accommodates SiPM architectures where a slow current-response pole exceeds $\tau_{rec}$, in which case the coefficient becomes negative. Regardless of the sign distribution across poles, the physical non-negativity of the macroscopic current ($I_{dyn}(t) \ge 0$) is guaranteed by its definition as a convolution of two non-negative functions \eqref{eq:i_dyn} (a formal proof is provided in Appendix~\ref{app:nonneg_proof}). Any algebraically negative contributions that appear in the integration-by-parts decomposition \eqref{eq:i_dyn_explicit} are exactly compensated by the boundary term $I_{cell}^{macro} N_{busy}(t; z)$.

To verify consistency, we assess the limit $\tau_{rec} \to \infty$. The transient dynamic state bounds to the static model $N_{busy}(t; z) \to N_{fired}(t; z)$, and the inverse recovery time constant approaches $\frac{1}{\tau_{rec}} \to 0$. By substituting these limits into the dynamic current equation \eqref{eq:i_dyn_explicit} and evaluating the boundary closures (derived in Appendix~\ref{app:dynamic_to_static}), the dynamic expectation structurally reduces to the exact static expectation:
\begin{equation}
    \lim_{\tau_{rec} \to \infty} I_{dyn}(t; z) = I_{stat}(t; z)
    \label{eq:idyn_limit}
\end{equation}
This confirms mathematical equivalence to the precise closed-form relation defined in \eqref{eq:i_stat}.

For low-energy depositions in the linear regime, the static activation rate simplifies to $\lambda(t; z) \approx q \cdot r_{ph}(t; z)$. This simplification is quantitatively valid when the effective activation parameter satisfies $\alpha n_{ph}(t; z) \ll 1$. Recognizing that the triggering parameter scales as $\alpha = -\ln(1-q/M) \approx q/M$, this linear limit includes both the Photon Detection Efficiency (PDE) and microcell density. Enforcing the constraint $\alpha n_{ph}(t; z) \approx q \cdot n_{ph}(t; z) / M < 0.1$ restricts the algebraic divergence to less than $\sim 5\%$ relative to the first-order exponential Taylor expansion ($1 - e^{-x} \approx x$). Under this assumption:
\begin{align}
    I_{lin}(t; z) & = \int_{-\infty}^t \left[ q \cdot r_{ph}(t'; z) \right] i_{ser}(t-t') dt' \nonumber \\
                  & = q \cdot (r_{ph}(\cdot; z) * i_{ser})(t)
    \label{eq:ilin_conv}
\end{align}

By expanding the macroscopic photon rate $r_{ph}(t; z) = k_{trans} (Y_{mod} * f_{TTS}(\cdot; z))(t)$, and iteratively applying the commutative and associative sequence scaling properties of LTI convolutions (expanded explicitly in Appendix~\ref{app:commutative_lti}), the geometric spatial spread $f_{TTS}(t; z)$ can be isolated and decoupled externally:
\begin{equation}
    I_{lin}(t; z) = (I_{ideal} * f_{TTS}(\cdot; z))(t)
    \label{eq:ilin_final}
\end{equation}
where $I_{ideal}(t) = q k_{trans} (Y_{mod} * i_{ser})(t)$ acts as the intrinsic base kernel assuming an idealized instantaneous optical transit. This commutative decoupling establishes a distinct computational advantage: it theoretically isolates the geometry-dependent spatial blur $f_{TTS}(z)$ from the intrinsic crystal-sensor kinetics $I_{ideal}(t)$, allowing detector designers to computationally evaluate different crystal lengths without recalculating the underlying kinetic convolutions~\cite{Roncali2013Application, vanDam2013Scintillation}.

To compute $I_{ideal}(t) = \int_0^t [q k_{trans} Y_{mod}(t')] i_{ser}(t-t') dt'$ (where the lower integration limit starts at $0$ due to the Heaviside step function $u(t')$ embedded in $Y_{mod}$), we define a global scale factor $C = q k_{trans} C_{gen} I_{cell}^{macro}$, and introduce two coupling time constants: $\tilde{\tau}_{x,eff} = \left(\frac{1}{\tau_x} - \frac{1}{\tau_{eff}}\right)^{-1}$ and $\tilde{\tau}_{x,r} = \left(\frac{1}{\tau_x} - \frac{1}{\tau_r}\right)^{-1}$. Analytically expanding and distributing this LTI convolution (refer to Appendix~\ref{app:ideal_kernel} for algebraic details), this resolves into the continuous response kernel for $t \ge 0$:
\begin{align}
    I_{ideal}(t) & = C \sum_{x} A_x \Big[ \tilde{\tau}_{x,eff} \left(e^{-\frac{t}{\tau_{eff}}} - e^{-\frac{t}{\tau_x}}\right) \nonumber \\
                                        & \quad \quad \quad \quad \quad \quad - \tilde{\tau}_{x,r} \left(e^{-\frac{t}{\tau_r}} - e^{-\frac{t}{\tau_x}}\right) \Big] u(t)
    \label{eq:i_ideal_final}
\end{align}

If the microcell response pole matches the intrinsic parameters (e.g., $\tau_x \to \tau_{eff}$), substituting $\tilde{\tau}_{x,eff}$ yields an indeterminate $0/0$ fraction. By applying an algebraic variable mapping $u = 1/\tau_x$ over the derivative limit (Appendix~\ref{app:ideal_kernel}), this singularity structurally collapses into a physically bounded resonance profile $t e^{-t/\tau_{eff}}$. Similarly, if a single-cell response pole matches the thermalization cascade constant ($\tau_x \to \tau_r$), the bi-exponential coupling term resolves into the critically damped resonance profile $t e^{-t/\tau_r}$.

The macroscopic output pulse, incorporating TTS geometric transport, is obtained via the Gaussian convolution $I_{lin}(t; z) = \int_{-\infty}^{\infty} I_{ideal}(s) f_{TTS}(t-s; z) ds$. Convolving a causal continuous exponential decay $e^{-s/\tau}u(s)$ with a full-domain Gaussian distribution generates an Exponentially Modified Gaussian (EMG) function valid for all $t \in (-\infty, \infty)$. It should be noted that the EMG function defined here is not a normalized probability density function; its integral over all time yields $\tau$ rather than $1$ (proven in Appendix~\ref{app:emg_normalization}), preserving dimensional consistency for the current expressions. Computing the convolution of a generic causal exponential $e^{-s/\tau} u(s)$ with the Gaussian kernel $f_{TTS}(t-s; z)$ expands algebraically by merging the exponents and completing the square. By substituting the exponent argument and analytically completing the square with respect to $s$ (detailed in Appendix~\ref{app:emg_square}), the convolution reduces to a standard integral over the complementary error function---which arises specifically because the Heaviside causality constraint $u(s)$ truncates the Gaussian integration domain to $s \in [0, \infty)$, retaining only one tail of the standard normal integral. Multiplying this back into the prefactor, we define the resultant EMG kernel (note that this function is inherently asymmetric):
\begin{equation}
    \operatorname{EMG}(t; \mu, \sigma, \tau) = \frac{1}{2} e^{\frac{\sigma^2}{2\tau^2} - \frac{t-\mu}{\tau}} \operatorname{erfc}\left( \frac{\mu - t + \sigma^2/\tau}{\sqrt{2}\sigma} \right)
    \label{eq:emg_def}
\end{equation}
Mapping into this formulation relies on the extension of the optical Gaussian tail to $s \to -\infty$ justified earlier in Section II-B. Utilizing the spatial abbreviations $\mu_z = \mu_{TTS}(z)$, $\sigma_z = \sigma_{TTS}(z)$, and $\operatorname{EMG}_z(\tau) = \operatorname{EMG}(t; \mu_z, \sigma_z, \tau)$ for brevity, the linear current superposition becomes:
\begin{equation}
    \begin{aligned}
        I_{lin}(t; z)
            &= C \sum_{x \in \{d, p1, p2\}} A_x \\
            &\quad \times \bigl[
                \tilde{\tau}_{x,eff}
                \bigl( \operatorname{EMG}_z(\tau_{eff}) - \operatorname{EMG}_z(\tau_x) \bigr) \\
            &\qquad\qquad
                - \tilde{\tau}_{x,r}
                \bigl( \operatorname{EMG}_z(\tau_r) - \operatorname{EMG}_z(\tau_x) \bigr)
                \bigr]
    \end{aligned}
    \label{eq:ilin_emg}
\end{equation}
This analytical expansion bypasses numerical error function boundary matching, providing a closed-form forward model for the linear operational regime.

\subsection{Asymptotic Limiting Forms}
The derived intrinsic continuous linear kernel $I_{ideal}(t)$ clarifies how commonly used phenomenological fit models emerge as controlled asymptotic reductions of the physics-based pulse model. Considering the classical limit in which the intrinsic thermalization cascade is treated as instantaneous ($\tau_r \to 0$), the coupled transient branch proportional to $\tilde{\tau}_{x,r} \bigl(e^{-t/\tau_r} - e^{-t/\tau_x}\bigr)$ vanishes in the exact kernel limit (Appendix~\ref{app:ideal_kernel}). Simultaneously, limits over the generation scale factor simplify dimensionally. Redefining this collapsed global scale factor as $C_{inst} = q k_{trans} \frac{N_0(1-\epsilon a)}{\tau} I_{cell}^{macro}$, the continuous LTI base kernel $I_{ideal}(t)$ presented in \eqref{eq:i_ideal_final} reduces to:
\begin{align}
    \lim_{\tau_r \to 0} I_{ideal}(t) & = \sum_{x} C_{inst} A_x \tilde{\tau}_{x,eff} \nonumber                             \\
                                     & \quad \times \left( e^{-\frac{t}{\tau_{eff}}} - e^{-\frac{t}{\tau_x}} \right) u(t)
    \label{eq:i_ideal_limit}
\end{align}

When the primary fast avalanche discharge pole dominates the SiPM response ($A_d \gg A_{p1}, A_{p2}$), a dominant-pole approximation becomes useful. For the representative parameter set used in Section~\ref{sec:verification}, where $A_d = 0.98$, $A_{p1} = 0.015$, and $A_{p2} = 0.005$, neglecting the secondary current-response poles ($x = p1, p2$) yields the leading reduced kernel:
\begin{equation}
    I_{ideal}(t) \approx A_{macro} \left( e^{-\frac{t}{\tau_{eff}}} - e^{-\frac{t}{\tau_d}} \right) u(t)
    \label{eq:double_exp}
\end{equation}
where the macroscopic amplitude is defined as $A_{macro} = C_{inst} A_d \tilde{\tau}_{d,eff}$. Consequently, guided by the LTI mapping \eqref{eq:ilin_final}, the macroscopic observable pulse $I_{lin}(t; z)$ reduces to a difference of two EMG terms governed by the slow decay pole $\tau_{eff}$ and the fast discharge pole $\tau_{d}$; the corresponding reduction and its mapping into the experimental fitting forms are detailed in Appendix~\ref{app:fitting_biexp_derivation}. In the common regime $\tau_d \ll \tau_{eff}$, the late-time tail is therefore controlled primarily by the slow-decay EMG branch. This asymptotic reduction identifies the physical origin of the dominant temporal poles appearing in standard EMG-based macroscopic scintillation pulse fits~\cite{Seifert2012Comprehensive}.

\subsection{Intrinsic Poisson Noise Modeling}
\label{sec:poisson_noise_modeling}
While the continuous kernels derived above define the deterministic mean pulse shape, the timing resolution is governed by stochastic fluctuations of the underlying Poisson mechanisms, which we now characterize. Physically, the steady-state dark count rate only triggers available (non-busy) microcells, suggesting a strictly modulated dynamic primary triggering rate $\lambda_{tot}(t; z) = \lambda_{dyn}(t; z) + \nu_{DCR} \frac{M - N_{busy}(t; z)}{M}$. For $\nu_{DCR} \sim 1$~MHz and a $\sim 1~\mu\mathrm{s}$ recorded waveform window, the expected dark-count contribution is on the order of one avalanche per waveform, while scintillation-driven firings are on the order of thousands for a 511~keV event. The availability modulation of the dark-count term is therefore small relative to the scintillation-driven rate, although it is not identically zero. We therefore approximate $\lambda_{tot}(t; z) \approx \lambda_{dyn}(t; z) + \nu_{DCR}$ (utilizing the dynamic rate $\lambda_{dyn}(t; z)$ rather than the static $\lambda(t; z)$ to capture primary variance suppression caused by optical photon microcell starvation in deep pileups) when evaluating the variance envelope $\sigma_I^2(t; z)$.

To incorporate the compound variance scaling induced by correlated secondary avalanches (e.g., optical crosstalk), we analyze the statistical cascade distribution, encapsulated in the Excess Noise Factor ($F_{ENF}$). It is essential to map the compound variance scaling strictly to the generalized Campbell's formula. While deterministic expectations linearly scale with the mean cascade gain via the equivalent macroscopic amplitude $I_{cell}^{macro} = \langle G \rangle I_{cell}^{single}$ (where $G$ denotes the discrete cascade size per primary triggering event), the variance of a compound point-process inherently depends on the second moment of the cascaded avalanche amplitude $\mathbb{E}[(G \cdot I_{cell}^{single})^2]$. Utilizing the definition of the Excess Noise Factor $F_{ENF} = \frac{\mathbb{E}[G^2]}{\langle G \rangle^2}$, this second moment gives $\mathbb{E}[G^2] (I_{cell}^{single})^2 = F_{ENF} \langle G \rangle^2 (I_{cell}^{single})^2 = F_{ENF} (I_{cell}^{macro})^2$.

For a recursive branching Poisson process, the total number of avalanches $G$ triggered by a single primary photoelectron can be modeled via a Borel probability distribution~\cite{Tanner1961Derivation, Vinogradov2012Analytical}. If each avalanche independently triggers a Poisson-distributed number of secondary avalanches with mean branching parameter $P_{ct}$, the probability mass function for the total cascade size of $\ell_G \ge 1$ avalanches is given by:
\begin{equation}
    P(G=\ell_G) = \frac{(\ell_G P_{ct})^{\ell_G-1} e^{-\ell_G P_{ct}}}{\ell_G!}
    \label{eq:borel_pmf}
\end{equation}
Provided the mathematical stability boundary $P_{ct} < 1$ holds, the generating function of the Borel distribution yields an expected cascade mean $\langle G \rangle = 1/(1-P_{ct})$ and a variance of $\operatorname{Var}(G) = P_{ct}/(1-P_{ct})^3$. This unbounded Borel branching distribution should be interpreted as a mean-field approximation: it assumes an effectively unlimited reservoir of available microcells for crosstalk cascades and neglects nearest-neighbor spatial correlations and physical array boundaries. In real arrays, crosstalk propagates primarily to adjacent cells, which may already be refractory during high-rate events, thereby truncating the cascade and compressing the empirical $F_{ENF}$ below the Borel prediction. For operating conditions where $P_{ct} \lesssim 15\%$, the error of this unbounded approximation is expected to remain small; the numerical examples in Section~\ref{sec:verification} intentionally retain $P_{ct} = 0.20$ as a realistic near-edge stress test of the noise model. The resulting $F_{ENF}$ should therefore be interpreted as a near-edge, mildly conservative estimate rather than as a precision crosstalk law. Under the unbounded Borel approximation, the theoretical noise scale factor evaluates to $F_{ENF} = \frac{\mathbb{E}[G^2]}{\langle G \rangle^2} = 1 + \frac{\operatorname{Var}(G)}{\langle G \rangle^2} = \frac{1}{1-P_{ct}}$ (Appendix~\ref{app:borel_enf}).

Conversely, a simplified non-recursive additive geometric chain model (which constrains avalanches to a non-branching sequential cascade) yields $F_{ENF} = 1+P_{ct}$~\cite{Gallego2013Modeling}. The relationship $F_{ENF} \approx 1 + P_{ct}$ operates as an exact match for the non-branching geometric model alone, and serves as a converging first-order Taylor approximation for the true recursive Borel branching model restricted to domains where $P_{ct} \ll 1$. To maintain broad theoretical validity without imposing specific avalanche topological constraints, we synthesize these multiplicative cascade effects within a generalized $F_{ENF}$ variable. Applying the non-stationary Campbell theorem to the filtered primary-event process, and absorbing the independent cascade marks into the second-moment factor $F_{ENF}$ through $\mathbb{E}[(G I_{cell}^{single})^2] = F_{ENF}(I_{cell}^{macro})^2$ (Appendix~\ref{app:campbell_extension})~\cite{Rice1944Mathematical, Snyder1972Filtering}:
\begin{align}
    \sigma_I^2(t; z) & \approx F_{ENF} \int_{-\infty}^{t} \lambda_{tot}(t_j; z) \nonumber                         \\
                     & \quad \times \left[ I_{cell}^{macro} \sum_{x} A_x e^{-\frac{t-t_j}{\tau_x}} \right]^2 dt_j
    \label{eq:poisson_noise}
\end{align}

Linearly separating the pre-event steady-state baseline noise power $\sigma_{base}^2$ generated by $\nu_{DCR}$ permits analytical treatment. Expanding the square of the summed exponential components introduces dual sum indices $x$ and $y$:
\begin{equation}
    \sigma_{base}^2 = F_{ENF} \nu_{DCR} \int_{-\infty}^t \left[ I_{cell}^{macro} \sum_{x} A_x e^{-\frac{t-t_j}{\tau_x}} \right]^2 dt_j
    \label{eq:sigma_base_expand}
\end{equation}
Computing this bounded quadratic integral expands the macroscopic bi-exponential covariance. By mapping the integration boundaries relative to the impulse trigger ($u = t-t_j$) and evaluating the algebraically converging spatial limits (demonstrated formally in Appendix~\ref{app:integral_variance}; an explicit numerical expansion of the individual cross-terms confirming that the fast discharge pole dominates the baseline variance is given in Appendix~\ref{app:baseline_crossterms}), the baseline variance simplifies to an exact continuous scaling sum:
\begin{equation}
    \sigma_{base}^2 = F_{ENF} \nu_{DCR} \left(I_{cell}^{macro}\right)^2 \sum_{x,y} \frac{A_x A_y \tau_x \tau_y}{\tau_x + \tau_y}
    \label{eq:sigma_base}
\end{equation}

Adding the transient dynamic component, the total instantaneous variance tracks as:
\begin{align}
    \sigma_I^2(t; z) & = \sigma_{base}^2 + F_{ENF} \left(I_{cell}^{macro}\right)^2 \nonumber                                                                      \\
                     & \quad \times \int_{-\infty}^{t} \lambda_{dyn}(t_j; z) \sum_{x,y} A_x A_y e^{-(t-t_j)\left(\frac{1}{\tau_x} + \frac{1}{\tau_y}\right)} dt_j
    \label{eq:sigma_total}
\end{align}

Defining a composite decay constant for the squared impulse response $\tau_{xy} = \left(\frac{1}{\tau_x} + \frac{1}{\tau_y}\right)^{-1}$, the integral collapses into a convolution:
\begin{align}
    \sigma_I^2(t; z) & = \sigma_{base}^2 + F_{ENF} \left(I_{cell}^{macro}\right)^2 \nonumber                                    \\
                     & \quad \times \sum_{x,y} A_x A_y \left( \lambda_{dyn}(\cdot; z) * e^{-\frac{\cdot}{\tau_{xy}}}u(\cdot) \right)(t)
    \label{eq:sigma_conv}
\end{align}

Campbell's formula assumes a Poisson point process characterized by statistically independent events. In the low-energy linear regime, microcell firings are largely independent, making this application accurate. However, once the instantaneous occupancy becomes appreciable (heuristically, $\alpha n_{ph} \gtrsim 0.1$--$0.2$, corresponding to about 10--18\% occupancy in the static closure), microcell dead-time introduces a deterministic refractory period that transitions the temporal avalanche sequence from an independent Poisson process into a deterministically correlated sub-Poissonian renewal process. This dead-time acts as an anti-bunching mechanism, suppressing the physical variance below the independent Poisson prediction~\cite{Ramilli2010Photon}. Consequently, the unconstrained moment-matching variance derived here acts as an analytical upper bound for timing jitter under strong pileup conditions.

Furthermore, this baseline formulation omits delayed correlated noise components, specifically after-pulsing. Given typical modern SiPM after-pulsing probabilities extending around $1\%$--$5\%$ with subsequent microsecond-scale physical trap delays~\cite{Piemonte2019Overview}, their omission primarily neglects a heavy-tailed delayed variance structure that offsets macroscopic continuous integration baselines, thereby minimally impacting the critical sub-nanosecond primary leading-edge triggering jitter essential for ultra-fast coincidence timing.

Finally, while this derivation encapsulates the physical intrinsic avalanche generation fluctuations, the total macroscopic temporal variance observable in practical scintillation front-end digitizers must superimpose an independent continuous electronic variance baseline $\sigma_{elec}^2$. This captures the Gaussian thermal and transimpedance amplifier noise spectra, so that the comprehensive instrument timing variance is an additive closure: $\sigma_{total}^2(t; z) = \sigma_I^2(t; z) + \sigma_{elec}^2$.

\subsection{Physical Limits of Time-of-Flight Resolution}
Building upon the derived continuous expectation models and the generalized non-stationary Poisson variance envelope, we analytically estimate lower bounds on Time-of-Flight (ToF) resolution.

In conventional scintillation timing systems utilizing Leading-Edge Discrimination (LED), an architecture-dependent timing estimate can be obtained from first-order error propagation, provided the threshold-crossing fluctuations remain small and the mean waveform has a non-zero local slope. At a triggering threshold $I_{th}$ and threshold-crossing time $t_{th}$ defined by $I_{dyn}(t_{th}; z) = I_{th}$, the LED timing variance is approximated by:
\begin{equation}
    \sigma_{t, LED}^2(t_{th}; z) \approx \frac{\sigma_{total}^2(t_{th}; z)}{\left( \left. \frac{\partial I_{dyn}(t; z)}{\partial t} \right|_{t=t_{th}} \right)^2}
    \label{eq:sigma_led}
\end{equation}
The corresponding coincidence timing resolution estimate for a pair of identical detectors is therefore $\mathrm{CTR} \approx 2.355 \sqrt{2 \sigma_{t, LED}^2(t_{th}; z)}$.

Equation \eqref{eq:sigma_led} summarizes the engineering trade-offs: minimizing timing variance requires maximizing the macroscopic current gradient while simultaneously suppressing early variance. However, the framework shows that this gradient is limited by the front-end physics. As demonstrated in \eqref{eq:ymod_limit} and the Gaussian TTS convolutions, the initial macroscopic current growth is bounded by the intrinsic thermalization cascade $\tau_r$ and the spatial optical dispersion $\sigma_{TTS}(z)$. Because the continuous bi-exponential generation mechanism enforces $Y_{mod}(0^+) = 0$ while retaining a finite initial derivative $\left.\partial_t Y_{mod}\right|_{0^+}=C_{gen}(\tau_r^{-1}-\tau_{eff}^{-1})$ in the non-degenerate case, the leading edge cannot jump to a finite amplitude. This shows that arbitrarily lowering thresholds does not unconditionally optimize timing jitter, as very early crossings sample a small signal level with limited local slew rate after TTS and readout smoothing.

Furthermore, while optical crosstalk ($P_{ct}$) marginally amplifies the macroscopic current and its gradient, the intrinsic variance formulations in \eqref{eq:sigma_base} and \eqref{eq:sigma_total} show that, under the unbounded Borel approximation of Section~\ref{sec:poisson_noise_modeling}, the Excess Noise Factor $F_{ENF} = 1/(1-P_{ct})$ directly scales the Poisson fluctuations of the entire avalanche cascade. This correlated variance penalty counteracts the steepened slope, establishing a physical boundary on leveraging geometric crosstalk for SPTR (Single Photon Time Resolution) enhancement.

To determine an ideal event-level lower bound on ToF resolution---independent of specific discriminator architectures and in the idealized limit of negligible readout electronics noise ($\sigma_{elec}^2 \to 0$)---we formulate the Cram\'er-Rao Lower Bound (CRLB) directly from the Poisson statistics governing individual microcell avalanche triggering events~\cite{Seifert2012Lower}. This lower bound treats the interaction time as the sole unknown parameter while the forward-model waveform parameters are assumed known.

Within the mean-field framework developed above, and conditional on fixed DOI $z$ and known detector parameters, the sequence of SiPM microcell firings is modeled as a non-homogeneous Poisson point process with deterministic intensity $\lambda_{dyn}(t; z)$ as derived in \eqref{eq:dynamic_ode}. The arrival-time parameter to be estimated is the gamma-ray interaction instant $t_{int}$, which shifts the entire rate function: $\lambda_{dyn}(t; z) \to \lambda_{dyn}(t - t_{int}; z)$. Including the irreducible dark count rate $\nu_{DCR}$ (independent of $t_{int}$), the total observable rate is $\lambda_{tot}(t; z) = \lambda_{dyn}(t - t_{int}; z) + \nu_{DCR}$. For this non-homogeneous Poisson shift model, the exact Fisher Information for estimating $t_{int}$ is derived from the Poisson likelihood in Appendix~\ref{app:fisher_info_derivation}~\cite{Snyder1991Random}:
\begin{equation}
    \mathcal{I}(z) = \int_{0}^{\infty} \frac{\left[\dot{\lambda}_{dyn}(t; z)\right]^2}{\lambda_{dyn}(t; z) + \nu_{DCR}} \, dt
    \label{eq:fisher_info}
\end{equation}
where $\dot{\lambda}_{dyn} \equiv \partial\lambda_{dyn}/\partial t$. The absence of the single-photoelectron response kernel $i_{ser}(t)$ from \eqref{eq:fisher_info} is deliberate rather than accidental: \eqref{eq:fisher_info} is the Fisher Information of the underlying avalanche-event time sequence under an ideal point-process observation model, in which the estimator is assumed to access the triggering times themselves and therefore depends only on the shifted intensity $\lambda_{dyn}(t-t_{int}; z)$. In the layered formulation of Section~II-D, $i_{ser}(t)$ acts only after those triggering events through the convolution $I_{dyn} = \lambda_{dyn} * i_{ser}$, so it shapes the analog current waveform and the filtered shot-noise covariance without altering the point-process intensity itself. A waveform-observation extension, in which $i_{ser}(t)$ enters both the mean waveform and the covariance kernel, is derived in Appendix~\ref{app:waveform_fisher}. The present CRLB should therefore be interpreted as an event-level photon-counting bound rather than a full analog front-end bound. This result requires no Gaussian approximation and is exact for the stated inhomogeneous Poisson shift model; under deep saturation, where microcell dead-time introduces sub-Poissonian correlations (see Section~\ref{sec:discussion}), the CRLB should be interpreted as an approximate lower bound whose tightness degrades with increasing occupancy. Importantly, the photon arrival rate $r_{ph}(t; z)$ entering the ODE \eqref{eq:dynamic_ode} must incorporate the TTS convolution \eqref{eq:rph_tts}: the finite rise-time/TTS smoothing softens the leading-edge onset of $\lambda_{dyn}$, and together with the dark-count floor $\nu_{DCR}$ prevents the ultra-early Fisher integrand from becoming singular.

For any unbiased single-detector estimator of $t_{int}$, the timing variance is lower-bounded by the inverse Fisher Information. We denote this bound by:
\begin{equation}
    \sigma_{CRLB}^2(z) = \mathcal{I}^{-1}(z) = \biggl[ \int_{0}^{\infty} \frac{\dot{\lambda}_{dyn}^2(t; z)}{\lambda_{dyn}(t; z) + \nu_{DCR}} \, dt \biggr]^{\!-1}
    \label{eq:crlb_bound}
\end{equation}
For a coincidence measurement between two identical, statistically independent detectors, the combined timing variance doubles: $\sigma_{coinc}^2(z) = 2\sigma_{CRLB}^2(z)$. Converting the corresponding timing standard deviation to the experimentally reported equivalent-Gaussian Full Width at Half Maximum (FWHM) via the standard relation $\mathrm{FWHM} = 2\sqrt{2\ln 2}\,\sigma \approx 2.355\,\sigma$ (derived in Appendix~\ref{app:ctr_crlb_derivation}), the event-level coincidence CTR bound is:
\begin{equation}
    \mathrm{CTR}_{CRLB}(z) = 2\sqrt{2\ln 2} \cdot \sqrt{2\sigma_{CRLB}^2(z)} = 2\sqrt{2\ln 2} \cdot \sqrt{\frac{2}{\mathcal{I}(z)}}
    \label{eq:ctr_crlb}
\end{equation}
This expression makes explicit the analytical chain: the Poisson Fisher Information integral $\mathcal{I}(z)$ from \eqref{eq:fisher_info}---which includes the scintillation kinetics, TTS, dynamic saturation, and dark count rate---sets the corresponding coincidence timing bound.

In this model, the ToF lower bound is governed by how rapidly the TTS-convolved dynamic triggering rate rises against the Poisson counting noise floor. For the representative LYSO-SiPM parameter set evaluated in Section~\ref{sec:val_crlb}, the Fisher Information integrand is concentrated within the first 0.5~ns of the rising edge, where the signal gradient $\dot{\lambda}_{dyn}$ is largest relative to the accumulated rate $\lambda_{dyn}$. By incorporating TTS, saturation, and dark counts into this formulation, the framework provides a convergent timing-performance bound for scintillator-SiPM configurations within the stated assumptions.

The same point-process sensitivity construction also provides an energy-resolution bound when the time-shift derivative $\dot{\lambda}_{dyn}$ is replaced by the deposited-energy derivative $\partial_E\lambda_{dyn}$. Appendix~\ref{app:energy_resolution_fisher} derives this companion bound for an arbitrary reference energy $E_0$ and separates the ideal primary-trigger limit from charge- and waveform-observation limits.

\section{Numerical Verification}
\label{sec:verification}
To verify the mathematical consistency of the derived analytical reductions and the linear-regime closed-form solutions, we perform numerical checks using a physically grounded, representative numerical parameter set for a LYSO:Ce crystal coupled to a SiPM.

\textit{Scintillator parameters.} The intrinsic decay constant $\tau = 40$~ns and thermalization rise time $\tau_r = 70$~ps are representative of standard LYSO:Ce crystals~\cite{Lecoq2017Pushing, Gundacker2016High}. The self-absorption product $\epsilon a = 0.12$ and re-emission quantum yield $\eta = 0.83$ yield an effective decay constant $\tau_{eff} = \tau/(1 - \epsilon a \eta) = 44.4$~ns, consistent with the experimentally observed $40$--$50$~ns range for Ce$^{3+}$-activated orthosilicates. The initial scintillation photon number $N_0 = 16000$ corresponds to the LYSO light yield of $\sim$30~photons/keV at 511~keV~\cite{Schaart2021Time}.

\textit{Optical transport parameters.} The crystal-to-SiPM light transfer efficiency $k_{trans} = 0.40$ accounts for solid-angle coverage, surface reflectivities, and coupling-medium losses in a $3.9 \times 3.9 \times 20~\mathrm{mm}^3$ wrapped crystal geometry. The mean optical transit time $\mu_{TTS} = 150$~ps and its standard deviation $\sigma_{TTS} = 40$~ps are representative values for the depth-of-interaction-dependent photon travel time spread in elongated orthosilicate crystals~\cite{Gundacker2020Experimental, Schaart2021Time}.

\textit{SiPM parameters.} A representative SiPM parameter set is adopted with $M = 7100$ microcells, a photon detection efficiency $q = 0.50$ at 420~nm, a microcell RC-recharge time constant $\tau_{rec} = 15$~ns, and an optical crosstalk probability $P_{ct} = 0.20$. Although this crosstalk probability sits slightly above the $P_{ct} \lesssim 15\%$ comfort range discussed in Section~\ref{sec:poisson_noise_modeling} for the unbounded Borel approximation, it remains well below the mathematical stability limit $P_{ct} < 1$ and therefore serves as a realistic near-edge stress test of the noise model. The multi-exponential single-cell current response is parameterized following the analytical model of Marano~\textit{et al.}~\cite{marano_accurate_2014}, with a dominant fast discharge pole $\tau_d = 0.5$~ns ($A_d = 0.98$) and two secondary current-response poles at $\tau_{p1} = 1.5$~ns ($A_{p1} = 0.015$) and $\tau_{p2} = 4$~ns ($A_{p2} = 0.005$). These slow-pole time constants satisfy $\tau_x \ll \tau_{eff}$ to avoid resonant amplification of the coupling constant $\tilde{\tau}_{x,eff} = \tau_x \tau_{eff}/(\tau_{eff} - \tau_x)$.

\subsection{Dynamic Microcell Population}
\begin{figure}[htbp]
    \centering
    \includegraphics[width=0.95\columnwidth]{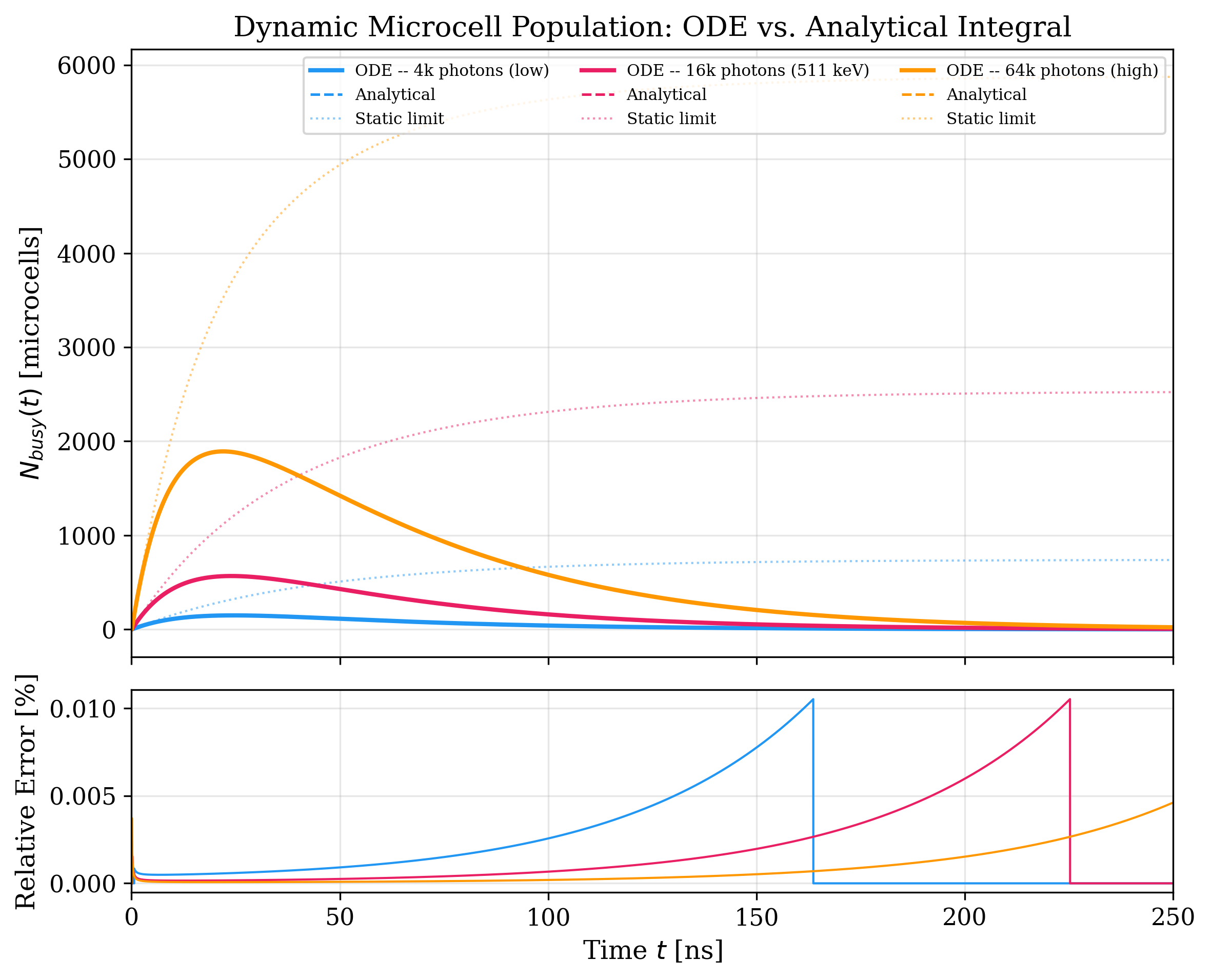}
    \caption{Upper panel: Transient recovering microcell population $N_{busy}(t)$ computed via numerical ODE integration (solid), analytical integral \eqref{eq:analytical_dynamic} (dashed), and the static binomial limit \eqref{eq:nfired} (dotted), for three representative photon loads. Lower panel: Relative error between the analytical and numerical ODE solutions.}
    \label{fig:val_Nbusyt}
\end{figure}

We compare the analytical integral solution for $N_{busy}(t; z)$ derived in \eqref{eq:analytical_dynamic} against direct numerical integration of the governing ODE \eqref{eq:dynamic_rec} using an adaptive Runge-Kutta (RK45) solver with strict tolerance settings ($\mathrm{rtol} = 10^{-11}$, $\mathrm{atol} = 10^{-13}$)~\cite{Dormand1980Family}. The comparison is performed at three representative energy depositions spanning over one order of magnitude: $N_0 = 4000$ ($\sim$130~keV, representative of low-energy Compton scatter), $N_0 = 16000$ ($\sim$511~keV, the PET annihilation photopeak), and $N_0 = 64000$ ($\sim$2~MeV, probing the deep-saturation regime). As shown in Fig.~\ref{fig:val_Nbusyt}, the analytical solution and numerical ODE integration are visually indistinguishable across all three regimes, with maximum relative deviations on the order of $0.01\%$. The relative error (lower panel) exhibits a slow monotonic rise during the recovery tail, attributable to the cumulative truncation error of the trapezoidal quadrature used to evaluate the analytical integral \eqref{eq:analytical_dynamic}; at late times, the error increases sharply as $N_{busy}(t)$ decays toward zero, amplifying the relative metric before the signal becomes physically negligible. At $N_0 = 64000$, the static binomial occupancy model \eqref{eq:nfired} overestimates the instantaneous microcell population by $\sim 86\%$ at the dynamic-population peak, confirming the necessity of the intra-pulse recovery correction for high-energy depositions~\cite{Vinogradov2015SiPM}.

\subsection{Macroscopic Pulse Shapes Under Saturation}
\begin{figure}[htbp]
    \centering
    \includegraphics[width=0.95\columnwidth]{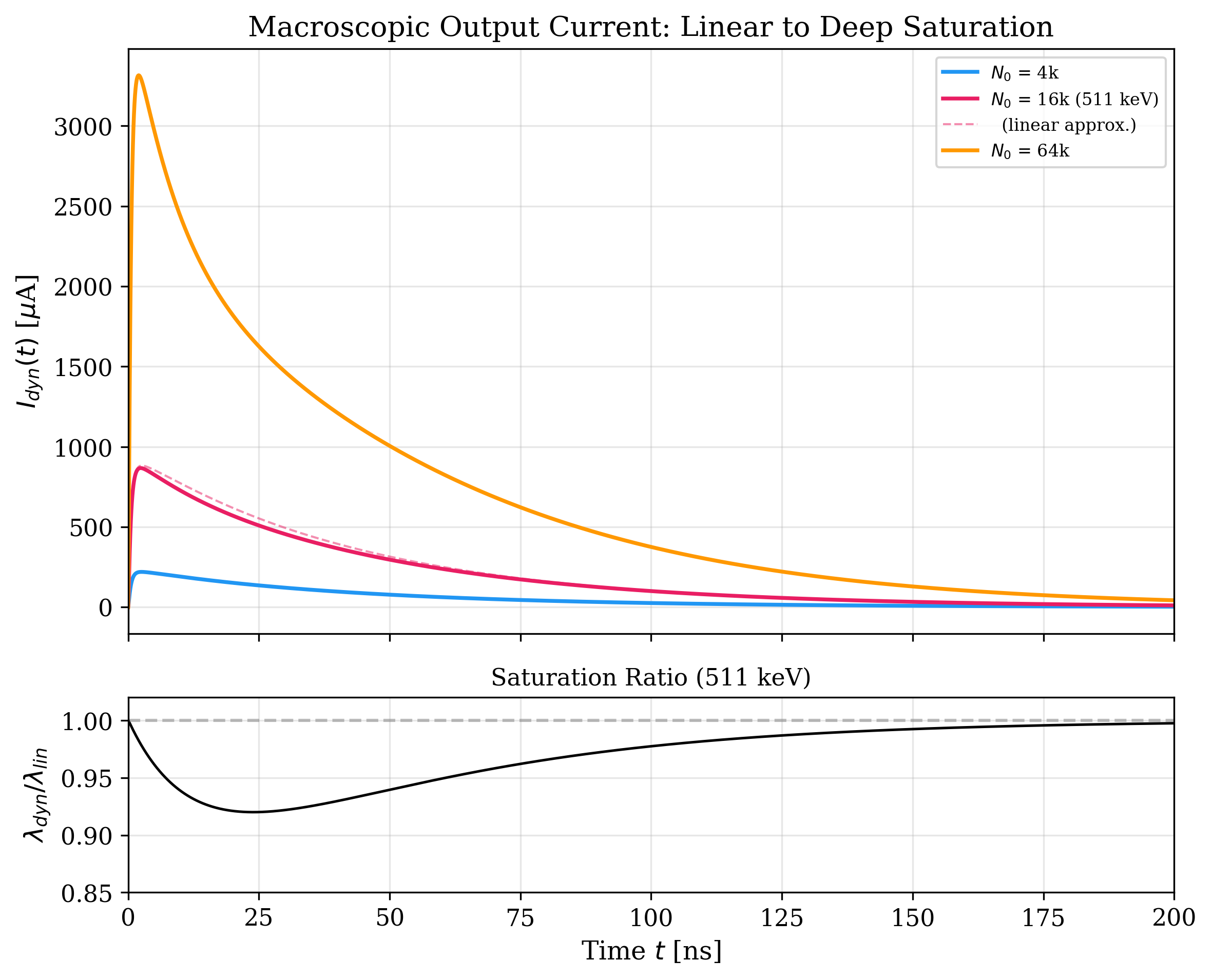}
    \caption{Upper panel: Macroscopic output current $I_{dyn}(t)$ under three photon loads ($N_0 = 4000$, $16000$, $64000$), computed with the intrinsic pre-TTS photon input to isolate saturation. The dashed line for $N_0 = 16000$ shows the corresponding unsaturated linear approximation. Lower panel: Saturation ratio $\lambda_{dyn}/\lambda_{lin}$ for the $511$~keV reference case.}
    \label{fig:val_saturation}
\end{figure}

To illustrate the continuous transition from the unsaturated linear regime to deep detector saturation, we evaluate the macroscopic output current $I_{dyn}(t)$ via numerical convolution of the dynamic triggering rate $\lambda_{dyn}(t)$ with the multi-exponential single-cell response $i_{ser}(t)$. This saturation-only comparison uses the intrinsic pre-TTS photon rate so that microcell depletion is isolated from optical transit-time broadening; the TTS-convolved cascade is examined separately in Fig.~\ref{fig:val_cascade}. Fig.~\ref{fig:val_saturation} displays the resulting pulse family across the same three initial photon populations used above: $N_0 = 4000$, $16000$, and $64000$. At $N_0 = 16000$ (corresponding to $511$~keV), the saturation ratio $\lambda_{dyn}/\lambda_{lin}$ remains close to unity at the photon-rate maximum but later drops to a minimum of approximately $0.92$ during the pulse, indicating a $\sim 8\%$ non-linear compression of the instantaneous triggering rate relative to the linear reference. This mild saturation effect progressively distorts the macroscopic pulse shape, suppressing the peak amplitude and broadening the temporal profile, consistent with the theoretical dead-time analysis in \eqref{eq:dead_phys}.

\subsection{EMG Closed-Form Verification}
\begin{figure}[htbp]
    \centering
    \includegraphics[width=0.95\columnwidth]{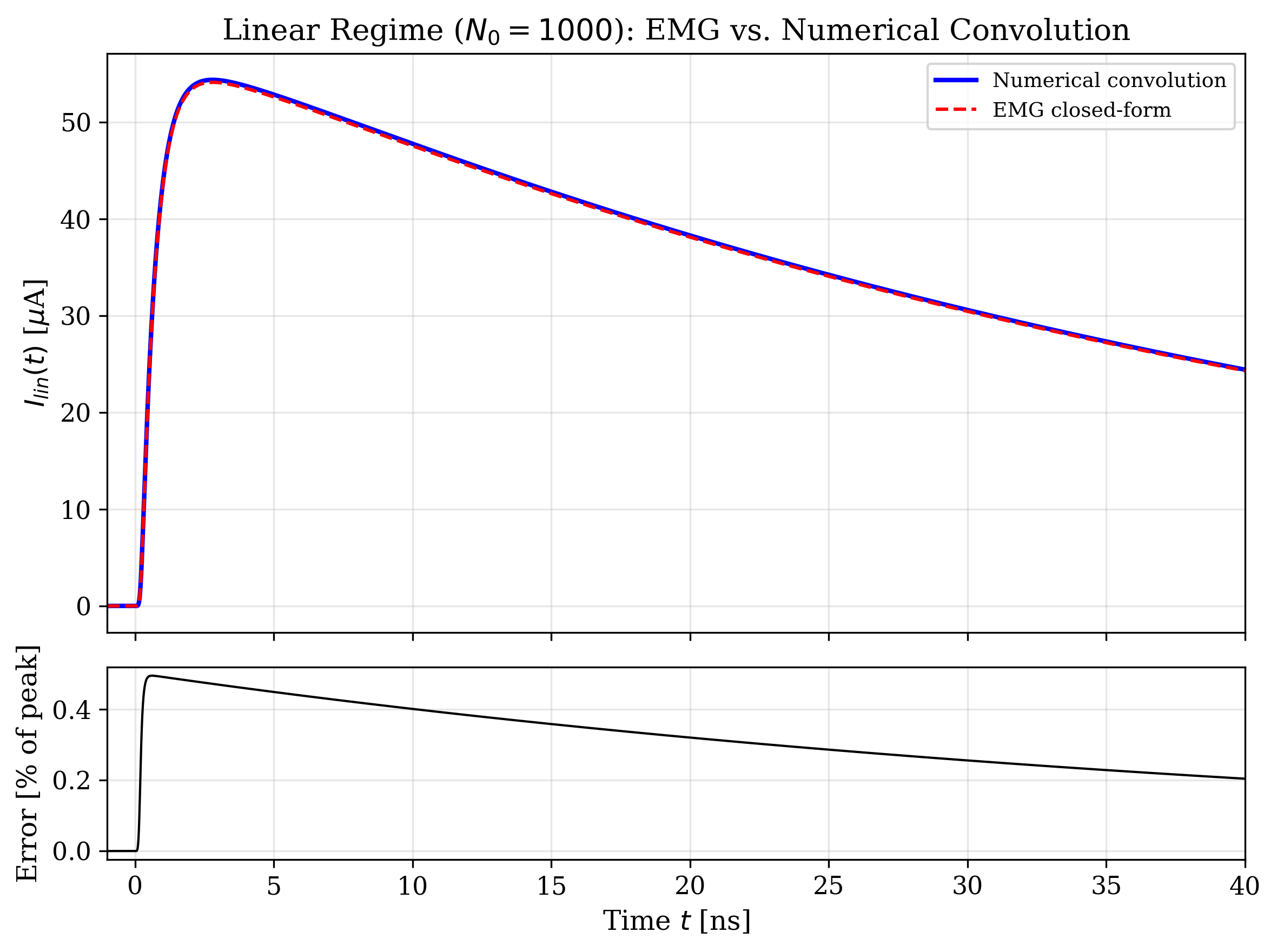}
    \caption{Upper panel: Linear-regime macroscopic current $I_{lin}(t)$ computed via brute-force numerical convolution (solid) and closed-form EMG superposition \eqref{eq:ilin_emg} (dashed). Lower panel: Deviation between the two methods, expressed as a percentage of peak amplitude.}
    \label{fig:val_EMG}
\end{figure}

In the linear regime ($N_0 = 1000$), we verify the Exponentially Modified Gaussian (EMG) superposition \eqref{eq:ilin_emg} against brute-force numerical convolution of the intrinsic generation profile $Y_{mod}(t)$, the single-cell response $i_{ser}(t)$, and the Gaussian TTS kernel $f_{TTS}(t; z)$ using the optical transport parameters defined at the beginning of this section ($\mu_{TTS} = 150$~ps, $\sigma_{TTS} = 40$~ps). As shown in Fig.~\ref{fig:val_EMG}, the closed-form EMG expression reproduces the numerical convolution to within $0.5\%$ of peak amplitude across the entire temporal domain. The deviation (lower panel) peaks near the rising edge where the signal gradient is steepest, maximizing the finite-step truncation error of the discrete FFT-based convolution; it then decays monotonically along the exponential tail as the signal varies slowly and the discretization error diminishes proportionally. This confirms that the residual discrepancy originates from the numerical convolution rather than the analytical EMG expression. The result further validates extending the Gaussian TTS integral to $s \to -\infty$ (as justified in Section~II-B), demonstrating that the associated probability leakage into non-physical negative time induces negligible error for practical TTS parameters.

\subsection{Cascade Model Comparison}
\begin{figure*}[htbp]
    \centering
    \includegraphics[width=0.95\textwidth]{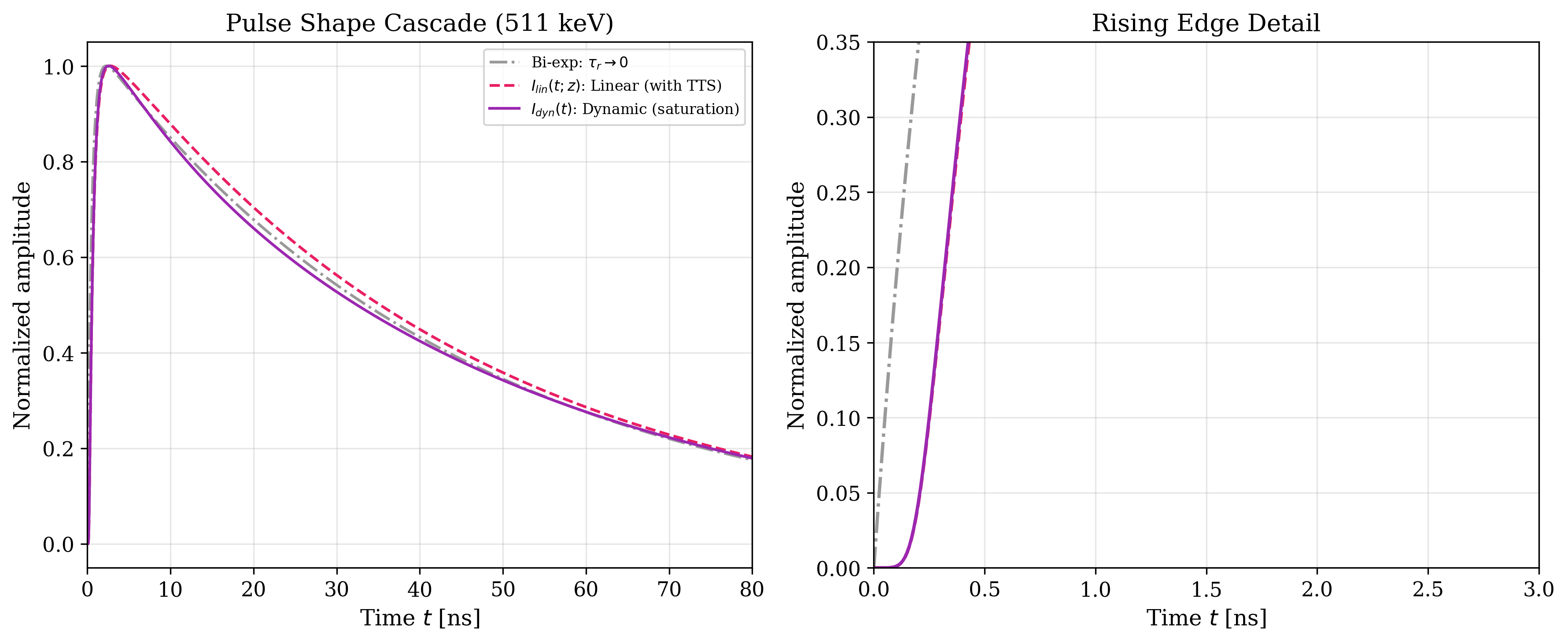}
    \caption{Normalized macroscopic current profiles at 511~keV comparing three model tiers: asymptotic bi-exponential \eqref{eq:double_exp} (gray dash-dot), linear-regime $I_{lin}(t;z)$ with TTS (red dashed), and dynamic $I_{dyn}(t)$ with saturation (purple solid). Left: full profile ($0$--$80$~ns). Right: rising-edge expansion ($0$--$3$~ns).}
    \label{fig:val_cascade}
\end{figure*}

To visualize how each analytical layer modifies the observable pulse shape, Fig.~\ref{fig:val_cascade} overlays three macroscopic current profiles at $N_0 = 16000$ (511~keV), normalized to their respective maxima: (1)~the asymptotic bi-exponential \eqref{eq:double_exp} obtained under the classical limits $\tau_r \to 0$ and dominant-pole approximation, (2)~$I_{lin}(t; z)$ with TTS, and (3)~the full dynamic current $I_{dyn}(t)$ incorporating microcell saturation. The left panel presents the full temporal domain; the right panel expands the first 3~ns to resolve the rising-edge structure.

On the full-domain scale, the bi-exponential approximation (gray dash-dot) closely tracks the full model in both peak position and decay profile, with discrepancies emerging primarily in the 10--40~ns range where the secondary current-response poles ($\tau_{p1}$, $\tau_{p2}$) contribute residual current absent from the single-pole approximation. This agreement supports the dominant-pole reduction derived in \eqref{eq:double_exp} as a useful first-order model. The rising-edge detail (right panel) reveals the key structural differences: the bi-exponential exhibits an instantaneous onset at $t=0$ (because $\tau_r \to 0$ eliminates the thermalization delay), whereas the finite-$\tau_r$ linear model with TTS (red dashed) rises later. The dynamic saturation curve (purple solid) is nearly coincident with the TTS-convolved linear curve on the first nanosecond but departs slightly in the tail, an effect small in the bulk waveform but important for timing resolution as quantified in Section~\ref{sec:val_crlb}.

\subsection{Cram\'er-Rao Timing Bounds}
\label{sec:val_crlb}
To numerically evaluate the timing lower bound predicted by the framework, we compute the Poisson Fisher Information integral \eqref{eq:fisher_info} using the dynamic triggering rate $\lambda_{dyn}(t; z)$ obtained from the ODE \eqref{eq:dynamic_ode} with TTS-convolved photon input. All scintillator and SiPM parameters are identical to those defined at the beginning of Section~\ref{sec:verification}, and with $N_0 = 16000$ (511~keV in LYSO). The dark count rate is set to $\nu_{DCR} = 1$~MHz ($10^{-3}$/ns), a representative value for moderately cooled SiPM operation, consistent with the typical reduction in DCR by approximately a factor of two per $8$--$10\,^{\circ}$C of cooling~\cite{Piemonte2019Overview}. In any case, the DCR contribution to the Fisher Information denominator is small compared to the scintillation-driven rate $\lambda_{dyn} \gg \nu_{DCR}$ during the signal window. The numerical time step ($\Delta t = 2$~ps) is verified to yield a fully converged Fisher Information integral (relative change $<0.1\%$ upon further refinement).

\begin{figure*}[htbp]
    \centering
    \includegraphics[width=0.95\textwidth]{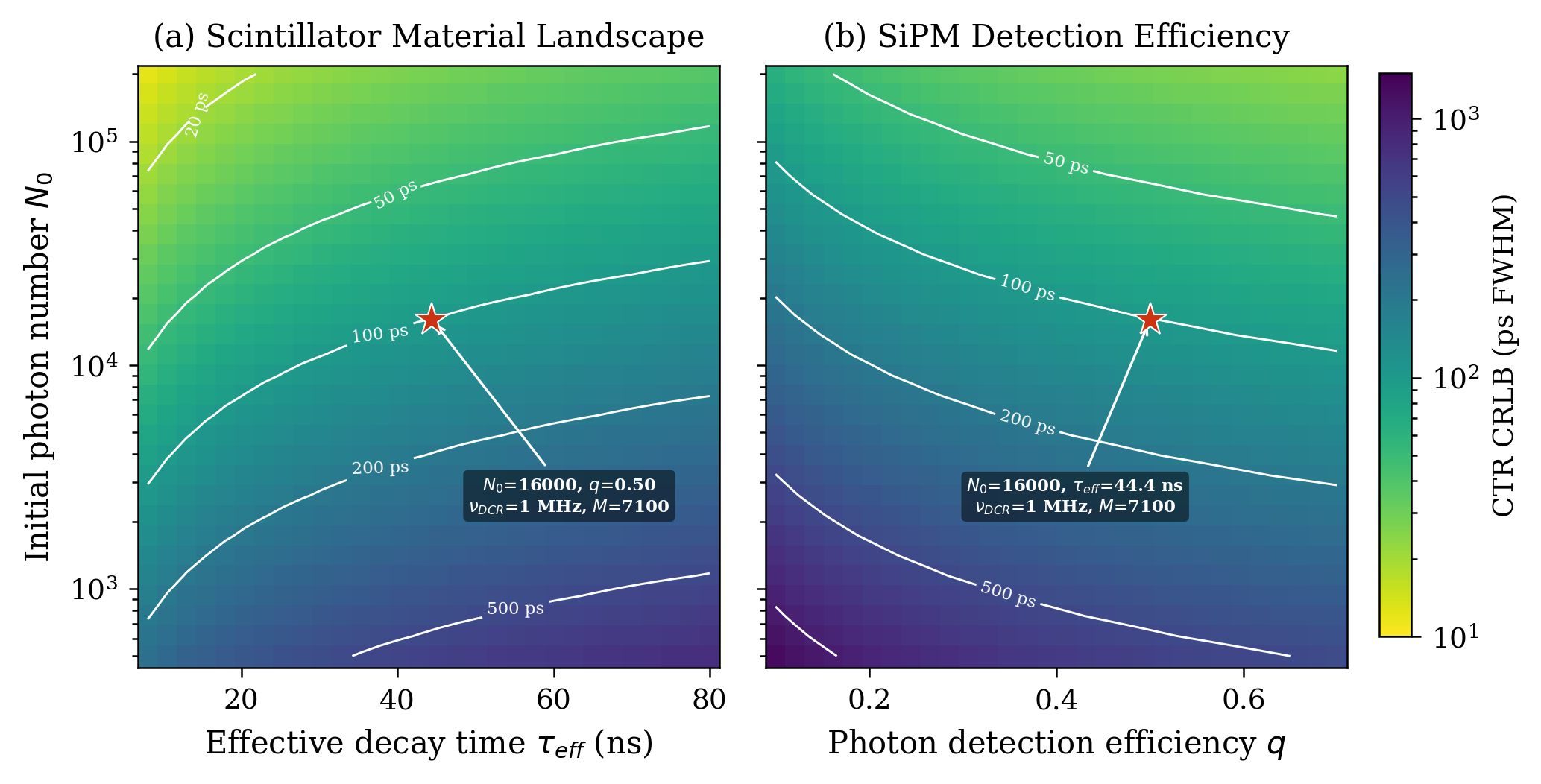}
    \caption{Coincidence CTR CRLB phase diagrams computed via the Poisson Fisher Information \eqref{eq:fisher_info}. (a)~Scintillator material landscape: effective decay time $\tau_{eff}$ vs.\ initial photon number $N_0$. (b)~SiPM detection efficiency landscape: photon detection efficiency $q$ vs.\ $N_0$, at fixed $\tau_{eff} = 44.4$~ns. White contour lines denote constant-CTR iso-surfaces. Red stars mark the LYSO operating point at 511~keV, yielding $\sim$100~ps FWHM.}
    \label{fig:val_CRLB}
\end{figure*}

Fig.~\ref{fig:val_CRLB} presents the coincidence CTR CRLB \eqref{eq:ctr_crlb} as two-dimensional phase diagrams jointly spanning scintillator and SiPM parameter spaces. Panel~(a) maps the ($\tau_{eff}$, $N_0$) landscape: for the 511~keV LYSO reference ($\tau_{eff} = 44.4$~ns, $N_0 = 16000$), the framework predicts a coincidence CTR bound of $\sim$100~ps FWHM (single-detector $\sim$70~ps FWHM), consistent with fundamental limits discussed in the scintillation timing community~\cite{Lecoq2017Pushing, Gundacker2020Experimental}. The contour topology reveals that reducing $\tau_{eff}$ yields diminishing CTR improvements beyond $\tau_{eff} \lesssim 10$~ns, where front-edge dispersion ($\tau_r$, $\sigma_{TTS}$) becomes the dominant bottleneck. Panel~(b) sweeps SiPM photon detection efficiency $q$ at constant $\tau_{eff}$: CTR improves approximately as $q^{-0.5}$ in the high-$N_0$ regime but more steeply at intermediate photon counts, reflecting the interplay between Poisson statistics and detection efficiency. Together, these maps enable direct assessment of the timing performance ceiling for any scintillator-SiPM configuration without Monte Carlo simulation.

\section{Experimental Validation}
\label{sec:exp_validation}

\subsection{Experimental Setup}
\label{sec:exp_setup}
A $^{22}$Na radioactive source emitting 511~keV and 1274~keV gamma rays was used to provide the excitation photons. Their interactions in the source environment, air, and scintillator produce a recorded deposited-energy/pulse-height continuum spanning from near zero up to the 1274~keV line. The detector module comprises a lutetium-yttrium oxyorthosilicate (LYSO) scintillator coupled to a SiPM and read out without external preamplification. The LYSO scintillator has dimensions of $3.9~\mathrm{mm} \times 3.9~\mathrm{mm} \times 20~\mathrm{mm}$. It is optically coupled to a SiPM. The SiPM output waveform---i.e., the macroscopic scintillation current pulse---is digitized directly by a high-bandwidth digital oscilloscope (50~GS/s real-time sampling rate, 16~GHz analog bandwidth) without any intermediate preamplifier stage. Channel~1 is configured with a rising-edge trigger at a threshold of 100~mV. Each acquisition captures a 1000~ns window at 50~GS/s, yielding 50,000 sample points per pulse. A total of 10,000 individual scintillation pulses were recorded across ten sequential \texttt{.wfm} files (1,000 pulses per file). The vertical scale is set to 200~mV/div and the horizontal scale to 100~ns/div. This direct-readout configuration preserves the intrinsic SiPM pulse shape and avoids the additional pole-zero structure that would be introduced by a shaping amplifier. The full 10,000-pulse acquisition serves as the source dataset for this section; the model-comparison statistics reported below are computed on one 100-pulse random sample and two additional 100-pulse subsets selected for high- and medium-amplitude saturation analysis.

\subsection{Validation of the Bi-Exponential Model}
\label{sec:val_biexp}
As derived in Section~II, the asymptotic bi-exponential waveform \eqref{eq:double_exp} emerges from first-principles analysis of the LYSO-SiPM cascade under the unsaturated regime and dominant-pole approximation. It is also the most widely adopted empirical model in scintillation detector engineering. In practice, however, the recorded waveform is shaped not only by the scintillation-SiPM physics but also by back-end electronic artifacts: baseline drift from DC-coupled readout, finite-bandwidth filtering from the oscilloscope front-end and cabling, impedance mismatches at connector interfaces, and parasitic capacitances in the SiPM bond wires and PCB traces. To quantify these effects, we systematically augment the bi-exponential template with additional physical correction terms and evaluate whether the augmented models yield statistically superior fits. Three models of increasing complexity are compared:

\noindent\textbf{1) Bi-exponential model} (5 parameters: $A, t_0, \tau_d^{\mathrm{fit}}, \tau_r^{\mathrm{fit}}, b$). The fitted waveform model is denoted by $I_{biexp}(t)$:
\begin{equation}
    I_{biexp}(t) = A \left( e^{-\frac{t-t_0}{\tau_d^{\mathrm{fit}}}} - e^{-\frac{t-t_0}{\tau_r^{\mathrm{fit}}}} \right) u(t - t_0) + b
    \label{eq:fit_biexp}
\end{equation}
where $A$ is the amplitude, $t_0$ is the pulse onset time, $\tau_d^{\mathrm{fit}}$ and $\tau_r^{\mathrm{fit}}$ are the macroscopic fitted decay and rise time constants, and $b$ is the baseline offset. The superscript ``fit'' distinguishes these macroscopic apparent parameters from the intrinsic physical constants $\tau_d$ (SiPM fast discharge pole, $0.5$~ns) and $\tau_r$ (thermalization rise time, $70$~ps) defined in Section~II.

\noindent\textbf{2) Linear-baseline bi-exponential model} (6 parameters: $A, t_0, \tau_d^{\mathrm{fit}}, \tau_r^{\mathrm{fit}}, k, b_0$). The corresponding fitted waveform $I_{linbase}(t)$ replaces the constant baseline by a linear drift to account for slow DC wander in the readout chain:
\begin{equation}
    I_{linbase}(t) = A \left( e^{-\frac{t-t_0}{\tau_d^{\mathrm{fit}}}} - e^{-\frac{t-t_0}{\tau_r^{\mathrm{fit}}}} \right) u(t - t_0) + kt + b_0
    \label{eq:fit_linbase}
\end{equation}

\noindent\textbf{3) Ringing-corrected model} (11 parameters: $A, t_0, \tau_d^{\mathrm{fit}}, \tau_r^{\mathrm{fit}}, \sigma, k, b_0, A_{ring}, \tau_{ring}, \omega, \phi$). The corresponding fitted waveform $I_{ring}(t)$ first convolves the bi-exponential core with a Gaussian kernel of width $\sigma$ to model the combined effect of TTS and finite electronic bandwidth, then augments it with a damped sinusoidal term capturing the parasitic LC resonance:
\begin{align}
    I_{ring}(t) & = A \left[ \mathcal{E}\!\left(t; t_0, \tau_d^{\mathrm{fit}}, \sigma\right) - \mathcal{E}\!\left(t; t_0, \tau_r^{\mathrm{fit}}, \sigma\right) \right] + kt + b_0 \nonumber \\
                & \quad + A_{ring} \, e^{-\frac{t-t_0}{\tau_{ring}}} \sin\!\left(\omega(t-t_0) + \phi\right) \cdot u(t - t_0)
    \label{eq:fit_ringing}
\end{align}
where
\begin{equation}
    \mathcal{E}(t; t_0, \tau, \sigma) = \frac{1}{2} \exp\!\left(\frac{\sigma^2}{2\tau^2} - \frac{t-t_0}{\tau}\right) \operatorname{erfc}\!\left(\frac{\sigma}{\sqrt{2}\,\tau} - \frac{t-t_0}{\sqrt{2}\,\sigma}\right)
    \label{eq:ecg_kernel}
\end{equation}
is the Exponentially Modified Gaussian (EMG) kernel derived earlier in \eqref{eq:emg_def}. The damped sinusoidal component $A_{ring} \, e^{-(t-t_0)/\tau_{ring}} \sin(\omega(t-t_0) + \phi)$ models the under-damped LC oscillation arising from the series inductance of bond wires and the parasitic junction capacitance of the SiPM at the readout interface~\cite{Corsi2007Modelling}. The complete derivation tracing equations~\eqref{eq:fit_biexp}--\eqref{eq:fit_ringing} from the first-principles linear pulse model~\eqref{eq:ilin_emg}, including the dominant-pole reduction, DOI averaging, and the physical origin of each correction term, is provided in Appendix~\ref{app:fitting_biexp_derivation}.

\begin{figure*}[htbp]
    \centering
    \includegraphics[width=0.95\textwidth]{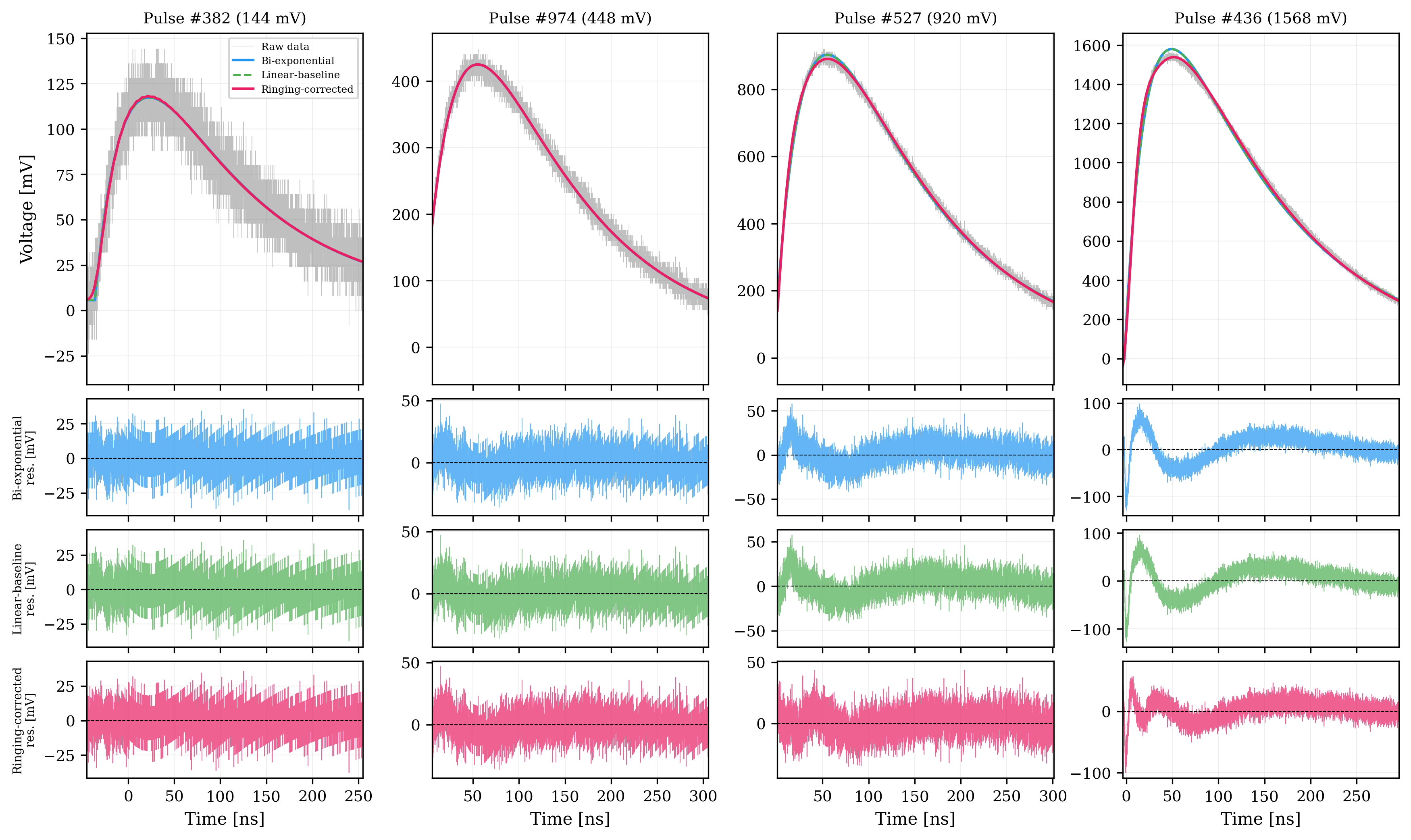}
    \caption{Three-model comparison on four representative scintillation pulses spanning the observed pulse-amplitude range of the Na-22 acquisition (144--1568~mV peak amplitude). Top row: raw waveform (gray) overlaid with bi-exponential (blue), linear-baseline (green dashed), and ringing-corrected (pink) fits. Rows~2--4: separated residuals for each model. The ringing-corrected model produces visibly smaller and more uniform residuals across all amplitude regimes.}
    \label{fig:exp_model}
\end{figure*}

For the three-model comparison, 100 scintillation pulses were randomly sampled from the 10,000-pulse Na-22 source dataset (fixed random seed for reproducibility). Each waveform was independently fitted to all three models using nonlinear least-squares optimization~\cite{Marquardt1963Algorithm}. Goodness-of-fit is quantified by four complementary metrics. Let $r_i = y_i - \hat{y}_i$ denote the residual at sample $i$, $n$ the number of samples, and $p_{\mathrm{fit}}$ the number of free fit parameters. The root-mean-square error, coefficient of determination, Akaike Information Criterion~\cite{Akaike1974New}, and Durbin--Watson statistic~\cite{Durbin1950Testing, Durbin1951Testing} are defined as:
\begin{align}
    \mathrm{RMSE} &= \sqrt{\frac{1}{n}\sum_{i=1}^{n} r_i^2}
    \label{eq:rmse} \\
    R^2 &= 1 - \frac{\sum_{i=1}^{n} r_i^2}{\sum_{i=1}^{n}(y_i - \bar{y})^2}
    \label{eq:r2} \\
    \mathrm{AIC} &= n \ln\!\left(\frac{1}{n}\sum_{i=1}^{n} r_i^2\right) + 2p_{\mathrm{fit}}
    \label{eq:aic} \\
    \mathrm{DW} &= \frac{\sum_{i=2}^{n}(r_i - r_{i-1})^2}{\sum_{i=1}^{n} r_i^2}
    \label{eq:dw}
\end{align}
Lower RMSE and AIC indicate better fit quality; $R^2 \to 1$ indicates complete variance capture within the sampled waveform; $\mathrm{DW} = 2$ corresponds to uncorrelated white-noise residuals, whereas $\mathrm{DW} \ll 2$ signals systematic positive autocorrelation due to unmodeled structure. The results are summarized in Table~\ref{tab:model_comparison}, and representative pulse overlays are shown in Fig.~\ref{fig:exp_model}.

\begin{table}[!t]
    \centering
    \caption{Goodness-of-fit statistics for 100 randomly sampled scintillation pulses. Mean $\pm$ standard deviation are reported where applicable. The AIC-optimal model is recorded per pulse; the ``Best AIC Count'' row shows how many pulses each model wins.}
    \label{tab:model_comparison}
    \begin{tabular}{l@{\hspace{4pt}}c@{\hspace{4pt}}c@{\hspace{4pt}}c}
        \hline
        Metric              & Bi-exponential         & Linear-baseline        & Ringing-corrected               \\
        \hline
        Parameters          & 5                      & 6                      & 11                              \\
        Converged Fits      & 100/100                & 100/100                & 100/100                         \\
        Best AIC Count      & 0                      & 5                      & \textbf{95}                     \\
        \hline
        Mean RMSE           & $1.195 \times 10^{-2}$ & $1.179 \times 10^{-2}$ & $\mathbf{1.089 \times 10^{-2}}$ \\
        Mean $R^2$          & 0.9910                 & 0.9913                 & \textbf{0.9915}                 \\
        Mean AIC            & $-4.515 \times 10^5$   & $-4.521 \times 10^5$   & $\mathbf{-4.593 \times 10^5}$   \\
        Mean DW             & 1.014                  & 1.020                  & \textbf{1.136}                  \\
        \hline
    \end{tabular}
\end{table}

The ringing-corrected model achieves the lowest AIC in 95 out of 100 pulses, indicating that parasitic LC oscillations constitute a measurable systematic artifact in the direct-readout configuration. The improvement in RMSE ($1.09 \times 10^{-2}$ vs.\ $1.19 \times 10^{-2}$, an $\approx 8.5\%$ reduction) and AIC ($\Delta\mathrm{AIC} \approx 7800$) is substantial, and the Durbin--Watson statistic rises from 1.01 to 1.14, indicating reduced residual autocorrelation.

\begin{figure*}[htbp]
    \centering
    \includegraphics[width=\textwidth]{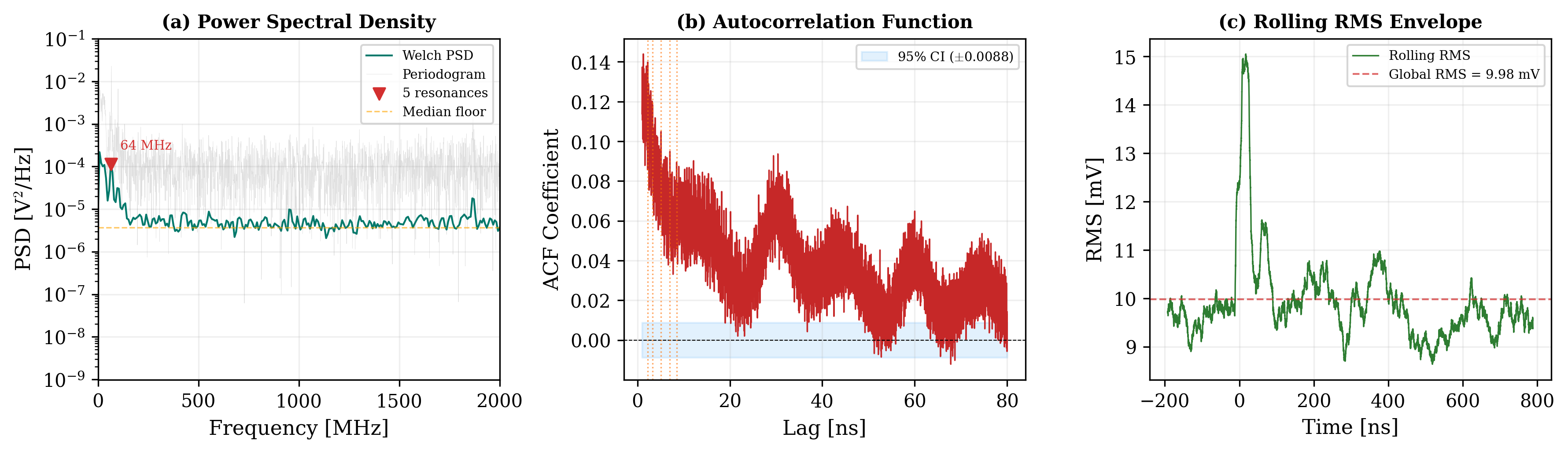}
    \caption{Residual diagnostics of the ringing-corrected bi-exponential model on a representative pulse (\#209, 816~mV peak). (a)~Welch power spectral density identifying a parasitic LC resonance at 64~MHz. (b)~Autocorrelation function showing persistent correlation above the 95\% confidence band. (c)~Rolling RMS envelope revealing non-stationary residual amplitude near the pulse peak.}
    \label{fig:residual_diag}
\end{figure*}

Although the ringing-corrected model provides the best statistical fit, further residual diagnostics (formal metric definitions are provided in Appendix~\ref{app:residual_diagnostics}) reveal that even this 11-parameter model does not fully reduce the residuals to white noise (Fig.~\ref{fig:residual_diag}). Power spectral density (PSD) analysis of the fit residuals (panel~a) identifies a dominant parasitic resonance at $\sim$64~MHz~\cite{Welch1967Use}---approximately 30$\times$ above the median noise floor---attributable to the under-damped LC oscillation formed by the SiPM bond-wire inductance and the junction capacitance at the direct-readout interface~\cite{Corsi2007Modelling}. The autocorrelation function (ACF, panel~b) of the residuals shows persistent correlation well above the 95\% confidence interval out to $\sim$70~ns lag, with structured echo peaks at lags of $\sim$2.2, 3.3, 5.1, and 7.0~ns. Interpreting these echo periods as round-trip reflections with a propagation velocity of $0.18c$ gives centimeter-scale path lengths ($\sim$3--19~cm, depending on which echo is assigned to the round trip), consistent with connector and cable discontinuities in the direct-readout chain~\cite{Paulter2001Assessment}. Finally, the rolling root-mean-square (RMS) envelope (panel~c) reveals pronounced non-stationarity near the pulse peak, where the residual RMS spikes to $\sim$1.5$\times$ the global level; this is expected because the higher instantaneous signal amplitude amplifies the absolute magnitude of multiplicative parasitic artifacts. These diagnostics collectively indicate that back-end electronic effects---parasitic capacitances, impedance mismatches at connector interfaces, and finite-bandwidth filtering---impress systematic, low-level signatures onto the recorded waveform that are not fully absorbed by the 11-parameter phenomenological model.

\subsection{Validation of the Dynamic $I_{dyn}(t)$ with Saturation}
\label{sec:val_saturation}
The preceding analysis demonstrates that back-end electronic artifacts---baseline drift, parasitic LC ringing, and finite-bandwidth filtering---largely dominate the residual structure for a typical scintillation pulse. These effects mask the fine physical differences between linear and saturating detector regimes, which is precisely why the simple bi-exponential template is effective for the majority of pulses across the Na-22 energy spectrum.

However, the dynamic saturation model derived in Section~II predicts that the bi-exponential approximation should progressively deteriorate as the pulse amplitude increases. At high photon flux, the instantaneous microcell occupancy $N_{busy}(t)$ becomes a significant fraction of the total microcell count $M$, causing the effective firing rate $\lambda_{dyn}(t) = \alpha \, r_{ph}(t) \left(M - N_{busy}(t)\right)$ to deviate from its linear-regime counterpart $\lambda_{lin}(t)$. This manifests as a suppressed peak and a subtly modified decay profile compared to the purely exponential tail.

To test this prediction, we construct a \emph{ringing-corrected dynamic model} (12 parameters) that replaces the bi-exponential core with a saturation-modulated pulse:
\begin{equation}
    I_{dyn}(t) = A \bigl[ g(t\!-\!t_0) \cdot \tfrac{M - N_{busy}(t)}{M} \bigr] \!\otimes\! G_\sigma + kt + b_0 + R_{ring}(t)
    \label{eq:fit_dynamic}
\end{equation}
where $g(t) = (e^{-t/\tau_d^{\mathrm{fit}}} - e^{-t/\tau_r^{\mathrm{fit}}})\,u(t)$ is the linear bi-exponential template, $\otimes G_\sigma$ denotes Gaussian convolution with width $\sigma$, and $N_{busy}(t)$ is governed by the recovery ODE
\begin{equation}
    \frac{dN_{busy}}{dt} = \beta \, g(t - t_0) \left(M - N_{busy}\right) - \frac{N_{busy}}{\tau_{rec}}
    \label{eq:ode_fit}
\end{equation}
with $\beta$ the saturation coupling strength (the sole additional parameter compared to the ringing-corrected model; units are inverse time when $g$ is dimensionless). The additive ringing term is $R_{ring}(t)=A_{ring}e^{-(t-t_0)/\tau_{ring}}\sin[\omega(t-t_0)+\phi]u(t-t_0)$, identical to the damped sinusoidal component in~\eqref{eq:fit_ringing}. The complete derivation tracing \eqref{eq:ode_fit} from the first-principles dynamic recovery ODE~\eqref{eq:dynamic_ode}, including all intermediate simplifications and their physical justification, is provided in Appendix~\ref{app:fitting_ode_derivation}.

\textbf{Fitting procedure.}
Because the 12-parameter dynamic model couples a stiff nonlinear ODE with damped-sinusoidal corrections, the optimization landscape contains numerous local minima. We therefore adopt a staged initialization strategy that progressively increases model complexity:

\begin{enumerate}
    \item \emph{Preprocessing.} Each 50{,}000-point waveform (50~GS/s $\times$ 1{,}000~ns) is uniformly downsampled by a factor of~25 to 2{,}000 points ($\Delta t \approx 0.5$~ns) for the nonlinear least-squares iterations. Goodness-of-fit statistics (RMSE, AIC, DW) are subsequently evaluated on the \emph{full-resolution} waveform to preserve high-frequency residual information.
    \item \emph{Progressive initialization.} The parameter vector is bootstrapped through four successive fits of increasing complexity: plain bi-exponential (5~params) $\to$ linear-baseline bi-exponential (6~params) $\to$ Gaussian-convolved bi-exponential (7~params) $\to$ ringing-corrected model~\eqref{eq:fit_ringing} (11~params). At each stage, the converged parameters seed the next model, ensuring that the final 11-parameter fit starts near the global basin of attraction.
    \item \emph{ODE integration.} The microcell recovery ODE~\eqref{eq:ode_fit} is integrated via a classical fourth-order Runge--Kutta (RK4) scheme at the downsampled time step $\Delta t \approx 0.5$~ns. This yields fourth-order truncation accuracy ($\mathcal{O}(\Delta t^4)$) at the adopted temporal resolution; the lower-order forward Euler scheme ($\mathcal{O}(\Delta t)$) produces appreciable discretization error under these conditions.
    \item \emph{Multi-start $\beta$ initialization.} The 12-parameter dynamic model is fitted five times with $\beta$ initialized at $\{0.05, 0.2, 0.5, 1.0, 2.0\}$ (all other parameters seeded from the converged 11-parameter ringing-corrected fit). The solution yielding the lowest residual sum of squares is retained. This multi-start strategy largely eliminates convergence failures attributable to the $\beta$-dependent nonlinearity.
    \item \emph{Convergence filter.} A pulse is classified as a convergence failure and excluded from aggregate statistics if the dynamic model's RMSE exceeds $1.5\times$ that of the ringing-corrected model, indicating that the optimizer settled in a pathological local minimum rather than the physically meaningful basin. Failed or timed-out candidates are replaced until each reported cohort contains 100 converged pulse fits.
\end{enumerate}

For the saturation comparison, this model was fitted alongside the ringing-corrected bi-exponential to two targeted 100-pulse cohorts drawn from the source dataset: (i) the 100 highest-amplitude pulses, and (ii) 100 medium-amplitude pulses centered on the spectral median. The results are summarized in Table~\ref{tab:saturation_comparison}.

\begin{table}[!t]
    \centering
    \caption{Goodness-of-fit comparison between the ringing-corrected bi-exponential (11 parameters) and the dynamic saturation model (12 parameters) on high-amplitude ($\sim$1584--1816\,mV) and medium-amplitude ($\sim$736--768\,mV) 100-pulse cohorts drawn from the source acquisition.}
    \label{tab:saturation_comparison}
    \begin{tabular}{lcc}
        \hline
        Metric         & Ringing-corrected      & Dynamic saturation              \\
        \hline
        Parameters     & 11                     & 12                              \\
        \hline
        \multicolumn{3}{l}{\textit{High-amplitude cohort (100 pulses)}}            \\
        \hline
        Converged Fits & 100/100                & 100/100                         \\
        Best AIC count & 0                      & \textbf{100}                    \\
        Mean RMSE      & $1.383 \times 10^{-2}$ & $\mathbf{9.531 \times 10^{-3}}$ \\
        Mean $R^2$     & 0.9991                 & \textbf{0.9996}                 \\
        Mean DW        & 0.594                  & \textbf{1.220}                  \\
        Mean AIC       & $-4.285 \times 10^5$   & $\mathbf{-4.653 \times 10^5}$   \\
        Mean $\beta$ [ns$^{-1}$] & ---          & $0.052 \pm 0.007$               \\
        \hline
        \multicolumn{3}{l}{\textit{Medium-amplitude cohort (100 pulses)}}          \\
        \hline
        Converged Fits & 100/100                & 100/100                         \\
        Best AIC count & 2                      & \textbf{98}                     \\
        Mean RMSE      & $9.878 \times 10^{-3}$ & $\mathbf{9.517 \times 10^{-3}}$ \\
        Mean $R^2$     & 0.9979                 & \textbf{0.9980}                 \\
        Mean DW        & 1.141                  & \textbf{1.227}                  \\
        Mean AIC       & $-4.618 \times 10^5$   & $\mathbf{-4.654 \times 10^5}$   \\
        Mean $\beta$ [ns$^{-1}$] & ---          & $0.023 \pm 0.009$               \\
        \hline
    \end{tabular}
\end{table}

\begin{figure*}[htbp]
    \centering
    \includegraphics[width=0.95\textwidth]{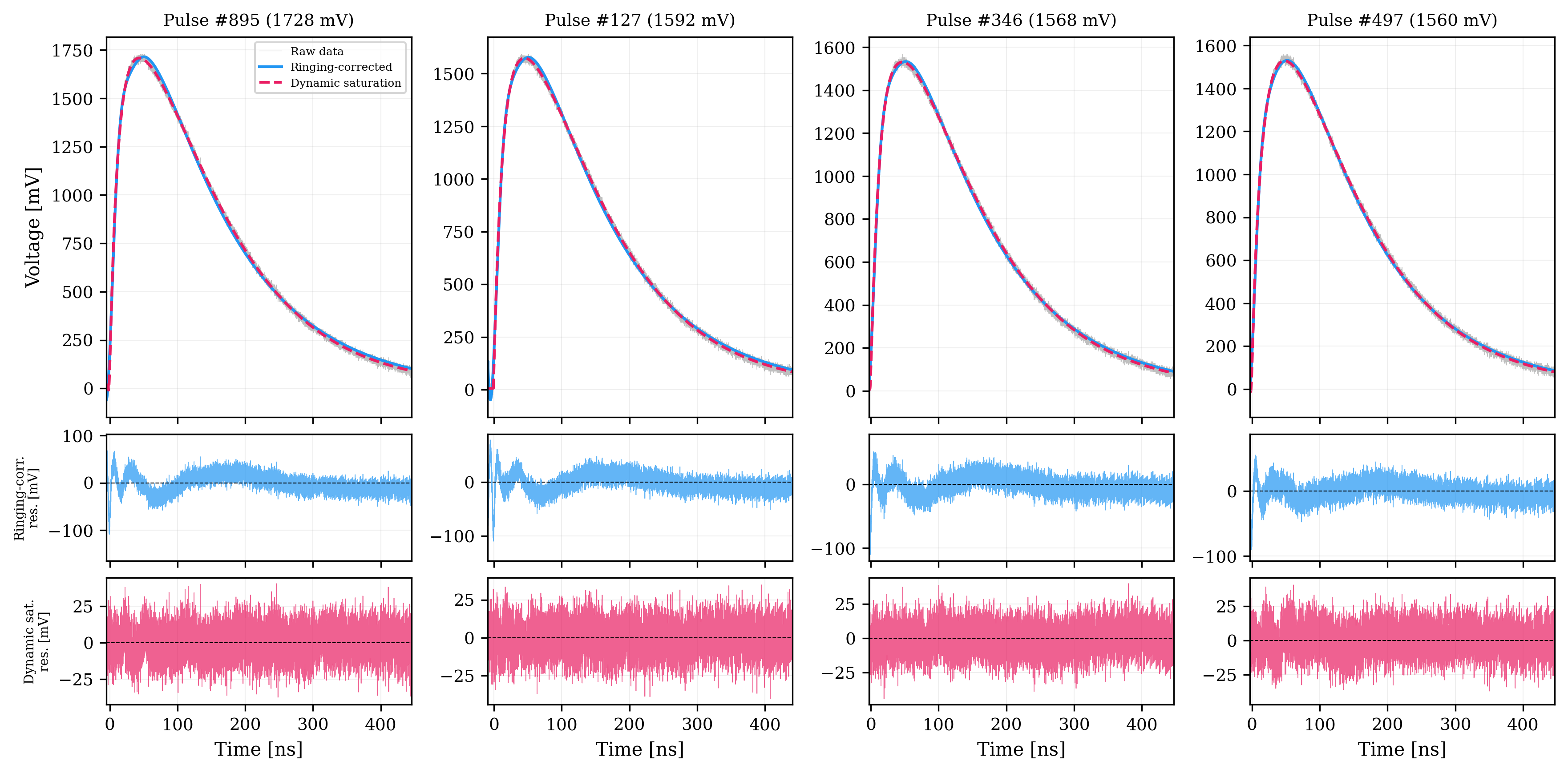}
    \caption{Saturation model comparison on four representative high-amplitude scintillation pulses ($\sim$1560--1728~mV). Top row: raw waveform (gray) overlaid with ringing-corrected (blue) and dynamic saturation (pink dashed) fits. Row~2: ringing-corrected residuals. Row~3: dynamic saturation residuals. The dynamic model consistently reduces residual structure near the peak and in the 50--150~ns decay region across all four representative pulses.}
    \label{fig:exp_saturation}
\end{figure*}

Two physical trends are apparent. First, the mean saturation coupling $\bar{\beta}$ is approximately $2.3\times$ larger for the high-amplitude cohort ($0.052$~ns$^{-1}$) than the medium-amplitude cohort ($0.023$~ns$^{-1}$), consistent with the theoretical expectation that more intense gamma-ray depositions produce greater instantaneous microcell occupancy. Second, the dynamic saturation model achieves a lower (better) mean AIC than the ringing-corrected model in both cohorts (Table~\ref{tab:saturation_comparison}), indicating that even moderate-amplitude pulses experience a detectable degree of saturation with 50~GS/s sampling resolution. With the expanded 100-pulse cohorts and a staged initialization pipeline (including RK4 ODE integration and multi-start $\beta$ optimization), the dynamic saturation model wins the per-pulse AIC comparison in all 100 high-amplitude pulses and 98 of 100 medium-amplitude pulses, providing strong statistical evidence that the saturation ODE captures physical structure beyond what the ringing-corrected template can absorb.

Fig.~\ref{fig:exp_saturation} illustrates this effect on four representative high-amplitude pulses. The bi-exponential residuals (middle panel) exhibit a characteristic systematic undershoot near the peak followed by an overshoot in the 50--150~ns range---precisely the signature of an overestimated peak and underestimated tail predicted by the saturation-ignorant model. The dynamic model residuals (bottom panel), in contrast, are substantially reduced and approach the noise floor throughout the entire pulse duration, corroborating the physical validity of the microcell-depletion ODE~\eqref{eq:dynamic_ode} derived from first principles in Section~II.

The Durbin--Watson (DW) statistic further quantifies the residual autocorrelation. For high-amplitude pulses, the dynamic saturation model raises the mean DW from 0.594 to 1.220---a factor of $2\times$ improvement that reflects the reduction of the systematic peak-undershoot/tail-overshoot pattern characteristic of saturation-ignorant models. For medium-amplitude pulses, the improvement is more modest (1.141 to 1.227), consistent with the smaller saturation coupling $\beta$ at lower photon flux. We note that the baseline DW of 1.141 for the medium-amplitude ringing-corrected model is comparable to the mean DW of 1.136 reported in Table~\ref{tab:model_comparison} for the same model across the full energy spectrum. Although both values remain below the ideal DW~$= 2$ expected for white-noise residuals, the persistent deviation reflects remaining back-end electronic artifacts (parasitic LC ringing, impedance mismatches, and finite-bandwidth filtering) identified in Section~\ref{sec:val_biexp}, which introduce systematic autocorrelation patterns not fully absorbed by these finite-parameter scintillation models.

\begin{figure*}[htbp]
    \centering
    \includegraphics[width=\textwidth]{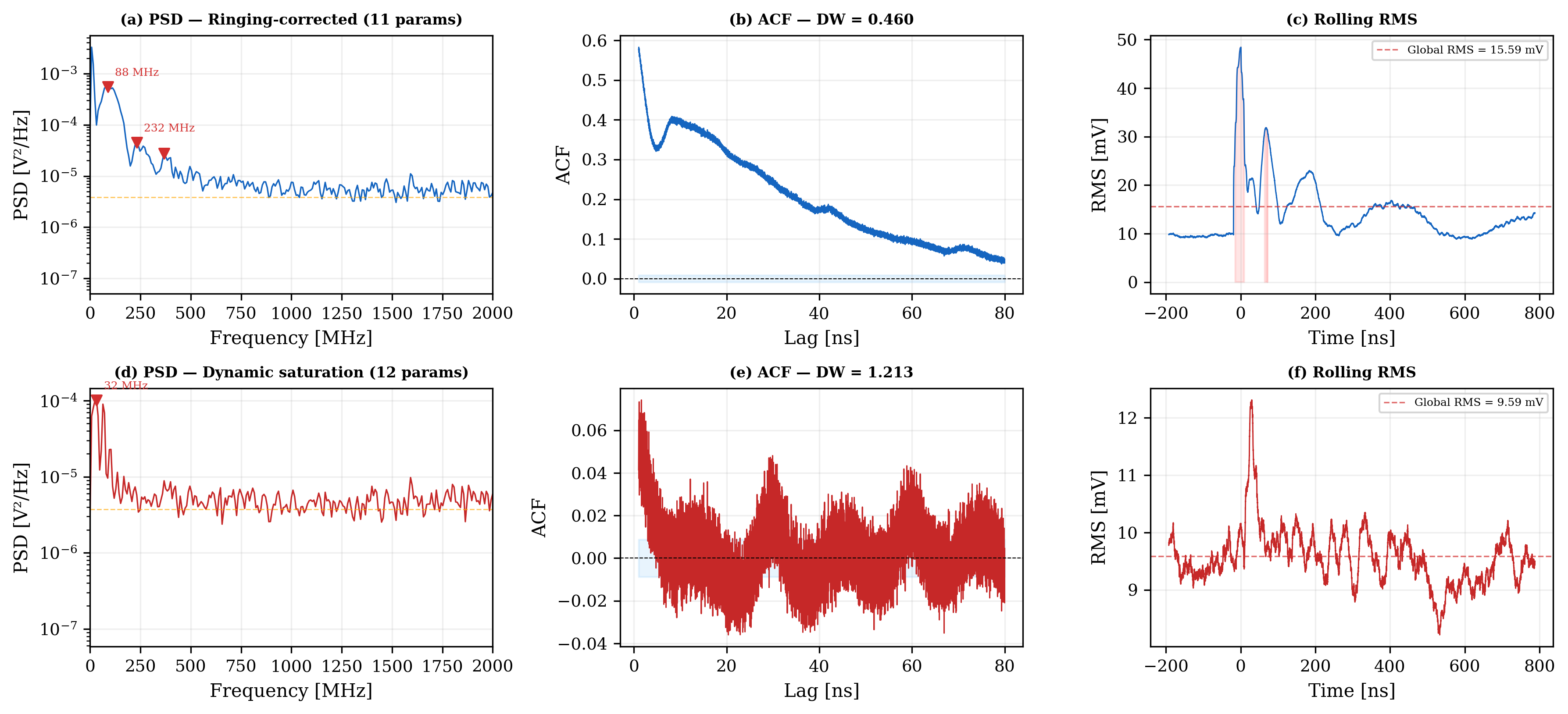}
    \caption{Comparative residual diagnostics on a high-amplitude diagnostic pulse (\#895, 1728~mV). Top row: ringing-corrected model (11 params). Bottom row: dynamic saturation model (12 params). (a,~d)~Welch PSD; (b,~e)~autocorrelation function with 95\% CI; (c,~f)~rolling RMS envelope. The dynamic saturation model reduces both the spectral resonance amplitude and the temporal autocorrelation structure, indicating that the saturation ODE captures pulse-shape modulation beyond parasitic electronic artifacts.}
    \label{fig:residual_diag_dynamic}
\end{figure*}

To further characterize the spectral and temporal structure of the residual improvement, Fig.~\ref{fig:residual_diag_dynamic} presents a side-by-side residual diagnostic comparison between the ringing-corrected and dynamic saturation models on a high-amplitude diagnostic pulse (\#895, 1728~mV). Power spectral density analysis (panels~a,~d) reveals that the dominant resonance peak appears at $\sim$88~MHz in this high-amplitude pulse---upshifted from the $\sim$64~MHz observed in the moderate-amplitude pulse (\#209, 816~mV; Fig.~\ref{fig:residual_diag}). This amplitude-dependent frequency shift is physically consistent with the voltage-dependent junction capacitance of the SiPM: at higher instantaneous photon flux, a larger fraction of microcells are simultaneously in the discharged state, reducing the effective parallel junction capacitance $C_j^{eff}$ and consequently increasing the parasitic LC resonance frequency via $f_{res} = 1/(2\pi\sqrt{L_{\mathrm{RLC}}\,C_j^{eff}})$, where $L_{\mathrm{RLC}}$ is the effective series inductance. The 88~MHz resonance reaches $142\times$ the median noise floor in the ringing-corrected residuals, while the largest residual feature in the dynamic model appears near 32~MHz at $27\times$~SNR. This reduction indicates that part of the low-frequency spectral energy attributed to ``parasitic ringing'' in Section~\ref{sec:val_biexp} is consistent with saturation structure that the 11-parameter model absorbs into its damped-sinusoidal component. The autocorrelation function (panels~b,~e) shows a corresponding DW improvement from 0.460 to 1.213: the pronounced long-range correlation structure visible in the ringing-corrected ACF is reduced once the saturation ODE absorbs the peak-suppression physics. The rolling RMS envelope (panels~c,~f) shows that the non-stationary residual amplification near the pulse peak---where the ringing-corrected model exhibits its characteristic overshoot---is attenuated by the dynamic model, bringing the peak-region RMS closer to the global noise floor.

In summary, the experimental validation supports two central predictions of the analytical framework: (i)~the ringing-corrected model, derived from the linear EMG reduction (Appendix~\ref{app:fitting_biexp_derivation}), captures the dominant pulse-shape features and substantially outperforms a plain bi-exponential across the reported statistical metrics (Table~\ref{tab:model_comparison}); and (ii)~the dynamic saturation ODE, derived from first principles in Section~II-C, reduces the systematic peak-suppression signature that persists in the ringing-corrected residuals for high-amplitude pulses (Table~\ref{tab:saturation_comparison}). Together, these results indicate that the theoretical framework provides a physically interpretable and quantitatively useful description of LYSO-SiPM waveforms across the energy range accessible with $^{22}$Na.

\section{Discussion}
\label{sec:discussion}
The central implication of this work is that scintillation-detector waveform formation and timing limits can be understood as outcomes of a single coupled physical cascade, rather than as separate problems of scintillation kinetics, optical transport, SiPM response, and timing statistics. Within this unified view, finite thermalization, depth-dependent optical transit-time spread, and microcell occupancy dynamics shape not only the observed waveform but also the attainable timing precision. A key theoretical consequence is the temporal decoupling of intrinsic quantum cascade thermalization ($\tau_r$) from geometric optical transit time spread ($\sigma_{TTS}$).

\subsection{Variance Decomposition and Timing Implications}
From a statistical signal-processing perspective, the macroscopic light pulse $Y_{mod}$ is defined structurally as the LTI convolution of an intrinsic generation mechanism ($h_{rise}(t) \propto e^{-t/\tau_r}u(t)$) and a macroscopic decay mechanism ($h_{decay}(t) \propto e^{-t/\tau_{eff}}u(t)$). A standard property of LTI convolutions is that the temporal variance of the output equals the sum of the variances of the individual component kernels, provided each kernel variance is computed from its amplitude-normalized shape. Since the normalized form of each causal exponential kernel ($e^{-t/\tau}u(t)/\tau$) is a standard exponential probability density with variance $\tau^2$, the overall intrinsic temporal variance of $Y_{mod}$ evaluates to $\tau_r^2 + \tau_{eff}^2$. When cascaded with the Gaussian optical dispersion $f_{TTS}(t)$, the temporal variance of the incident optical profile becomes $\sigma_{profile}^2(z) = \tau_{eff}^2 + \tau_r^2 + \sigma_{TTS}^2(z)$, where $\sigma_{profile}^2$ is a temporal-profile variance and is distinct from the current-noise variance $\sigma_{total}^2(t;z)$ in \eqref{eq:sigma_led}.

However, it is crucial to distinguish between the variance of the overall macroscopic distribution and the specific temporal jitter governing leading-edge triggering. While variance additivity under convolution describes the structural broadening of the total pulse, practical Coincidence Timing Resolution (CTR) is dominated by the earliest, steepest portion of the photon-arrival process, so the slow tail governed by $\tau_{eff}$ contributes much less to leading-edge timing than to the bulk pulse variance. As formalized by the Poisson Fisher Information \eqref{eq:fisher_info}, the quantitative timing jitter is bottlenecked by the rate gradient $\dot{\lambda}_{dyn}$ against the instantaneous Poisson counting noise $\lambda_{dyn} + \nu_{DCR}$, rather than the bulk statistical width of the whole waveform. Consequently, the heuristic summation $\sigma_{front}^2(z) \approx \tau_r^2 + \sigma_{TTS}^2(z)$ serves as an intuitive order-of-magnitude scale for early dispersion, rather than a rigorous formula for CTR.

Because the representative intrinsic thermalization constant used here ($\tau_r = 70$~ps) and the optical-spread parameter used in the numerical examples ($\sigma_{TTS} = 40$~ps) lie on the same sub-nanosecond scale, ignoring either mechanism leads to underestimation of early slope constraints. By incorporating the DOI-dependent optical parameterizations \eqref{eq:mu_tts} and \eqref{eq:sigma_tts}, our framework isolates these respective components, explaining why the apparent pulse rise-time fluctuates with the gamma-ray absorption depth.

Moreover, TTS is not merely a linear time shift; it also modulates non-linear early pileup mechanics. As quantified in the dynamic ODE \eqref{eq:dynamic_ode}, TTS temporal smearing dilutes the instantaneous peak photon flux $r_{ph}(t; z)$ at the sensor surface. Spatial dispersion thereby acts as a temporal buffer against instantaneous microcell starvation, suppressing severe early-stage non-linear signal clipping that degrades conventional leading-edge discriminators (LED).

\subsection{Comparison with Existing CRLB Estimates}
The $\sim$100~ps coincidence CTR predicted by our Poisson CRLB analysis at 511~keV is consistent with $\sim$100~ps limits discussed in the scintillation timing community~\cite{Lecoq2017Pushing, Lecoq2020Roadmap, Gundacker2020Experimental}. By directly employing the non-homogeneous Poisson Fisher Information \eqref{eq:fisher_info}---the same statistical framework used by Seifert~\textit{et al.}~\cite{Seifert2012Lower} and Vinogradov~\cite{Vinogradov2012Analytical}---our result is directly comparable to prior theoretical bounds without relying on a Gaussian timing-noise approximation. The present formulation additionally incorporates dynamic microcell saturation via the ODE \eqref{eq:dynamic_ode} and DOI-dependent TTS, extending the photon-counting CRLB to account for detector-level non-linearities absent in earlier treatments.

\subsection{Framework Limitations}
Several approximations in the present framework merit explicit acknowledgment:

\textit{1) Single-exponential thermalization:} The lumped $\tau_r$ model assumes a single rate-limiting step dominates the cascade. If multiple intermediate kinetic steps have comparable time constants ($\sim$100--300~ps), the single-exponential approximation may underestimate the leading-edge dispersion.

\textit{2) Binary recovery ODE:} As quantified in \eqref{eq:dead_phys}--\eqref{eq:dead_ode}, the binary microcell model underestimates the effective dead-time by $0.5\tau_{rec}$ per recovery cycle. Under the quadratic partial-recovery envelope used here, this biases the dynamic current high in dense pileup and therefore gives the corresponding timing predictions an optimistic tendency.

\textit{3) Campbell variance upper bound:} Under deep saturation ($\alpha n_{ph} \gtrsim 0.2$), microcell dead-time introduces sub-Poissonian correlations that suppress the physical variance below the independent Poisson prediction used here. The Campbell-based variance therefore constitutes an upper bound whose tightness degrades with increasing photon flux.

\textit{4) Poisson independence assumption:} The Fisher Information \eqref{eq:fisher_info} assumes that individual microcell firings are statistically independent conditional on the rate $\lambda_{dyn}$. Under deep saturation, microcell recovery correlations introduce renewal-process structure that is not captured by this Poisson likelihood. The resulting event-level CRLB should therefore be interpreted as a Poisson-observer benchmark whose tightness, and not merely its numerical value, must be reassessed if dead-time correlations become dominant.

\textit{5) Gaussian TTS model:} The Gaussian TTS approximation is justified for intermediate-to-long crystals ($L/d \gtrsim 5$) but may fail for short, highly polished crystals where direct and low-order specular pathways dominate.

\subsection{Physical and Design Implications}
The primary value of the framework is explanatory: by separating deterministic pulse formation from stochastic variance buildup, it clarifies which physical mechanisms control waveform shape and which limit timing precision. The same separation also provides practical design guidance. Several concrete implications emerge from the framework:

\textit{Crystal length--TTS trade-off:} The DOI-dependent parameterizations \eqref{eq:mu_tts}--\eqref{eq:sigma_tts} directly predict how increasing crystal length improves detection efficiency but degrades timing through enhanced $\sigma_{TTS}(z)$~\cite{Nemallapudi2015Sub}. For a given SiPM microcell count, the framework can identify the optimal crystal length that minimizes the DOI-averaged CRLB.

\textit{Microcell density requirements:} The dynamic saturation analysis quantifies the minimum microcell count $M$ needed to maintain linearity ($\alpha n_{ph} < 0.1$) at a given energy. Here $n_{ph}$ denotes the cumulative photon count incident on the SiPM surface as defined in \eqref{eq:nph_tts}; for 511~keV depositions, $n_{ph}(\infty) = N_0 \frac{1-\epsilon a}{1-\epsilon a \eta} k_{trans} \approx 6300$ under the parameters of Section~\ref{sec:verification}. With $q = 0.50$, enforcing $\alpha \, n_{ph}(\infty) < 0.1$ yields $M \gtrsim q \, n_{ph}(\infty) / 0.1 \approx 31000$ microcells---far above the reference value $M = 7100$ adopted here. This criterion is deliberately conservative; the numerical results in Section~\ref{sec:verification} show only mild time-distributed compression at 511~keV, while the experimental high-amplitude cohort confirms that saturation becomes readily detectable as pulse amplitude increases.

\textit{Crosstalk penalty quantification:} The excess noise factor $F_{ENF} = 1/(1-P_{ct})$ directly scales the Poisson variance in \eqref{eq:poisson_noise}--\eqref{eq:sigma_total}, providing an explicit penalty function for evaluating the timing cost of elevated crosstalk probability.

\subsection{Waveform-Based DOI Inference}
The DOI-dependent transport parameterizations \eqref{eq:mu_tts}--\eqref{eq:sigma_tts} suggest a principled inverse extension of the present framework: rather than treating $z$ only as a nuisance variable affecting timing, the same forward model can, in principle, be used for waveform-based DOI estimation. At the waveform-observation level, Appendix~\ref{app:waveform_fisher} gives the exact sampled-Gaussian mean/covariance formulation. For a practical DOI estimator, we adopt the same diagonal reduction as \eqref{eq:wf_white_noise}: after analog bandwidth smoothing, and assuming that off-diagonal covariance terms vary weakly over the narrow DOI-sensitive leading-edge window, the sampled points are treated as conditionally independent Gaussian variables with parameter-dependent variances. The resulting objective is therefore an approximate waveform likelihood rather than an exact full-covariance maximum-likelihood estimator. A full derivation of this diagonal likelihood and its IRLS implementation is given in Appendix~\ref{app:doi_irls_derivation}.

In the linear regime, the sampled waveform $y_n \equiv y(t_n)$ is modeled as
\begin{equation}
    \begin{aligned}
        y_n &= g_{ro}\,I_{lin}(t_n - t_{int}; z) + k t_n + b_0 + \varepsilon_n \\
        \varepsilon_n &\sim \mathcal{N}\!\left(0, \sigma_{total}^2(t_n; z)\right)
    \end{aligned}
    \label{eq:doi_obs_linear}
\end{equation}
where $I_{lin}(t; z)$ is the forward model of \eqref{eq:ilin_emg}, $g_{ro}$ is an event-level readout scale factor, $t_{int}$ is the interaction time, and $k t_n + b_0$ captures slow baseline drift. After calibrating the finite detector parameter set
\begin{equation}
    \begin{aligned}
        \Theta_{det} = \{&\mu_0, \sigma_0, v_{eff}, k_{disp}, k_{trans}, \tau_r, \tau_{eff}, M, \alpha, \tau_{rec},\\
        &I_{cell}^{macro}, (\tau_x, A_x)_{x \in \{d,p1,p2\}}, \sigma_{elec}, \nu_{DCR}\}
    \end{aligned}
    \label{eq:theta_det}
\end{equation}
from DOI-tagged reference data, and defining $\boldsymbol{\psi}_{lin} = (z, t_{int}, g_{ro}, k, b_0)$, a per-pulse depth estimate can be obtained by minimizing the approximate negative log-likelihood
\begin{equation}
    \begin{split}
        \hat{\boldsymbol{\psi}}_{lin}
            &= \arg\min_{\substack{\boldsymbol{\psi}_{lin}:\\ 0 \le z \le L_{cry}}} \mathcal{L}_{lin} \\
        \mathcal{L}_{lin}
            &= \frac{1}{2}\sum_n \log\!\bigl(\sigma_{total}^2(t_n; z)\bigr) \\
            &\quad + \frac{1}{2}\sum_n \frac{\left[y_n - g_{ro} I_{lin}(t_n - t_{int}; z) - k t_n - b_0\right]^2}{\sigma_{total}^2(t_n; z)}
    \end{split}
    \label{eq:doi_wls_linear}
\end{equation}
which exploits both the monotonic delay shift induced by $\mu_{TTS}(z)$ and the leading-edge broadening induced by $\sigma_{TTS}(z)$. If the absolute front-end gain is independently calibrated, one may fix $g_{ro}=1$; otherwise, estimating $g_{ro}$ jointly prevents residual gain mismatch from being spuriously absorbed into $z$.

For higher photon flux, the same observation architecture can be retained by replacing the linear template with the dynamic current of \eqref{eq:i_dyn_explicit} and jointly fitting the event light-yield parameter $N_0$:
\begin{equation}
    \begin{aligned}
        y_n &= g_{ro}\,I_{dyn}(t_n - t_{int}; z, N_0) + k t_n + b_0 + \varepsilon_n \\
        \varepsilon_n &\sim \mathcal{N}\!\left(0, \sigma_{total}^2(t_n; z, N_0)\right)
    \end{aligned}
    \label{eq:doi_obs_dyn}
\end{equation}
With $\boldsymbol{\psi}_{dyn} = (z, t_{int}, N_0, g_{ro}, k, b_0)$,
\begin{equation}
    \begin{split}
        \hat{\boldsymbol{\psi}}_{dyn}
            &= \arg\min_{\substack{\boldsymbol{\psi}_{dyn}:\\ 0 \le z \le L_{cry}}} \mathcal{L}_{dyn} \\
        \mathcal{L}_{dyn}
            &= \frac{1}{2}\sum_n \log\!\bigl(\sigma_{total}^2(t_n; z, N_0)\bigr) \\
            &\quad + \frac{1}{2}\sum_n \frac{1}{\sigma_{total}^2(t_n; z, N_0)} \\
            &\qquad\qquad \times \Bigl( y_n - g_{ro} I_{dyn}(t_n - t_{int}; z, N_0) \\
            &\qquad\qquad\qquad\quad - k t_n - b_0 \Bigr)^2
    \end{split}
    \label{eq:doi_wls_dyn}
\end{equation}
so that DOI, light yield, and residual gain mismatch are inferred within the same microcell-depletion model used in the forward timing analysis. If the variance terms in \eqref{eq:doi_wls_linear} or \eqref{eq:doi_wls_dyn} are frozen from an initial forward pass, the resulting optimization reduces to the usual iteratively reweighted least-squares (IRLS) form.

In practice, numerical stability can be improved through a two-stage initialization. An initial EMG fit provides a Gaussian width estimate $\sigma^{\mathrm{fit}}$ and an onset estimate $t_0$. Under the same diagonal approximation, the width-only DOI seed follows from inverting $(\sigma^{\mathrm{fit}})^2 \approx \sigma_0^2 + k_{disp} z + \sigma_{elec}^2$:
\begin{equation}
    z_{\sigma}^{(0)} = \frac{(\sigma^{\mathrm{fit}})^2 - \sigma_{elec}^2 - \sigma_0^2}{k_{disp}}
    \label{eq:doi_init_sigma}
\end{equation}
with clipping to the physical interval $[0, L_{cry}]$. This seed can then be refined jointly with $t_{int}$ using \eqref{eq:doi_wls_linear} or \eqref{eq:doi_wls_dyn}. If an external timing reference or double-ended readout provides a preliminary interaction-time estimate $t_{int}^{ref}$, the mean-delay relation \eqref{eq:mu_tts} yields a second initializer
\begin{equation}
    z_{\mu}^{(0)} \approx v_{eff}(t_0 - t_{int}^{ref} - \mu_0)
    \label{eq:doi_init_mu}
\end{equation}
which reduces the otherwise strong single-ended correlation between $z$ and $t_{int}$.

The present Fisher-information treatment also admits a direct extension from scalar timing estimation to joint timing-depth inference under the same Poisson-observer assumptions used in \eqref{eq:fisher_info}. Defining $\boldsymbol{\theta} = (t_{int}, z)$ and $\lambda_{dyn}(t; \boldsymbol{\theta}) \equiv \lambda_{dyn}(t-t_{int}; z)$, the corresponding $2 \times 2$ Fisher information matrix has entries
\begin{equation}
    [\boldsymbol{\mathcal{I}}(\boldsymbol{\theta})]_{ij} = \int
    \frac{\partial_{\theta_i}\lambda_{dyn}(t; \boldsymbol{\theta})\,\partial_{\theta_j}\lambda_{dyn}(t; \boldsymbol{\theta})}
    {\lambda_{dyn}(t; \boldsymbol{\theta}) + \nu_{DCR}}\,dt
    \label{eq:doi_fim}
\end{equation}
from which the attainable DOI variance for any locally unbiased estimator follows as $\sigma_z^2 \ge [\boldsymbol{\mathcal{I}}^{-1}(\boldsymbol{\theta})]_{zz}$. In that sense, the present framework provides the forward model and information-theoretic benchmark required for waveform-based DOI reconstruction, DOI-resolution prediction, and detector-geometry optimization, while leaving the final calibration and validation to future experiments.

\subsection{Future Experimental Validation}
The present experimental validation demonstrates that the dynamic saturation model captures measurable pulse-shape distortions at high photon flux. However, several predictions of the framework remain to be tested:

\textit{1)} DOI-resolved timing measurements using a position-sensitive readout or a collimated pencil beam could directly verify the predicted $\sigma_{TTS}(z)$ dependence and the DOI-dependent CRLB landscape.

\textit{2)} Ultrafast optical or single-photon time resolution (SPTR) measurements could help isolate the intrinsic $\tau_r$ contribution from the macroscopic rise time, enabling independent validation of the thermalization model.

\textit{3)} Energy-gated pulse-shape analysis across the Compton continuum could test the predicted amplitude-dependent saturation coupling $\beta(N_0)$ and quantify the transition from linear to saturated regimes.

\textit{4)} A waveform-based DOI reconstruction experiment using a collimated pencil beam or position-tagged irradiation at known depths could calibrate $\Theta_{det}$, estimate per-pulse depth via \eqref{eq:doi_wls_linear} or \eqref{eq:doi_wls_dyn}, and compare the empirical DOI error against the joint Fisher benchmark $\sigma_z^2 \ge [\boldsymbol{\mathcal{I}}^{-1}(\boldsymbol{\theta})]_{zz}$ from \eqref{eq:doi_fim}. Such a study would directly test whether the DOI-dependent pulse-shape signatures predicted here remain identifiable once front-end gain variation, baseline drift, and waveform covariance approximations are all treated explicitly.

\subsection{Multi-Event Pileup Extension}
While the dynamic ODEs \eqref{eq:dynamic_ode} in this framework track an isolated gamma interaction initialized with a zero-state $N_{busy}(-\infty; z) = 0$, the identical state-dependent architecture naturally extends to continuous multi-event pileup conditions. For high-count-rate analysis, the terminal microcell state after one interaction can be carried forward as the initial condition for the next, while the incident photon-rate history is updated by adding the new event's shifted $r_{ph}(t-t_0;z)$. Thus, the framework scales to pileup analysis through state propagation and rate superposition rather than structural reformulation.

\section{Conclusion}
In this work, we developed a unified analytical framework for LYSO-SiPM scintillation detectors that links scintillation kinetics, depth-dependent optical transport, SiPM microcell occupancy dynamics, and timing statistics within a single forward description of waveform formation. The framework yields closed-form exponentially modified Gaussian pulse expressions in the linear regime and stable state-dependent integral solutions in saturation, while recovering the conventional bi-exponential pulse model as a limiting case.

By coupling the dynamic triggering rate to generalized non-stationary compound Poisson statistics, including dark-count and crosstalk-induced variance terms, the model predicts current-variance envelopes and Fisher-information-based timing limits, including an intrinsic coincidence timing resolution lower bound of about 100 ps FWHM for a reference 511-keV LYSO-SiPM configuration.

Experimental validation on a directly digitized 10,000-pulse Na-22 dataset shows that the dynamic saturation model captures amplitude-dependent pulse distortion and is favored by Akaike information criterion over a matched ringing-corrected bi-exponential baseline in 100/100 high-amplitude pulses and 98/100 medium-amplitude pulses. Taken together, these results deepen the physical understanding of scintillation-detector waveform formation and timing limits by clarifying how scintillation kinetics, optical transport, and SiPM microcell dynamics jointly shape the observed response.

\clearpage

\begin{table*}[!p]
    \setlength{\abovecaptionskip}{0pt}
    \setlength{\belowcaptionskip}{2pt}
    \caption{Summary of Mathematical Notation (page 1 of 2)}
    \label{tab:notation}
    \centering
    \scriptsize
    \setlength{\tabcolsep}{2pt}
    \renewcommand{\arraystretch}{0.86}
    \begin{tabular}{@{}p{0.22\textwidth}p{0.74\textwidth}@{}}
        \hline\hline
        \textbf{Symbol} & \textbf{Definition or Role} \\
        \hline
        $N_0$ & Initial excitation-center population generated by the gamma-ray deposition \\
        $E, E_0$ & Deposited energy and reference energy or photopeak center for energy-resolution bounds \\
        $\bar N_0(E), Y_0$ & Mean energy-dependent excitation-center population and local proportional light-yield coefficient \\
        $R(t)$ & Thermalization generation rate feeding the emitting centers \\
        $\epsilon, a, \eta$ & Spectral-overlap probability, self-absorption probability, and re-emission quantum yield; $\epsilon a$ is the net spectral self-absorption probability \\
        $\tau_r$ & Intrinsic thermalization bottleneck time constant \\
        $\tau$ & Primary intrinsic decay time of idealized localized luminescence \\
        $\tau_{eff}$ & Effective macroscopic decay time inclusive of recursive self-absorption \\
        $h(t)$ & Intrinsic escaping-photon impulse response per initial excitation center \\
        $Y_{mod}(t)$ & Modified intrinsic photon emission rate after thermalization convolution \\
        $Y_{total}$ & Total detectable internal photon yield after recursive self-absorption \\
        $u(t)$ & Heaviside step function enforcing causality \\
        $t, s, t', t_j$ & Time argument, convolution variable, generic integration or recharge elapsed time, and stochastic event time \\
        $t_n, y_n, \hat{y}_i, \bar{y}, r_i, \varepsilon_n$ & Sample times, waveform samples, fitted waveform values, sample mean, fit residuals, and additive noise samples in waveform fitting or DOI inversion \\
        $i,j,k,m,n,\ell$ & Local integer indices or counters; each meaning is defined in its immediate equation block and is not used as a global physical parameter \\
        $\xi$ & Dimensionless recharge time $\xi=t'/\tau_{rec}$ used in the dead-time integrals \\
        $z$ & Longitudinal distance from interaction vertex to the photodetector \\
        $L_{cry}$ & Crystal length along the DOI axis, bounding $0 \le z \le L_{cry}$ in DOI estimation \\
        $\mu_{TTS}(z)$ & DOI-dependent mean Optical Transit Time Spread \\
        $\sigma_{TTS}(z)$ & DOI-dependent standard deviation of optical dispersion \\
        $\mu_0, \sigma_0$ & Near-surface mean delay and standard deviation of optical transport \\
        $c, n$ & Speed of light in vacuum and crystal refractive index \\
        $n_1, n_2$ & Refractive indices of the crystal and coupling medium in the TIR correction \\
        $v_{eff}, k_{disp}$ & Effective longitudinal optical velocity and spatial variance accumulation rate \\
        $\theta, \theta_c$ & Internal photon angle and TIR critical angle \\
        $L/d$ & Crystal aspect ratio used to classify optical transport regimes \\
        $\Phi$ & Standard normal cumulative distribution function \\
        $f_{TTS}(t; z)$ & DOI-dependent optical transport time-spread kernel \\
        $k_{trans}$ & Bulk optical transfer efficiency from crystal to SiPM \\
        $r_{ph}(t; z)$ & Macroscopic incident photon arrival rate \\
        $\psi_{ph}(t; z)$ & Incident photon-rate shape per initial excitation center in energy-parametrized bounds \\
        $n_{ph}(t; z)$ & Cumulative macroscopic incident photon count \\
        \parbox[t]{0.21\textwidth}{$\rho_{esc}$, $\bar n_{ph,\infty}(E,z)$,\\ $\bar N_{pe}(E,z)$, $\bar N_{pe,lin}(E,z)$} & Escape fraction, total incident photon count, mean primary photoelectron count, and its low-occupancy linear limit in energy-resolution bounds \\
        $M, q$ & Total microcell count and global Photon Detection Efficiency (PDE) \\
        $\alpha$ & Effective per-photon triggering parameter in the static occupancy closure ($-\ln(1-q/M)$) [photon$^{-1}$] \\
        $N_{fired}(t; z)$ & Static cumulative activated microcell population \\
        $N_{busy}(t; z)$ & Transient recovering microcell population \\
        $S_E(t; E,z)$ & Energy sensitivity of the busy-cell population, $\partial_E N_{busy}(t;E,z)$ \\
        $\lambda(t; z)$ & Static macroscopic activation rate \\
        $\lambda_{dyn}(t; z)$ & Dynamic activation rate under recovery and saturation \\
        $\tau_{rec}$ & Internal SiPM microcell RC-recharging constant \\
        $V_{OV}$ & SiPM overvoltage above breakdown; sets PDE ($q$), crosstalk ($P_{ct}$), and DCR ($\nu_{DCR}$) \\
        $\Delta V(t')$ & Recovering microcell overvoltage after elapsed recharge time $t'$ \\
        \parbox[t]{0.21\textwidth}{$Y_{phys}(t')$,\\ $Y_{ODE}(t')$} & Physical and binary-ODE recovery-yield proxies \\
        \parbox[t]{0.21\textwidth}{$\tau_{dead, phys}$,\\ $\tau_{dead, ODE}$} & Effective dead-times in the partial-recovery and binary-ODE models \\
        $\mu(t; z)$ & Integrating factor of the dynamic microcell recovery ODE \\
        $i_{ser}(t)$ & Causal single-cell macroscopic current impulse response \\
        $I_{cell}^{single}$ & Physical peak current of a single primary avalanche \\
        $I_{cell}^{macro}$ & Equivalent peak current including mean crosstalk gain ($\langle G \rangle I_{cell}^{single}$) \\
        $\tau_x, A_x$ & Multi-exponential current decay poles and non-negative normalized amplitudes \\
        $x,y; d,p1,p2$ & Pole-sum indices; fast discharge branch and two slower post-pulse current branches \\
        $I_{stat}(t; z)$ & Static saturation-regime macroscopic current \\
        $I_{dyn}(t; z)$ & Dynamic recovery-corrected macroscopic current \\
        $I_{lin}(t; z)$ & Linear-regime macroscopic current \\
        $\lambda_{lin}(t; z)$ & Low-occupancy linear triggering rate, $q r_{ph}(t;z)$ \\
        $I_{ideal}(t)$ & Intrinsic linear current kernel before TTS convolution \\
        $C_{gen}, C, C_{inst}$ & Generation, linear-current, and instantaneous-rise scales \\
        $\tilde{\tau}_{x,eff}, \allowbreak \tilde{\tau}_{x,r}, \allowbreak \tau_{xy}$ & Coupling times and squared-response pair decay constant \\
        $A_{macro}, A_{eff}$ & Dominant-pole amplitudes for the reduced ideal kernel and reduced linear pulse \\
        $\operatorname{EMG}$ & Exponentially Modified Gaussian kernel used for linear closed-form pulses \\
        $\mu_z, \sigma_z, \operatorname{EMG}_z$ & Spatial shorthand for $\mu_{TTS}(z)$, $\sigma_{TTS}(z)$, and $\operatorname{EMG}(t;\mu_z,\sigma_z,\tau)$ \\
        $\operatorname{erfc}$ & Complementary error function appearing in the EMG kernel \\
        \hline\hline
    \end{tabular}
\end{table*}

\begin{table*}[!p]
    \addtocounter{table}{-1}
    \setlength{\abovecaptionskip}{0pt}
    \setlength{\belowcaptionskip}{2pt}
    \caption[]{Summary of Mathematical Notation (continued, page 2 of 2)}
    \centering
    \scriptsize
    \setlength{\tabcolsep}{2pt}
    \renewcommand{\arraystretch}{0.86}
    \begin{tabular}{@{}p{0.22\textwidth}p{0.74\textwidth}@{}}
        \hline\hline
        \textbf{Symbol} & \textbf{Definition or Role} \\
        \hline
        $\nu_{DCR}$ & Steady-state Dark Count Rate (DCR) \\
        \parbox[t]{0.21\textwidth}{$P_{ct}, \langle G \rangle$,\\ $G$} & Optical crosstalk probability, mean cascade gain, and avalanche multiplicity \\
        $F_{ENF}$ & Excess Noise Factor compounding optical crosstalk cascades \\
        $\lambda_{tot}(t; z)$ & Total triggering rate including dark counts \\
        $\dot{\lambda}_{dyn}$ & Time derivative $\partial\lambda_{dyn}/\partial t$ used in Fisher Information \\
        $\sigma_I^2(t; z)$ & Intrinsic current variance envelope from filtered avalanche statistics \\
        $\sigma_{base}^2, \sigma_{elec}, \sigma_{elec}^2$ & Steady dark-count variance, electronic-noise standard deviation, and electronic variance \\
        $\sigma_{total}^2(t; z)$ & Total instantaneous current variance used for discriminator timing \\
        $\sigma_{profile}^2(z)$ & Temporal-profile variance of incident optical waveform, $\tau_{eff}^2 + \tau_r^2 + \sigma_{TTS}^2(z)$; distinct from $\sigma_{total}^2$ \\
        $\sigma_{front}^2(z)$ & Heuristic front-edge dispersion scale, $\approx \tau_r^2 + \sigma_{TTS}^2(z)$ \\
        $I_{th}, t_{th}, t_{int}$ & LED threshold, threshold-crossing time, and interaction-time shift \\
        $t_{int}^{ref}$ & External preliminary estimate of the interaction time used in DOI initialization \\
        $\sigma_{t,LED}^2, \sigma_{coinc}^2$ & LED timing variance and coincidence timing variance \\
        $\mathcal{I}(z)$ & Fisher Information for estimating the gamma interaction time \\
        $\mathcal{I}_{wf}(z), \mathcal{I}_{wf}^{(1)}(z)$ & Exact and reduced waveform-observation Fisher Information incorporating $i_{ser}(t)$; Appendix~\ref{app:waveform_fisher} \\
        \parbox[t]{0.21\textwidth}{$\mathcal{I}_E(E,z)$,\\ $\mathcal{I}_{E,stat}$, $\mathcal{I}_{E,Q}$} & Energy-estimation Fisher Information for primary-trigger, static-binomial, and charge observations; Appendix~\ref{app:energy_resolution_fisher} \\
        \parbox[t]{0.21\textwidth}{$\mathcal{I}_{E,wf}$,\\ $\mathcal{I}_{E,wf}^{(1)}$, $\mathcal{I}_{E,det}$} & Exact waveform, reduced waveform, and generic detector-level energy Fisher Information \\
        \parbox[t]{0.21\textwidth}{$R_{E_0,primary}$,\\ $R_{E_0,wf}$, $R_{E_0,det}$} & Relative FWHM energy-resolution bounds at reference energy $E_0$ for primary-trigger, waveform, and generic detector observations \\
        \parbox[t]{0.21\textwidth}{$R_{E,\mathrm{FWHM}}$,\\ $R_{E,stat}$, $R_{E,\mathrm{charge}}$} & Energy-dependent relative FWHM, static-saturation, and charge-statistics energy-resolution bounds \\
        \parbox[t]{0.21\textwidth}{$Q_{cell}^{macro}$, $\mu_Q$,\\ $\sigma_Q^2$, $\sigma_{Q,elec}^2$} & Mean-gain-mapped single-cell charge, charge mean, charge variance, and electronic charge variance \\
        \parbox[t]{0.21\textwidth}{$F_{sc}$, $\sigma_{np}^2(E)$,\\ $\sigma_{E,scint}^2(E)$, $\sigma_{E,coll}^2(E,z)$} & Scintillator excess fluctuation, nonproportionality spread, propagated scintillator energy variance, and residual collection-nonuniformity variance \\
        $\boldsymbol{\theta}, \boldsymbol{\mathcal{I}}(\boldsymbol{\theta})$ & Joint timing-depth parameter vector and Fisher information matrix \\
        $\sigma_{CRLB}^2(z)$ & Single-detector Cram\'er-Rao timing variance bound \\
        $\sigma_z^2$ & DOI variance bound obtained from the inverse joint Fisher information matrix \\
        $\mathrm{CTR}_{CRLB}\allowbreak{}(z)$ & Coincidence timing resolution bound expressed as FWHM \\
        $\mathbb{E}[\cdot], \operatorname{Var}(\cdot), P(\cdot), \Pi(\varsigma)$ & Expectation, variance, probability mass, and probability-generating function used in the crosstalk cascade model \\
        $\mathcal{N}(\cdot,\cdot)$ & Normal distribution notation for sampled waveform likelihoods \\
        $\delta(\cdot), \operatorname{Tr}(\cdot)$ & Dirac delta distribution and matrix trace operator \\
        $w(t), W(\zeta), k_i$ & Completion-time density, its Laplace-domain representation with local transform variable $\zeta$, and microscopic cascade rate constants in Appendix~\ref{app:rate_limiting} \\
        $\beta$ & Experimental saturation coupling in \eqref{eq:ode_fit}; inverse-time units if $g(t)$ is dimensionless \\
        $A, t_0, b, b_0, k, p_{\mathrm{fit}}$ & Local experimental fit amplitude, onset time, constant baseline, baseline intercept, baseline slope, and number of fit parameters \\
        $\tau_d^{\mathrm{fit}}, \tau_r^{\mathrm{fit}}$ & Macroscopic fitted decay and rise time constants in the experimental templates \\
        \parbox[t]{0.21\textwidth}{$I_{biexp}(t)$, $I_{linbase}(t)$,\\ $I_{ring}(t)$} & Bi-exponential, linear-baseline, and ringing-corrected experimental waveform models \\
        $\mathrm{RMSE}, R^2, \mathrm{AIC}, \mathrm{DW}$ & Root-mean-square error, coefficient of determination, Akaike information criterion, and Durbin--Watson fit statistic \\
        $A_{ring}, \tau_{ring}, \omega, \phi$ & Ringing amplitude, damping time, angular frequency, and phase in the damped-sinusoidal correction \\
        \parbox[t]{0.21\textwidth}{$L_{\mathrm{RLC}}, R_{\mathrm{RLC}}, C_j$,\\ $C_j^{eff}, f_{res}$} & Effective parasitic inductance, damping resistance, junction capacitance, effective junction capacitance, and LC resonance frequency \\
        $g(t\!-\!t_0)$ & Bi-exponential fitting template, $(e^{-(t-t_0)/\tau_d^{\mathrm{fit}}} - e^{-(t-t_0)/\tau_r^{\mathrm{fit}}})u(t\!-\!t_0)$ \\
        $\mathcal{E}(t; t_0, \tau, \sigma)$ & EMG fitting kernel with explicit onset $t_0$; see \eqref{eq:ecg_kernel} \\
        $G_\sigma$ & Gaussian convolution kernel of width $\sigma$ modeling combined TTS and electronic bandwidth \\
        $\kappa_{EMG}$ & Shape-matching correction between the true photon-rate profile and bi-exponential proxy \\
        $\Theta_{det}$ & Finite calibrated detector-parameter set used by the DOI forward models \\
        $g_{ro}$ & Event-level readout scale mapping analytical current to waveform units \\
        $\boldsymbol{\psi}_{lin}, \boldsymbol{\psi}_{dyn}$ & Parameter vectors of the linear and dynamic DOI inference models \\
        $\mathcal{L}_{lin}, \mathcal{L}_{dyn}, \mathcal{L}(\boldsymbol{\varphi})$ & Approximate and generic negative log-likelihood objectives for DOI fitting \\
        $\vartheta, \mu_{\vartheta}, K_{\vartheta}, \boldsymbol{\mu}_{\vartheta}, \mathbf{\Sigma}_{\vartheta}$ & Local waveform-observation time-shift parameter, mean waveform, covariance kernel, sampled mean vector, and sampled covariance matrix \\
        $\boldsymbol{\varphi}, \boldsymbol{\eta}, \mathbf{a}, \mathbf{y}, \mathbf{x}_n, \mathbf{X}, \mathbf{W}, \mathbf{J}$ & Generic DOI-estimation parameter vector, nonlinear block, linear nuisance block, sampled waveform vector, design row, design matrix, weight matrix, and Jacobian in the IRLS derivation \\
        \parbox[t]{0.21\textwidth}{$m_n(\boldsymbol{\varphi})$, $s_n(\boldsymbol{\varphi})$,\\ $w_n^{(m)}$, $Q^{(m)}$} & Predicted sample mean, model variance, frozen IRLS weight, and frozen-variance surrogate objective in the DOI IRLS derivation \\
        $\hat{S}(f), \hat{\rho}(\ell), \mathrm{RMS}(i)$ & Welch residual power spectral density, residual autocorrelation, and rolling root-mean-square diagnostics \\
        $\sigma^{\mathrm{fit}}$ & Gaussian width from a preliminary EMG fit used for DOI initialization \\
        $z_{\sigma}^{(0)}, z_{\mu}^{(0)}$ & Width-based and mean-delay-based initial DOI estimates \\
        \hline\hline
    \end{tabular}

    \vspace{2pt}
    \parbox{0.96\textwidth}{\scriptsize Dummy variables, local substitutions, and summation indices are defined at first use; scalar, bold/vector, matrix, and calligraphic variants denote distinct objects.}
\end{table*}

\clearpage

\appendices
\numberwithin{equation}{section}
\section{Formal Rate-Limiting-Step Reduction for Single-Exponential Thermalization}
\label{app:rate_limiting}
The single-exponential thermalization cascade used in Section~II-A,
\begin{equation}
    R(t) = \frac{N_0}{\tau_r} e^{-t/\tau_r} u(t),
    \label{eq:T1}
\end{equation}
is most cleanly interpreted as a \emph{macroscopic completion-rate density}: if $w(t)$ denotes the normalized probability density that one initial excitation completes a sequential thermalization chain at time $t$, then $R(t) = N_0 w(t)$. For an $n$-step first-order cascade with sequential rate constants $k_1, k_2, \ldots, k_n$, the Laplace transform of this completion-time density, using the local transform variable $\zeta$, is:
\begin{equation}
    W(\zeta) \equiv \operatorname{Lap}\{w(t)\}(\zeta) = \prod_{i=1}^{n} \frac{k_i}{\zeta + k_i}
    \label{eq:T2}
\end{equation}
because the total waiting time is the convolution of the $n$ exponential step kernels, and convolution becomes multiplication in Laplace space. The normalization is explicit from $W(0)=1$, so $\int_0^\infty w(t)\,dt = 1$.

Assume now that one step $j$ is slower than the remaining unresolved intermediate steps, in the sense that the long-time or low-frequency regime of interest satisfies $|\zeta| \ll k_i$ for every $i \neq j$. Factoring out the slow pole gives:
\begin{equation}
    W(\zeta) = \frac{k_j}{\zeta + k_j} \prod_{i \neq j} \frac{1}{1 + \zeta/k_i}
    \label{eq:T3}
\end{equation}
For the fast factors, a first-order low-frequency expansion yields:
\begin{equation}
    \begin{aligned}
        \prod_{i \neq j} \frac{1}{1 + \zeta/k_i}
            &= 1 - \zeta \sum_{i \neq j} \frac{1}{k_i} + O\!\left( \max_{i \neq j} \frac{|\zeta|^2}{k_i^2} \right) \\
            &= 1 + O\!\left( \max_{i \neq j} \frac{|\zeta|}{k_i} \right)
    \end{aligned}
    \label{eq:T4}
\end{equation}
and therefore the full kernel reduces, to leading order, to the dominant slow pole:
\begin{equation}
    \begin{aligned}
           W(\zeta) &= \frac{k_j}{\zeta + k_j} \left[ 1 + O\!\left( \max_{i \neq j} \frac{|\zeta|}{k_i} \right) \right] \\
               &\approx \frac{k_j}{\zeta + k_j}
    \end{aligned}
    \label{eq:T5}
\end{equation}
Inverse Laplace transformation then gives the leading-order completion-time density
\begin{equation}
    w(t) \approx k_j e^{-k_j t} u(t).
    \label{eq:T6}
\end{equation}
Multiplying by the initial excitation population $N_0$ and identifying $\tau_r = 1/k_j$ recovers the source term used in the main text:
\begin{equation}
    R(t) = N_0 w(t) \approx N_0 k_j e^{-k_j t} u(t) = \frac{N_0}{\tau_r} e^{-t/\tau_r} u(t),
    \label{eq:T7}
\end{equation}
with automatic conservation of the initial population, $\int_0^\infty R(t)\,dt = N_0$.

The reduction should therefore be understood as a \emph{first-order dominant-pole approximation} to the full microscopic cascade, not as a claim that every microscopic substep in LYSO:Ce is exactly or asymptotically infinitely separated. For LYSO, hot-carrier relaxation occurs on sub-picosecond to picosecond scales, exciton migration typically occupies the $\sim$10--50~ps range, and the final trap-center capture / thermalization bottleneck is commonly parameterized at the order of $\sim$70~ps~\cite{Vasiliev2008Microscopic}. These values justify integrating out the faster unresolved stages when the measurable macroscopic rise is represented by a single effective constant $\tau_r$, while also clarifying the limit of validity: if ultrafast measurements resolve multiple comparable intermediate constants, a multi-exponential or more detailed non-Markov cascade would be required instead of \eqref{eq:T7}.

\section{Self-Absorption Cascade Derivation}
\label{app:self_absorption}
The effective decay envelope in Section~II-A is obtained by summing successive self-absorption generations in the spirit of Birowosuto \textit{et al.}~\cite{birowosuto_novel_2009}, but expressed here in the present $(\epsilon, a, \eta)$ notation. Consider an instantaneous excitation normalized to one initial emitting center. The intrinsic first-emission kernel is $\tau^{-1} e^{-t/\tau}$ for $t \ge 0$. Of these photons, the fraction $(1-\epsilon a)$ escapes directly, while the fraction $\epsilon a \eta$ is self-absorbed and recreated as a fresh excited Ce$^{3+}$ center. After $n$ recycling events, the temporal kernel is the $(n+1)$-fold self-convolution of $\tau^{-1} e^{-t/\tau}u(t)$, namely the Erlang kernel $\tau^{-1}(t/\tau)^n e^{-t/\tau}/n!$. The escaping-light response is therefore the cascade sum
\begin{align}
    h(t) &= \frac{1-\epsilon a}{\tau} e^{-t/\tau} \nonumber \\
         &\quad + \frac{1-\epsilon a}{\tau}(\epsilon a\eta)\left(\frac{t}{\tau}\right)e^{-t/\tau} \nonumber \\
         &\quad + \frac{1-\epsilon a}{\tau}\frac{(\epsilon a\eta)^2}{2!}\left(\frac{t}{\tau}\right)^2 e^{-t/\tau} + \cdots \nonumber \\
         &= \frac{1-\epsilon a}{\tau} e^{-t/\tau} \sum_{n=0}^{\infty} \frac{1}{n!} \left( \frac{\epsilon a \eta \, t}{\tau} \right)^n \nonumber \\
         &= \frac{1-\epsilon a}{\tau} \exp\!\left[-\frac{(1-\epsilon a\eta)t}{\tau}\right].
    \label{eq:U1}
\end{align}
Hence the entire recursive cascade is equivalent to a single exponential envelope with
\begin{equation}
    \tau_{eff} = \frac{\tau}{1-\epsilon a\eta}. \nonumber
\end{equation}

\section{Macroscopic Intrinsic Impulse Response Integral}
\label{app:ht_integral}
The macroscopic intrinsic impulse response $h(t)$ describes the escaping photon emission rate from the crystal per initial excitation center following an instantaneous ($\delta$-function) excitation. From the cascade summation in Appendix~\ref{app:self_absorption}, the escaping rate is
\begin{equation}
    h(t) = \frac{1-\epsilon a}{\tau} e^{-t/\tau_{eff}}u(t), \qquad t \ge 0 \nonumber
\end{equation}
Note that the prefactor is $(1-\epsilon a)/\tau$ rather than $1/\tau_{eff}$: the \emph{instantaneous rate of spontaneous emission attempts} is governed by the intrinsic radiative lifetime $\tau$, while the \emph{population decay envelope} follows the slower effective time constant $\tau_{eff}$ because re-absorption and re-emission recycling extend the apparent lifetime. The infinite temporal integral physically represents the net macroscopic photon escape probability fraction. Integrating:
\begin{align}
    \int_0^\infty h(t) dt & = \int_0^\infty \frac{1-\epsilon a}{\tau} e^{-\frac{t}{\tau_{eff}}} dt \nonumber                                                                           \\
                          & = \frac{1-\epsilon a}{\tau} \left[ -\tau_{eff} e^{-\frac{t}{\tau_{eff}}} \right]_0^\infty \nonumber                                                        \\
                          & = \frac{1-\epsilon a}{\tau} \left( \lim_{t \to \infty} \left(-\tau_{eff} e^{-\frac{t}{\tau_{eff}}}\right) - \left(-\tau_{eff} e^0\right) \right) \nonumber \\
                          & = \frac{1-\epsilon a}{\tau} \left( 0 - (-\tau_{eff}) \right) \nonumber                                                                                     \\
                          & = \frac{1-\epsilon a}{\tau} \tau_{eff} \nonumber                                                                                                           \\
                          & = \frac{1-\epsilon a}{\tau} \left( \frac{\tau}{1-\epsilon a \eta} \right) \nonumber                                                                        \\
                          & = \frac{1-\epsilon a}{1-\epsilon a \eta}
\end{align}
This gives the net macroscopic photon escape probability fraction.

\section{Convolution and Singularity Resolution of Intrinsic Emission}
\label{app:convolution_ymod}
Expanding the temporal convolution between the intrinsic impulse response and the continuous thermal cascade $R(t)$ algebraically requires isolating the integration variable $t'$ from boundary parameter $t$:
\begin{align}
    Y_{mod}(t) & = \int_0^t R(t') h(t-t') dt' \nonumber                                                                                                                      \\
               & = \frac{N_0(1-\epsilon a)}{\tau_r \tau} \int_0^t e^{-\frac{t'}{\tau_r}} e^{-\frac{t}{\tau_{eff}}} e^{\frac{t'}{\tau_{eff}}} dt' \nonumber                   \\
               & = \frac{N_0(1-\epsilon a)}{\tau_r \tau} e^{-\frac{t}{\tau_{eff}}} \int_0^t e^{t'\left(\frac{1}{\tau_{eff}} - \frac{1}{\tau_r}\right)} dt'
\end{align}
Resolving the boundaries from $t'=0$ to $t'=t$ and distributing the negative exponential factor algebraically enforces causality for $t>0$:
\begin{align}
    Y_{mod}(t) & = \frac{N_0(1-\epsilon a)}{\tau_r \tau} e^{-\frac{t}{\tau_{eff}}} \nonumber \\
               & \quad \times \left[ \frac{e^{t\left(\frac{1}{\tau_{eff}} - \frac{1}{\tau_r}\right)} - e^0}{\frac{\tau_r - \tau_{eff}}{\tau_{eff} \tau_r}} \right] \nonumber \\
               & = \frac{N_0(1-\epsilon a)\tau_{eff}}{\tau(\tau_r - \tau_{eff})} \left( e^{-\frac{t}{\tau_r}} - e^{-\frac{t}{\tau_{eff}}} \right) u(t)
\end{align}
At the pole matching condition ($\tau_r \to \tau_{eff}$), direct substitution yields a $0/0$ form. Rewriting the expression as
\begin{equation}
    Y_{mod}(t) = \frac{N_0(1-\epsilon a)\tau_{eff}}{\tau} \frac{e^{-t/\tau_{eff}} - e^{-t/\tau_r}}{\tau_{eff} - \tau_r} u(t) \nonumber
\end{equation}
and applying L'H\^opital's rule with respect to $\tau_r$ gives the critically damped limit:
\begin{align}
    \lim_{\tau_r \to \tau_{eff}} Y_{mod}(t) & = \frac{N_0(1-\epsilon a)\tau_{eff}}{\tau} \nonumber \\
                                            & \quad \times \lim_{\tau_r \to \tau_{eff}} \frac{\partial_{\tau_r} \left( e^{-t/\tau_{eff}} - e^{-t/\tau_r} \right)}{\partial_{\tau_r}(\tau_{eff} - \tau_r)} \nonumber \\
                                            & = \frac{N_0(1\!-\!\epsilon a)\tau_{eff}}{\tau} \lim_{\tau_r \to \tau_{eff}} \frac{-e^{-t/\tau_r} \cdot t\tau_r^{-2}}{-1} \nonumber \\
                                            & = \frac{N_0(1-\epsilon a)}{\tau \tau_{eff}} t e^{-\frac{t}{\tau_{eff}}} u(t)
\end{align}

\section{Verification of Photon Number Conservation Through Convolution}
\label{app:photon_conservation}
Section~II-A defines the modified intrinsic photon emission rate $Y_{mod}(t) = C_{gen}(e^{-t/\tau_{eff}} - e^{-t/\tau_r})\,u(t)$ with $C_{gen} = N_0(1-\epsilon a)\tau_{eff}/[\tau(\tau_{eff}-\tau_r)]$. We verify that the temporal integral of $Y_{mod}$ reproduces the total detectable photon yield $Y_{total} = N_0(1-\epsilon a)/(1-\epsilon a\eta)$:
\begin{align}
    \int_0^\infty Y_{mod}(t)\,dt &= C_{gen}\int_0^\infty \left(e^{-t/\tau_{eff}} - e^{-t/\tau_r}\right)dt \nonumber \\
    &= C_{gen}\left(\tau_{eff} - \tau_r\right) \nonumber \\
    &= \frac{N_0(1-\epsilon a)\tau_{eff}}{\tau(\tau_{eff}-\tau_r)} \cdot (\tau_{eff}-\tau_r) \nonumber \\
    &= \frac{N_0(1-\epsilon a)\tau_{eff}}{\tau}
    \label{eq:BB1}
\end{align}
Substituting $\tau_{eff} = \tau/(1-\epsilon a\eta)$:
\begin{equation}
    \int_0^\infty Y_{mod}(t)\,dt = \frac{N_0(1-\epsilon a)}{1-\epsilon a\eta} = Y_{total} \qquad
    \label{eq:BB2}
\end{equation}
This confirms that the convolution of the thermalization cascade $R(t)$ with the macroscopic impulse response $h(t)$ preserves the total photon number. For the critically damped limit $\tau_r \to \tau_{eff}$ (Eq.~\eqref{eq:ymod_limit}), the analogous integral evaluates as $\int_0^\infty \frac{N_0(1-\epsilon a)}{\tau\tau_{eff}} t\,e^{-t/\tau_{eff}}dt = \frac{N_0(1-\epsilon a)}{\tau\tau_{eff}} \cdot \tau_{eff}^2 = Y_{total}$, confirming photon conservation in the degenerate case as well.

\section{Geometrical Effective Velocity Expectation}
\label{app:geom_velocity}
Assuming uniform isotropic internal photon emission, photons are emitted with equal probability per unit solid angle. For the forward-propagating hemisphere ($0 \le \theta \le \pi/2$), the differential solid angle element is $d\Omega = \sin\theta\, d\theta\, d\phi$. After integrating over azimuthal symmetry ($\phi$), the probability measure for the polar angle is $p(\theta)\,d\theta \propto \sin\theta\,d\theta$, where the $\sin\theta$ factor arises from the solid angle geometry---not from a Lambert cosine projection. The mean longitudinal velocity component is:
\begin{align}
    \langle \cos \theta \rangle &= \frac{\int_0^{\pi/2} \cos\theta \sin\theta\, d\theta}{\int_0^{\pi/2} \sin\theta\, d\theta} \nonumber        \\
                                &= \frac{\left[ \frac{1}{2}\sin^2\theta \right]_0^{\pi/2}}{\left[ -\cos\theta \right]_0^{\pi/2}} = \frac{0.5 - 0}{0 - (-1)} = 0.5
\end{align}
Physically, this result states that the average projected speed along the crystal axis is half the total photon speed in the medium, providing the geometric basis for the effective velocity estimate $v_{eff} \approx 0.5 \times c/n$ discussed in Section~II-B for the isotropic forward-hemisphere case.

\section{Total Internal Reflection Correction to Effective Optical Velocity}
\label{app:tir_correction}
Section~II-B and Appendix~\ref{app:geom_velocity} derive $\langle\cos\theta\rangle = 0.5$ for isotropic emission over the full forward hemisphere. When total internal reflection (TIR) at the crystal--coupling interface is considered, the effective angular average is modified. For refractive indices $n_1$ (LYSO) and $n_2$ (optical coupling medium), the TIR critical angle is $\theta_c = \arcsin(n_2/n_1)$. Restricting the angular integration to the transmitted cone $[0, \theta_c]$:
\begin{equation}
    \langle\cos\theta\rangle_{TIR} = \frac{\int_0^{\theta_c} \cos\theta\sin\theta\,d\theta}{\int_0^{\theta_c}\sin\theta\,d\theta} = \frac{\frac{1}{2}\sin^2\theta_c}{1-\cos\theta_c}
    \label{eq:V1}
\end{equation}
For LYSO ($n_1 = 1.82$)~\cite{Mao2008Optical} coupled to optical grease ($n_2 \approx 1.50$): $\theta_c \approx 55.5^\circ$, $\cos\theta_c \approx 0.566$, yielding $\langle\cos\theta\rangle_{TIR} = (1+\cos\theta_c)/2 \approx 0.783$---approximately 57\% higher than the full-hemisphere value. This accounts for the empirical range $v_{eff} \approx 0.5$--$0.8 \times c/n$ noted in Section~II-B. In the main framework, $v_{eff}$ is treated as a free effective parameter that absorbs TIR effects, multiple reflections, and wrapping geometry.

\section{Physical Interpretation of the Triggering Parameter $\alpha$}
\label{app:alpha_derivation}
The effective triggering parameter $\alpha = -\ln(1-q/M)$ (Section~II-C) admits a rigorous combinatorial derivation. Consider $n$ photons impinging on a SiPM with photon detection efficiency $q$ and $M$ microcells. Under the uniform random allocation model, the probability that a specific microcell is \emph{not} triggered by any of the $n$ photons is:
\begin{equation}
    P(\text{not triggered}) = \left(1 - \frac{q}{M}\right)^n = e^{n\ln(1-q/M)} = e^{-\alpha n}
    \label{eq:W1}
\end{equation}
The parameter $\alpha$ therefore represents the effective logarithmic triggering parameter induced by one incident photon per microcell in the occupancy closure. In the dilute limit $q/M \ll 1$:
\begin{equation}
    \alpha = -\ln\!\left(1 - \frac{q}{M}\right) \approx \frac{q}{M} + \frac{1}{2}\left(\frac{q}{M}\right)^2 + \cdots \approx \frac{q}{M}
    \label{eq:W2}
\end{equation}
recovering the linear approximation used in Section~II-C. The logarithmic form preserves the exact binomial statistics at arbitrary $q/M$ ratios.

\section{Derivation of the Static Macroscopic Activation Rate $\lambda(t; z)$}
\label{app:lambda_derive}
Deriving the static macroscopic activation rate $\lambda(t; z)$ requires applying the elementary derivative chain rule progressively to the static binomial occupancy model:
\begin{align}
    \lambda(t; z) & = \frac{d}{dt} N_{fired}(t; z) \nonumber                                                                                            \\
                  & = \frac{d}{dt} \left\{ M \left[ 1 - \left(1 - \frac{q}{M}\right)^{n_{ph}(t; z)} \right] \right\} \nonumber                          \\
                  & = -M \frac{d}{dt} \left[ \left(1 - \frac{q}{M}\right)^{n_{ph}(t; z)} \right] \nonumber                                              \\
                  & = -M \left(1 - \frac{q}{M}\right)^{n_{ph}(t; z)} \ln\left(1 - \frac{q}{M}\right) \nonumber \\
                  & \quad \times \left( \frac{d}{dt} n_{ph}(t; z) \right) \nonumber \\
                  & = -M \ln\left(1 - \frac{q}{M}\right) \left(1 - \frac{q}{M}\right)^{n_{ph}(t; z)} r_{ph}(t; z)
\end{align}
Using the substitution $\alpha = -\ln(1-q/M)$, we algebraically extract the exact equivalence to the residual available fractional formulation:
\begin{align}
    \lambda(t; z) & = M \alpha \left( e^{-\alpha} \right)^{n_{ph}(t; z)} r_{ph}(t; z) \nonumber                      \\
                  & = M \alpha e^{-\alpha n_{ph}(t; z)} r_{ph}(t; z) \nonumber                                       \\
                  & = \alpha r_{ph}(t; z) \left[ M e^{-\alpha n_{ph}(t; z)} \right] \nonumber                        \\
                  & = \alpha r_{ph}(t; z) \left[ M - M \left( 1 - e^{-\alpha n_{ph}(t; z)} \right) \right] \nonumber \\
                  & = \alpha r_{ph}(t; z) \left[ M - N_{fired}(t; z) \right]
\end{align}

\section{Effective Dead-Time Integrals}
\label{app:dead_time_integrals}
By designating a normalized, dimensionless operational recovery ratio $\xi = t'/\tau_{rec}$, with differential $dt' = \tau_{rec}\,d\xi$, the macroscopic continuum effective dead-time is constructed by integrating the complementary lost detection capability over an unbounded microcell recovery period. Because physical avalanche output capacity is modeled as scaling quadratically with restored diode fractional overvoltage, solving analytically leads to:
\begin{align}
    \tau_{dead, phys} & = \int_0^\infty \left[ 1 - \left(1 - e^{-t'/\tau_{rec}}\right)^2 \right] dt' \nonumber \\
                      & = \tau_{rec} \int_0^\infty \left[ 1 - (1 - e^{-\xi})^2 \right] d\xi \nonumber                       \\
                      & = \tau_{rec} \int_0^\infty \left( 2e^{-\xi} - e^{-2\xi} \right) d\xi \nonumber                        \\
                      & = \tau_{rec} \left[ -2e^{-\xi} + \frac{1}{2}e^{-2\xi} \right]_0^\infty \nonumber \\
                      & = \tau_{rec} \left[(0 + 0) - \left( -2 + \frac{1}{2} \right)\right] = 1.5\tau_{rec}
    \label{eq:dead_phys}
\end{align}
Conversely, the standard piecewise binary operational ODE ensemble conceptually models states exclusively as active versus depleted, accumulating these sub-states proportionally:
\begin{align}
    \tau_{dead, ODE} & = \int_0^\infty \left[ 1 - \left(1-e^{-t'/\tau_{rec}}\right) \right] dt' \nonumber \\
                     & = \tau_{rec} \int_0^\infty e^{-\xi} d\xi \nonumber \\
                     & = \tau_{rec} \left[ -e^{-\xi} \right]_0^\infty = \tau_{rec}
    \label{eq:dead_ode}
\end{align}

\section{Complete Analytical Solution of the Dynamic Microcell ODE}
\label{app:ibp_ode}
To solve the nonlinear dynamic tracking ODE analytically, we multiply both sides by an integrating factor $\mu(t; z) = \exp \left( \alpha n_{ph}(t; z) + \frac{t}{\tau_{rec}} \right)$. Recognizing that the temporal derivative of the integrating factor expands via the chain rule to $\frac{d}{dt}\mu(t; z) = \mu(t; z) \left( \alpha r_{ph}(t; z) + \frac{1}{\tau_{rec}} \right)$, the left side structurally collapses via the mathematical product rule into an exact differential: $\frac{d}{dt} \left[ N_{busy}(t; z) \mu(t; z) \right] = \alpha M r_{ph}(t; z) \mu(t; z)$.
Integrating across the temporal domain with the initial boundary $N_{busy}(-\infty; z) = 0$ yields:
\begin{align}
     N_{busy}(t; z) \mu(t; z) &= \int_{-\infty}^{t} M e^{\frac{t'}{\tau_{rec}}} \nonumber \\
     &\quad \times \left( \alpha r_{ph}(t'; z) e^{\alpha n_{ph}(t'; z)} \right) dt'
\end{align}
To perform integration by parts ($\int u \, dv = u v - \int v \, du$) on this right hand side (RHS), we isolate the exact differential via the chain rule $dv(t') = d\left( e^{\alpha n_{ph}(t'; z)} \right) = \alpha r_{ph}(t'; z) e^{\alpha n_{ph}(t'; z)} dt'$ and assign $u(t') = M e^{t'/\tau_{rec}}$. Correspondingly, this yields the derivative component $du(t') = \frac{M}{\tau_{rec}} e^{t'/\tau_{rec}} dt'$ and the integral component $v(t') = e^{\alpha n_{ph}(t'; z)}$. Substituting these directly evaluates the integral:
\begin{align}
    \text{RHS} & = \int_{-\infty}^{t} M e^{\frac{t'}{\tau_{rec}}} d\left( e^{\alpha n_{ph}(t'; z)} \right) \nonumber                                                                       \\
               & = \bigl[ M e^{t'/\tau_{rec}} e^{\alpha n_{ph}(t'; z)} \bigr]_{-\infty}^t \nonumber \\
               & \quad - \int_{-\infty}^t e^{\alpha n_{ph}(t'; z)} \left( \frac{M}{\tau_{rec}} e^{\frac{t'}{\tau_{rec}}} \right) dt' \nonumber                                             \\
               & = M e^{\frac{t}{\tau_{rec}}} e^{\alpha n_{ph}(t; z)} \nonumber \\
               & \quad - \frac{M}{\tau_{rec}} \int_{-\infty}^t e^{\alpha n_{ph}(t'; z) + \frac{t'}{\tau_{rec}}} dt'
\end{align}
Dividing both sides by the integrating factor $\mu(t; z) = \exp\left(\alpha n_{ph}(t; z) + \frac{t}{\tau_{rec}}\right)$, the boundary term simplifies:
\begin{align}
    \frac{M e^{t/\tau_{rec}} e^{\alpha n_{ph}(t; z)}}{\mu(t; z)} &= \frac{M e^{t/\tau_{rec}} e^{\alpha n_{ph}(t; z)}}{e^{\alpha n_{ph}(t; z) + t/\tau_{rec}}} = M \nonumber
\end{align}
while the integral term transforms as:
\begin{align}
    &\frac{M}{\tau_{rec}} \frac{1}{\mu(t; z)} \int_{-\infty}^t e^{\alpha n_{ph}(t'; z) + t'/\tau_{rec}} dt' \nonumber \\
    &= \frac{M}{\tau_{rec}} \int_{-\infty}^t e^{\alpha n_{ph}(t'; z) + t'/\tau_{rec} - \alpha n_{ph}(t; z) - t/\tau_{rec}} dt' \nonumber \\
    &= \frac{M}{\tau_{rec}} \int_{-\infty}^t \exp\!\left( -\alpha [n_{ph}(t; z) - n_{ph}(t'; z)] - \frac{t-t'}{\tau_{rec}} \right) dt' \nonumber
\end{align}
Combining $N_{busy}(t; z) = M - \frac{M}{\tau_{rec}} \int_{-\infty}^t \exp\!\left( -\alpha \Delta n_{ph} - \frac{t-t'}{\tau_{rec}} \right) dt'$ (where $\Delta n_{ph} \equiv n_{ph}(t; z) - n_{ph}(t'; z) \ge 0$ for all $t' \le t$) completes the state-dependent derivation presented in Equation~\eqref{eq:analytical_dynamic}.

A direct differentiation check closes the derivation. Define the remaining integral factor as
\begin{equation}
    \begin{aligned}
        J(t;z)&=\int_{-\infty}^{t}\exp\!\left(-\alpha[n_{ph}(t;z)-n_{ph}(t';z)]\right. \\
        &\qquad\left.-\frac{t-t'}{\tau_{rec}}\right)dt'.
    \end{aligned}
    \nonumber
\end{equation}
Since $n_{ph}'(t;z)=r_{ph}(t;z)$ and the upper-limit kernel equals unity, Leibniz differentiation gives
\begin{equation}
    \frac{dJ}{dt}=1-\left(\alpha r_{ph}(t;z)+\frac{1}{\tau_{rec}}\right)J(t;z).
    \nonumber
\end{equation}
Substituting $N_{busy}=M-MJ/\tau_{rec}$ then yields
\begin{align}
    \frac{dN_{busy}}{dt}
        &= -\frac{M}{\tau_{rec}}\frac{dJ}{dt} \nonumber \\
        &= \alpha r_{ph}(t;z)\left[M-N_{busy}(t;z)\right]-\frac{N_{busy}(t;z)}{\tau_{rec}},
        \nonumber
\end{align}
which recovers the governing ODE exactly.

\section{Integrating Factor Method---Complete Intermediate Steps}
\label{app:integrating_factor_detail}
Appendix~\ref{app:ibp_ode} provides the complete ODE solution. Here we expand the key intermediate step connecting the integrating factor to the state-dependent integral. For the dynamic ODE~\eqref{eq:dynamic_ode}:
\begin{equation}
    \frac{d}{dt}N_{busy} + \underbrace{\left[\alpha r_{ph}(t;z) + \frac{1}{\tau_{rec}}\right]}_{P(t)} N_{busy} = \underbrace{\alpha M r_{ph}(t;z)}_{Q(t)}
    \label{eq:X1}
\end{equation}
the integrating factor is defined up to an arbitrary non-zero multiplicative constant. A convenient representative is $\mu(t;z) = \exp\{\alpha n_{ph}(t;z) + t/\tau_{rec}\}$, using the cumulative photon definition $n_{ph}(t;z)=\int_{-\infty}^{t}r_{ph}(t';z)\,dt'$. Its logarithmic derivative separates as:
\begin{align}
    \frac{d}{dt}\left[\alpha n_{ph}(t;z)+\frac{t}{\tau_{rec}}\right]
                        &= \alpha r_{ph}(t;z) + \frac{1}{\tau_{rec}} = P(t)
    \label{eq:X2}
\end{align}
where the first term follows directly from the cumulative photon count definition~\eqref{eq:nph_tts}. Multiplying the ODE by $\mu(t;z)$, the left-hand side becomes:
\begin{equation}
    \frac{d}{dt}\left[\mu(t;z) \cdot N_{busy}(t;z)\right] = \mu(t;z) \cdot \alpha M r_{ph}(t;z)
    \label{eq:X3}
\end{equation}
by the product rule. Integrating both sides from $-\infty$ to $t$ with the initial condition $N_{busy}(-\infty;z) = 0$ yields the integral representation in~\eqref{eq:analytical_dynamic}.

\section{Asymptotic Convergence to Static Binomial Occupancy}
\label{app:asymptotic_static_limit}
In the extremely static asymptotic regime where individual diode recovery halts indefinitely ($\tau_{rec} \to \infty$), the temporal exponential recovery functional bounds identically to $1$. Because the physical integrated event capability $d(e^{\alpha n_{ph}})$ remains fully bounded, the limit smoothly traverses the Lebesgue integration boundary spanning the event history:
\begin{align}
    \lim_{\tau_{rec} \to \infty} N_{busy}(t; z) & = M e^{-\alpha n_{ph}(t; z)} \nonumber \\
                                                & \quad \times \int_{-\infty}^t \lim_{\tau_{rec} \to \infty} e^{-\frac{t-t'}{\tau_{rec}}} d\left( e^{\alpha n_{ph}(t'; z)} \right) \nonumber \\
                                                & = M e^{-\alpha n_{ph}(t; z)} \left[ e^{\alpha n_{ph}(t'; z)} \right]_{-\infty}^t \nonumber                                                 \\
                                                & = M e^{-\alpha n_{ph}(t; z)} \bigl[ e^{\alpha n_{ph}(t; z)} \nonumber \\
                                                & \quad - e^{\alpha n_{ph}(-\infty; z)} \bigr]
\end{align}
Assigning the boundary constraint representing complete pre-event temporal inactivity $n_{ph}(-\infty; z) = 0$, leading intrinsically to $e^{\alpha n_{ph}(-\infty; z)} = 1$. Utilizing the algebraic relationship mapping static depletion $\alpha = -\ln(1-q/M)$, this mathematically recovers identically the fractional generalized summation sequence:
\begin{align}
    \lim_{\tau_{rec} \to \infty} N_{busy}(t; z) & = M \left( 1 - e^{-\alpha n_{ph}(t; z)} \right) \nonumber                                        \\
                                                & = M \left[ 1 - \exp\left( \ln\left(1 - \frac{q}{M}\right) n_{ph}(t; z) \right) \right] \nonumber \\
                                                & = M \left[ 1 - \left(1 - \frac{q}{M}\right)^{n_{ph}(t; z)} \right]
\end{align}

\section{Integration by Parts for Macroscopic SiPM Current}
\label{app:ibp_macroscopic}
For both static and dynamic currents, bypassing the numerical convolutions of fluctuating tracking rate equations necessitates extracting derivatives over the physical SiPM state populations $N_{fired}$ via continuous integrations by parts. Assigning general variables $u(t') = i_{ser}(t-t')$ and corresponding derivatives structurally demands application of the implicit internal chain rule representing delay: $\frac{\partial}{\partial t'} e^{-\frac{t-t'}{\tau_x}} = e^{-\frac{t-t'}{\tau_x}} \cdot (-\frac{1}{\tau_x}) \cdot (-1) = \frac{1}{\tau_x} e^{-\frac{t-t'}{\tau_x}}$. For the generalized $I_{stat}$, assigning $dv = dN_{fired}$ bounds mathematically into:
\begin{align}
    I_{stat}(t; z) & = \int_{-\infty}^t i_{ser}(t-t') \, d N_{fired}(t'; z) \nonumber                                                               \\
                   & = \Big[ N_{fired}(t; z) i_{ser}(0) \nonumber \\
                   & \quad - \lim_{t' \to -\infty} N_{fired}(t'; z) i_{ser}(t-t') \Big] \nonumber                                           \\
                   & \quad - \int_{-\infty}^t N_{fired}(t'; z) \nonumber \\
                   & \quad \times \left( I_{cell}^{macro} \sum_{x} \frac{A_x}{\tau_x} e^{-\frac{t-t'}{\tau_x}} \right) dt'
\end{align}
Canceling the normalized initial causal summation amplitude $i_{ser}(0) = I_{cell}^{macro}$ mathematically enforces explicit cancellation of boundaries providing \eqref{eq:i_stat}.

For the dynamic current, the convolution $I_{dyn}(t; z) = \int_{-\infty}^t \lambda_{dyn}(t'; z)\, i_{ser}(t-t')\, dt'$ is decomposed using the ODE identity $\lambda_{dyn}(t'; z) = \frac{\partial}{\partial t'} N_{busy}(t'; z) + \frac{N_{busy}(t'; z)}{\tau_{rec}}$ from \eqref{eq:dynamic_rec}. This splits the integral into two independent terms:
\begin{align}
    I_{dyn}(t; z) &= \underbrace{\int_{-\infty}^t \frac{\partial N_{busy}}{\partial t'} i_{ser}(t-t')\, dt'}_{\text{Term 1: derivative component}} \nonumber \\
                  &\quad + \underbrace{\int_{-\infty}^t \frac{N_{busy}(t'; z)}{\tau_{rec}} i_{ser}(t-t')\, dt'}_{\text{Term 2: recovery component}} \nonumber
\end{align}

\textit{Term~1} is evaluated via integration by parts with $u = i_{ser}(t-t')$ and $dv = \frac{\partial N_{busy}}{\partial t'} dt'$:
\begin{align}
    \text{Term 1} &= \left[ N_{busy}(t'; z)\, i_{ser}(t-t') \right]_{-\infty}^t \nonumber \\
                  &\quad - \int_{-\infty}^t N_{busy}(t'; z) \frac{\partial}{\partial t'} i_{ser}(t-t')\, dt' \nonumber
\end{align}
The boundary term evaluates to $N_{busy}(t; z)\, i_{ser}(0) - 0 = I_{cell}^{macro} N_{busy}(t; z)$, since $i_{ser}(0) = I_{cell}^{macro}$ and $N_{busy}(-\infty; z) = 0$. The derivative of the delayed response, applying the chain rule with $\frac{\partial}{\partial t'} (t-t') = -1$, gives:
\begin{equation}
    \frac{\partial}{\partial t'} i_{ser}(t-t') = I_{cell}^{macro} \sum_{x} \frac{A_x}{\tau_x} e^{-\frac{t-t'}{\tau_x}} \nonumber
\end{equation}
(note the positive sign: $\frac{\partial}{\partial t'} e^{-(t-t')/\tau_x} = +\frac{1}{\tau_x} e^{-(t-t')/\tau_x}$). Thus:
\begin{align}
    \text{Term 1} &= I_{cell}^{macro} N_{busy}(t; z) \nonumber \\
                  &\quad - I_{cell}^{macro} \int_{-\infty}^t N_{busy}(t'; z) \sum_{x} \frac{A_x}{\tau_x} e^{-\frac{t-t'}{\tau_x}} dt' \nonumber
\end{align}

\textit{Term~2} expands directly:
\begin{align}
    \text{Term 2} &= \frac{I_{cell}^{macro}}{\tau_{rec}} \int_{-\infty}^t N_{busy}(t'; z) \sum_{x} A_x e^{-\frac{t-t'}{\tau_x}} dt' \nonumber
\end{align}

Combining Terms 1 and 2, the two integrals merge under a common integrand:
\begin{align}
    I_{dyn}(t; z) &= I_{cell}^{macro} N_{busy}(t; z) - I_{cell}^{macro} \int_{-\infty}^t N_{busy}(t'; z) \nonumber \\
                  &\quad \times \sum_{x} A_x \underbrace{\left( \frac{1}{\tau_x} - \frac{1}{\tau_{rec}} \right)}_{\text{factored coefficient}} e^{-\frac{t-t'}{\tau_x}} dt'
\end{align}
which is the result stated in \eqref{eq:i_dyn_explicit}.

\section{Boundary Condition Validation for Current Convolution}
\label{app:boundary_validation}
The integration by parts in Appendix~\ref{app:ibp_macroscopic}, which converts the rate convolution $I_{stat}(t;z) = \int_{-\infty}^t \lambda(t';z)\,i_{ser}(t-t')\,dt'$ into the state-dependent form~\eqref{eq:i_stat}, requires:
\begin{equation}
    \lim_{t' \to -\infty} N_{fired}(t';z) \cdot i_{ser}(t-t') = 0
    \label{eq:Y1}
\end{equation}
This product vanishes through two independent physical mechanisms:
\begin{enumerate}
    \item \textbf{Causal onset}: Prior to the scintillation event, no photons have arrived at the SiPM, so $N_{fired}(-\infty;z) = 0$ (no microcells triggered).
    \item \textbf{Exponential decay}: On the causal branch $\Delta t = t - t' \ge 0$, the single-cell response satisfies $i_{ser}(\Delta t) \propto \sum_x A_x e^{-\Delta t/\tau_x} \to 0$ as $\Delta t \to \infty$, with the slowest single-cell pole $\tau_x = 4~\mathrm{ns}$ in the representative parameter set.
\end{enumerate}
In the present event-driven formulation, dark counts are not absorbed into $N_{fired}$; they are introduced later through $\lambda_{tot}(t;z)$. If one instead superimposed a stationary dark-count baseline onto the occupancy variable, the same exponential suppression of $i_{ser}$ would still ensure convergence. The upper boundary at $t' = t$ yields $N_{fired}(t;z) \cdot i_{ser}(0)$, which is finite and contributes the expected instantaneous term.

\section{Formal Proof of Non-Negativity of the Dynamic Current}
\label{app:nonneg_proof}
We prove $I_{dyn}(t;z) \ge 0$ for all $t$, as claimed in Section~II-D. By definition~\eqref{eq:i_dyn}:
\begin{equation}
    I_{dyn}(t;z) = \int_{-\infty}^t \lambda_{dyn}(t';z)\,i_{ser}(t-t')\,dt'
    \label{eq:CC1}
\end{equation}
The integrand is the product of two non-negative factors:
\begin{enumerate}
    \item \textbf{Dynamic triggering rate:} $\lambda_{dyn}(t';z) = \alpha\,r_{ph}(t';z)\,[M - N_{busy}(t';z)] \ge 0$, because $\alpha \ge 0$, $r_{ph} \ge 0$, and $0 \le N_{busy}(t';z) \le M$. These state bounds follow from the integral representation~\eqref{eq:analytical_dynamic}: the integral term is non-negative, giving $N_{busy}\le M$, while $n_{ph}(t;z)-n_{ph}(t';z)\ge0$ for $t'\le t$ implies the integrand is bounded above by $e^{-(t-t')/\tau_{rec}}$, so the integral is at most $\tau_{rec}$ and $N_{busy}\ge M-(M/\tau_{rec})\tau_{rec}=0$.
    \item \textbf{Single-cell response:} On the causal domain $t \ge t'$, $i_{ser}(t-t') = I_{cell}^{macro}\sum_x A_x e^{-(t-t')/\tau_x} \ge 0$, since $I_{cell}^{macro} > 0$, all $A_x \ge 0$, and all $\tau_x > 0$.
\end{enumerate}
Since the integrand is the product of two non-negative functions over the integration domain $t' \in (-\infty, t]$, the integral $I_{dyn}(t;z) \ge 0$. $\square$

The integration-by-parts form~\eqref{eq:i_dyn_explicit} introduces terms with coefficients $(1/\tau_x - 1/\tau_{rec})$ that may be positive or negative depending on the pole ordering. While individual terms in this decomposition can be algebraically negative, they are exactly compensated by the boundary term $I_{cell}^{macro} N_{busy}(t;z)$, ensuring that the total expression remains non-negative and identical to~\eqref{eq:CC1}.

\section{Asymptotic Equivalence of Dynamic Current to Static Current}
\label{app:dynamic_to_static}
To mathematically verify the consistency of the tracking equations, we assess the asymptotic limit where the microcell recovery halts indefinitely ($\tau_{rec} \to \infty$). In this limit, the transient dynamic state bounds strictly back to the static model $N_{busy}(t; z) \to N_{fired}(t; z)$, and the inverse recovery coupling coefficient approaches $\frac{1}{\tau_{rec}} \to 0$. Substituting these bounded limits into the expanded dynamic current expectation enforces its structural reduction to the static expectation:
\begin{align}
    \lim_{\tau_{rec} \to \infty} I_{dyn}(t; z) & = I_{cell}^{macro} \lim_{\tau_{rec} \to \infty} N_{busy}(t; z) \nonumber                         \\
                                               & \quad - I_{cell}^{macro} \int_{-\infty}^t \lim_{\tau_{rec} \to \infty} N_{busy}(t'; z) \nonumber \\
                                               & \quad \times \!\sum_{x} A_x \bigl( \tau_x^{-1} - 0 \bigr) e^{-(t-t')/\tau_x} dt' \nonumber       \\
                                               & = I_{cell}^{macro} N_{fired}(t; z) \nonumber                                                     \\
                                               & \quad - I_{cell}^{macro} \int_{-\infty}^t N_{fired}(t'; z) \nonumber                             \\
                                               & \quad \times \sum_{x} \frac{A_x}{\tau_x} e^{-(t-t')/\tau_x}\, dt'
\end{align}
As derived previously via integration by parts, this RHS bounds identically to $I_{stat}(t; z)$.

\section{Commutative Decoupling of LTI Spatial Convolution}
\label{app:commutative_lti}
Operating in the linear regime where $\alpha n_{ph}(t; z) \ll 1$, the macroscopic current transforms into a sequence of continuous Linear Time-Invariant convolutions. By expanding the macroscopic photon rate geometric dependence $r_{ph}(t; z) = k_{trans} (Y_{mod} * f_{TTS}(\cdot; z))(t)$, we demonstrate that LTI associative ordering permits decoupling of the static SiPM spatial properties from the optical geometric transfer:
\begin{align}
    I_{lin}(t; z) & = q k_{trans} \left[ (Y_{mod} * f_{TTS}(\cdot; z)) * i_{ser} \right](t) \nonumber                \\
                  & = q k_{trans} \left[ (Y_{mod} * i_{ser}) * f_{TTS}(\cdot; z) \right](t) \nonumber                \\
                  & = \left[ \left( q k_{trans} (Y_{mod} * i_{ser}) \right) * f_{TTS}(\cdot; z) \right](t) \nonumber \\
                  & = (I_{ideal} * f_{TTS}(\cdot; z))(t)
\end{align}

\section{Analytical Expansion and Asymptotic Reductions of the Ideal Kernel}
\label{app:ideal_kernel}
Computing $I_{ideal}(t) = \int_0^t [q k_{trans} Y_{mod}(t')] i_{ser}(t-t') dt'$ requires distributing the bi-exponential forms. Substituting $Y_{mod}(t') = C_{gen}(e^{-t'/\tau_{eff}} - e^{-t'/\tau_r})u(t')$ and $i_{ser}(t-t') = I_{cell}^{macro}\sum_x A_x e^{-(t-t')/\tau_x}$, and defining the global scale factor $C = q k_{trans} C_{gen} I_{cell}^{macro}$:
\begin{align}
    I_{ideal}(t) &= C \sum_{x} A_x \int_0^t \bigl( e^{-t'/\tau_{eff}} - e^{-t'/\tau_r} \bigr) \nonumber \\
                 &\quad \times e^{-(t-t')/\tau_x}\, dt' \nonumber \\
                 &= C \sum_{x} A_x e^{-t/\tau_x} \nonumber \\
                 &\quad \times \int_0^t \!\bigl( e^{t'(\tau_x^{-1} - \tau_{eff}^{-1})} - e^{t'(\tau_x^{-1} - \tau_r^{-1})} \bigr) dt' \nonumber
\end{align}
where we factored $e^{-(t-t')/\tau_x} = e^{-t/\tau_x} e^{t'/\tau_x}$ and combined exponents. Each integral evaluates as:
\begin{align}
    &\int_0^t e^{t'(\tau_x^{-1} - \tau_{eff}^{-1})} dt' = \frac{e^{t(\tau_x^{-1} - \tau_{eff}^{-1})} - 1}{\tau_x^{-1} - \tau_{eff}^{-1}} \nonumber \\
    &\quad = \tilde{\tau}_{x,eff}\bigl( e^{t(\tau_x^{-1} - \tau_{eff}^{-1})} - 1 \bigr) \nonumber
\end{align}
with $\tilde{\tau}_{x,eff} = (\tau_x^{-1} - \tau_{eff}^{-1})^{-1} = \frac{\tau_x \tau_{eff}}{\tau_{eff} - \tau_x}$ (and analogously for $\tilde{\tau}_{x,r}$). Multiplying back by $e^{-t/\tau_x}$:
\begin{align}
    &e^{-t/\tau_x} \cdot \tilde{\tau}_{x,eff}\left( e^{t(\tau_x^{-1} - \tau_{eff}^{-1})} - 1 \right) \nonumber \\
    &\quad = \tilde{\tau}_{x,eff}\left( e^{-t/\tau_x} e^{t(\tau_x^{-1} - \tau_{eff}^{-1})} - e^{-t/\tau_x} \right) \nonumber \\
    &\quad = \tilde{\tau}_{x,eff}\left( e^{-t/\tau_{eff}} - e^{-t/\tau_x} \right) \nonumber
\end{align}
since $e^{-t/\tau_x} \cdot e^{t(\tau_x^{-1} - \tau_{eff}^{-1})} = e^{-t/\tau_{eff}}$. Collecting both branches:
\begin{align}
    I_{ideal}(t) &= C \sum_{x} A_x \Big[ \tilde{\tau}_{x,eff} \left(e^{-t/\tau_{eff}} - e^{-t/\tau_x}\right) \nonumber \\
                 &\quad\quad\quad\quad\quad\quad - \tilde{\tau}_{x,r} \left(e^{-t/\tau_r} - e^{-t/\tau_x}\right) \Big] u(t) \nonumber
\end{align}
which confirms \eqref{eq:i_ideal_final}.

\textit{Singularity Resolution ($\tau_x \to \tau_{eff}$).} As $\tau_x \to \tau_{eff}$, the coupling constant $\tilde{\tau}_{x,eff} \to \infty$ while the bracket $(e^{-t/\tau_{eff}} - e^{-t/\tau_x}) \to 0$, forming an indeterminate $\infty \cdot 0$ product. To resolve this, we apply the algebraic variable mapping $u = 1/\tau_x$ (with $u_0 = 1/\tau_{eff}$) and recognize the derivative definition of the limit:
\begin{align}
    \lim_{\tau_x \to \tau_{eff}} & \tilde{\tau}_{x,eff} \left(e^{-t/\tau_{eff}} - e^{-t/\tau_x}\right) \nonumber \\
                                 & = \lim_{u \to u_0} \frac{e^{-t u_0} - e^{-t u}}{u - u_0} \nonumber \\
                                 & = - \lim_{u \to u_0} \frac{e^{-t u} - e^{-t u_0}}{u - u_0} \nonumber \\
                                 & = -\left. \frac{d}{du} e^{-tu} \right|_{u=u_0} = t e^{-t/\tau_{eff}} \nonumber
\end{align}
This bounded result replaces the divergent product with a physically meaningful critically damped resonance profile.

\textit{Singularity Resolution ($\tau_x \to \tau_r$).} The recovery-coupling branch is treated identically. With $u = 1/\tau_x$ and $u_0 = 1/\tau_r$:
\begin{align}
    \lim_{\tau_x \to \tau_r} & \tilde{\tau}_{x,r} \left(e^{-t/\tau_r} - e^{-t/\tau_x}\right) \nonumber \\
                               & = \lim_{u \to u_0} \frac{e^{-t u_0} - e^{-t u}}{u - u_0} \nonumber \\
                               & = - \lim_{u \to u_0} \frac{e^{-t u} - e^{-t u_0}}{u - u_0} \nonumber \\
                               & = -\left. \frac{d}{du} e^{-tu} \right|_{u=u_0} = t e^{-t/\tau_r} \nonumber
\end{align}
This establishes the critically damped resonance profile quoted in Section~II-D when a SiPM current-response pole matches the thermalization cascade constant.

\textit{Asymptotic Limit $\tau_r \to 0$ and $C_{inst}$ Derivation.} As $\tau_r \to 0$, both the coupling constant $\tilde{\tau}_{x,r}$ and the bracket $(e^{-t/\tau_r} - e^{-t/\tau_x})$ require careful joint analysis. Applying L'H\^opital's rule to the product by writing it as a ratio:
\begin{align}
    \lim_{\tau_r \to 0} \tilde{\tau}_{x,r} \left( e^{-t/\tau_r} - e^{-t/\tau_x} \right) &= \lim_{\tau_r \to 0} \frac{e^{-t/\tau_r} - e^{-t/\tau_x}}{\tau_x^{-1} - \tau_r^{-1}} \nonumber
\end{align}
As $\tau_r \to 0$: the numerator approaches $0 - e^{-t/\tau_x}$ (for $t > 0$, $e^{-t/\tau_r} \to 0$) and the denominator diverges to $-\infty$. Thus the ratio $\to 0$ for all $t > 0$. At $t=0$, the numerator is $1-1=0$ identically, confirming the product vanishes uniformly.

Simultaneously, the generation scale factor $C_{gen} = \frac{N_0(1-\epsilon a)\tau_{eff}}{\tau(\tau_{eff} - \tau_r)}$ simplifies as $\tau_r \to 0$:
\begin{equation}
    \lim_{\tau_r \to 0} C_{gen} = \frac{N_0(1-\epsilon a)\tau_{eff}}{\tau \cdot \tau_{eff}} = \frac{N_0(1-\epsilon a)}{\tau} \nonumber
\end{equation}
Therefore the collapsed global scale factor is:
\begin{align}
    C_{inst} &= q k_{trans} \!\left[\lim_{\tau_r \to 0} C_{gen}\right]\! I_{cell}^{macro} \nonumber \\
             &= q k_{trans} \frac{N_0(1-\epsilon a)}{\tau} I_{cell}^{macro} \nonumber
\end{align}
confirming the definition stated in Section~II-D.
\section{Normalization of the EMG Function}
\label{app:emg_normalization}
Section~II-D notes that the EMG function is not a normalized probability density. Here we prove $\int_{-\infty}^{\infty} \mathrm{EMG}(t;\mu,\sigma,\tau)\,dt = \tau$. By definition:
\begin{align}
    &\int_{-\infty}^{\infty} \mathrm{EMG}(t;\mu,\sigma,\tau)\,dt \nonumber \\
    &= \int_{-\infty}^{\infty}\left[\int_0^{\infty} e^{-s/\tau} f_{TTS}(t-s;\mu,\sigma)\,ds\right]dt
    \label{eq:Z1}
\end{align}
Since the integrand is non-negative everywhere, the Fubini--Tonelli theorem permits interchanging the order of integration:
\begin{align}
    &= \int_0^{\infty} e^{-s/\tau} \underbrace{\left[\int_{-\infty}^{\infty} f_{TTS}(t-s;\mu,\sigma)\,dt\right]}_{= 1 \text{ (Gaussian normalization)}}\,ds \nonumber \\
    &= \int_0^{\infty} e^{-s/\tau}\,ds = \tau
    \label{eq:Z2}
\end{align}
Physically, the integral $\int \mathrm{EMG}\,dt = \tau$ has dimensions of time, which is dimensionally consistent when $\mathrm{EMG}$ appears inside current expressions of the form $I(t) \propto \sum_x C_x \cdot \mathrm{EMG}(t;\mu,\sigma,\tau_x)$, where $C_x$ carries units of $[\mathrm{current}/\mathrm{time}]$.

\section{Derivation of the Exponentially Modified Gaussian (EMG) Kernel}
\label{app:emg_square}
Computing the convolution of a generic causal exponential $e^{-s/\tau} u(s)$ with the Gaussian kernel $f_{TTS}(t-s; z)$ requires completing the square in the exponent. By algebraically isolating the localized components, the combined exponent argument simplifies to:
\begin{align}
     & -\frac{s}{\tau} - \frac{(t-s-\mu)^2}{2\sigma^2} \nonumber                                                                                                               \\
     & = -\frac{1}{2\sigma^2} \left[ 2\sigma^2 \frac{s}{\tau} + (t-s-\mu)^2 \right] \nonumber                                                                                  \\
     & = -\frac{1}{2\sigma^2} \left[ 2\sigma^2 \frac{s}{\tau} + s^2 - 2s(t-\mu) + (t-\mu)^2 \right] \nonumber                                                                  \\
     & = -\frac{1}{2\sigma^2} \left[ s^2 - 2s\left(t-\mu - \frac{\sigma^2}{\tau}\right) + (t-\mu)^2 \right] \nonumber                                                          \\
     & = -\frac{1}{2\sigma^2} \Bigg[ s^2 - 2s\left(t-\mu - \frac{\sigma^2}{\tau}\right) + \left(t-\mu - \frac{\sigma^2}{\tau}\right)^2 \nonumber \\
                                  &\quad \quad \quad \quad - \left(t-\mu - \frac{\sigma^2}{\tau}\right)^2 + (t-\mu)^2 \Bigg] \nonumber                               \\
     & = -\frac{1}{2\sigma^2} \Bigg[ \left(s - \left(t-\mu - \frac{\sigma^2}{\tau}\right)\right)^2 \nonumber \\
                                  &\quad \quad \quad \quad - \left( (t-\mu)^2 - 2(t-\mu)\frac{\sigma^2}{\tau} + \frac{\sigma^4}{\tau^2} \right) + (t-\mu)^2 \Bigg] \nonumber   \\
     & = -\frac{1}{2\sigma^2} \left[ \left(s - \left(t-\mu - \frac{\sigma^2}{\tau}\right)\right)^2 + 2(t-\mu)\frac{\sigma^2}{\tau} - \frac{\sigma^4}{\tau^2} \right] \nonumber \\
     & = -\frac{\left( s - \left(t-\mu - \frac{\sigma^2}{\tau}\right) \right)^2}{2\sigma^2} - \frac{t-\mu}{\tau} + \frac{\sigma^2}{2\tau^2}
\end{align}
Because of the inherent causality imposed by the Heaviside step constraint $u(s)$, the integration domain is restricted to $s \in [0, \infty)$. Substituting this expanded exponent back into the convolution integral extracts the temporal decay components outside the integral:
\begin{align}
     & \int_{0}^{\infty} e^{-\frac{s}{\tau}} \frac{1}{\sqrt{2\pi}\sigma} e^{-\frac{(t-s-\mu)^2}{2\sigma^2}} ds \nonumber                                                   \\
     & = \int_{0}^{\infty} \frac{1}{\sqrt{2\pi}\sigma} \exp\Bigg( -\frac{\left( s - \left(t-\mu - \frac{\sigma^2}{\tau}\right) \right)^2}{2\sigma^2} \nonumber             \\
     & \quad \quad \quad \quad \quad \quad \quad - \frac{t-\mu}{\tau} + \frac{\sigma^2}{2\tau^2} \Bigg) ds \nonumber                                                       \\
     & = \exp\left( \frac{\sigma^2}{2\tau^2} - \frac{t-\mu}{\tau} \right) \nonumber                                                                                        \\
     & \quad \times \int_{0}^{\infty} \frac{1}{\sqrt{2\pi}\sigma} \exp\left( -\frac{\left( s - \left(t-\mu - \frac{\sigma^2}{\tau}\right) \right)^2}{2\sigma^2} \right) ds
\end{align}
By applying the integral variable substitution $v = \frac{s - (t - \mu - \sigma^2/\tau)}{\sqrt{2}\sigma}$ (which yields the explicit differential $ds = \sqrt{2}\sigma dv$ and maps the lower boundary $s=0$ to $v_0 = \frac{\mu - t + \sigma^2/\tau}{\sqrt{2}\sigma}$) and recognizing the complementary error function definition $\operatorname{erfc}(x) = \frac{2}{\sqrt{\pi}} \int_x^\infty e^{-v^2} dv$, the integral reduces to the standard functional form:
\begin{align}
    \int_{\frac{\mu - t + \sigma^2/\tau}{\sqrt{2}\sigma}}^{\infty} \frac{1}{\sqrt{2\pi}\sigma} e^{-v^2} (\sqrt{2}\sigma dv) & = \frac{1}{\sqrt{\pi}} \int_{\frac{\mu - t + \sigma^2/\tau}{\sqrt{2}\sigma}}^{\infty} e^{-v^2} dv \nonumber \\
                                                                                                                            & = \frac{1}{2} \operatorname{erfc}\left( \frac{\mu - t + \sigma^2/\tau}{\sqrt{2}\sigma} \right)
\end{align}

\section{Derivation of the Borel Cascade Excess Noise Factor}
\label{app:borel_enf}
We derive the moments of the Borel distribution from first principles using the probability generating function (PGF). Let $G$ denote the total cascade size initiated by a single primary avalanche. Each fired microcell independently triggers a Poisson-distributed number of daughter avalanches with mean $P_{ct}$. If $N_D \sim \mathrm{Poisson}(P_{ct})$ denotes the number of direct daughters produced by the initiating avalanche and $G_1, G_2, \ldots$ are independent copies of the full cascade size $G$, then the Galton--Watson recursion is
\begin{equation}
    G = 1 + \sum_{j=1}^{N_D} G_j
    \nonumber
\end{equation}
The PGF $\Pi(\varsigma) = \mathbb{E}[\varsigma^G]$ therefore satisfies the implicit functional equation:
\begin{equation}
    \Pi(\varsigma) = \varsigma \cdot \exp\!\big( P_{ct}[\Pi(\varsigma) - 1] \big)
    \nonumber
\end{equation}
where the factor $\varsigma$ accounts for the initiating primary avalanche and $\exp(P_{ct}[\Pi(\varsigma)-1])$ is the PGF of a Poisson$(P_{ct})$-distributed number of independent descendant sub-cascades.

\textit{Mean $\langle G \rangle$.} Differentiating both sides with respect to $\varsigma$:
\begin{align}
    \Pi'(\varsigma) &= e^{P_{ct}(\Pi-1)} \left[ 1 + \varsigma P_{ct} \Pi'(\varsigma) \right] \nonumber
\end{align}
Evaluating at $\varsigma=1$ (where $\Pi(1)=1$):
\begin{align}
    \Pi'(1) &= 1 \cdot \left[ 1 + P_{ct} \Pi'(1) \right] \nonumber \\
    \Pi'(1)(1 - P_{ct}) &= 1 \nonumber \\
    \langle G \rangle &= \Pi'(1) = \frac{1}{1-P_{ct}} \nonumber
\end{align}

\textit{Second factorial moment $\mathbb{E}[G(G-1)]$.} Differentiating $\Pi'(\varsigma)$ again and evaluating at $\varsigma=1$:
\begin{align}
    \Pi''(\varsigma) &= P_{ct} \Pi'(\varsigma) e^{P_{ct}(\Pi-1)} [1 + \varsigma P_{ct} \Pi'(\varsigma)] \nonumber \\
             &\quad + e^{P_{ct}(\Pi-1)} [P_{ct} \Pi'(\varsigma) + \varsigma P_{ct} \Pi''(\varsigma)] \nonumber
\end{align}
At $\varsigma=1$, substituting $\Pi(1)=1$ and $\Pi'(1) = \frac{1}{1-P_{ct}}$:
\begin{align}
    \Pi''(1) &= P_{ct} \Pi'(1) [1 + P_{ct} \Pi'(1)] \nonumber \\
    &\quad + P_{ct} \Pi'(1) + P_{ct} \Pi''(1) \nonumber \\
    \Pi''(1)(1 - P_{ct}) &= P_{ct} \Pi'(1) [2 + P_{ct} \Pi'(1)] \nonumber \\
    &= \frac{P_{ct}}{1-P_{ct}} \cdot \frac{2-P_{ct}}{1-P_{ct}} \nonumber \\
    \Pi''(1) &= \frac{P_{ct}(2-P_{ct})}{(1-P_{ct})^3} \nonumber
\end{align}

\textit{Variance.} Using $\mathbb{E}[G^2] = \Pi''(1) + \Pi'(1)$:
\begin{align}
    \operatorname{Var}(G) &= \mathbb{E}[G^2] - \langle G \rangle^2 \nonumber \\
    &= \frac{P_{ct}(2-P_{ct})}{(1-P_{ct})^3} + \frac{1}{1-P_{ct}} - \frac{1}{(1-P_{ct})^2} \nonumber \\
    &= \frac{P_{ct}(2-P_{ct}) + (1-P_{ct})^2 - (1-P_{ct})}{(1-P_{ct})^3} \nonumber
\end{align}
Expanding the numerator: $2P_{ct} - P_{ct}^2 + 1 - 2P_{ct} + P_{ct}^2 - 1 + P_{ct} = P_{ct}$, yielding:
\begin{equation}
    \operatorname{Var}(G) = \frac{P_{ct}}{(1-P_{ct})^3} \nonumber
\end{equation}

\textit{Excess Noise Factor.} Defining $F_{ENF} = \frac{\mathbb{E}[G^2]}{\langle G \rangle^2} = 1 + \frac{\operatorname{Var}(G)}{\langle G \rangle^2}$:
\begin{align}
    F_{ENF} &= 1 + \frac{P_{ct} / (1-P_{ct})^3}{1 / (1-P_{ct})^2} \nonumber    \\
            &= 1 + \frac{P_{ct}}{1-P_{ct}} \nonumber                           \\
            &= \frac{1-P_{ct} + P_{ct}}{1-P_{ct}} = \frac{1}{1-P_{ct}}
\end{align}

\section{Extension of Campbell's Theorem to Non-Stationary Poisson Processes}
\label{app:campbell_extension}
The variance expressions~\eqref{eq:poisson_noise} and~\eqref{eq:sigma_total} utilize the non-stationary generalization of Campbell's theorem~\cite{Rice1944Mathematical, Snyder1991Random}. For a stationary Poisson process with constant rate $\lambda_0$ and filter $h(t)$, the classical Campbell theorem gives:
\begin{equation}
    \mathrm{Var}\!\left[\sum_k h(t-t_k)\right] = \lambda_0 \int_{-\infty}^{\infty} h^2(u)\,du
    \label{eq:AA1}
\end{equation}
When the Poisson intensity becomes time-varying, $\lambda(t)$, the variance generalizes through conditional expectation over the inhomogeneous process~\cite{Snyder1972Filtering}:
\begin{equation}
    \sigma^2(t) = \int_{-\infty}^{t} \lambda(t_j)\,[h(t-t_j)]^2\,dt_j
    \label{eq:AA2}
\end{equation}
where $h(t-t_j)$ is the single-event response kernel. If each primary event additionally carries an independent multiplicative mark $m_j$ with $\mathbb{E}[m_j]=1$ and second moment $\mathbb{E}[m_j^2]$, the variance simply acquires the scalar factor $\mathbb{E}[m_j^2]$:
\begin{equation}
    \sigma^2(t) = \mathbb{E}[m_j^2] \int_{-\infty}^{t} \lambda(t_j)\,[h(t-t_j)]^2\,dt_j
    \nonumber
\end{equation}
In the present framework, the physical event amplitude is $G I_{cell}^{single}$, while the deterministic kernel is expressed in terms of the mean-mapped amplitude $I_{cell}^{macro} = \langle G \rangle I_{cell}^{single}$. Writing the normalized mark as $m_j = G_j/\langle G \rangle$ gives $\mathbb{E}[m_j^2] = \mathbb{E}[G^2]/\langle G \rangle^2 = F_{ENF}$, so the marked-process variance reduces exactly to the prefactor used in \eqref{eq:poisson_noise}. The remaining unmarked kernel is $h(t-t_j) = I_{cell}^{macro}\sum_x A_x e^{-(t-t_j)/\tau_x}$. This result requires conditional independence of primary events given $\lambda(t)$ and conditional independence of the multiplicative marks from the primary event times. As noted in Section~\ref{sec:poisson_noise_modeling}, these assumptions become progressively violated once occupancy-driven dead-time correlations become strong. The resulting sub-Poissonian statistics yield actual variances \emph{below} the Campbell prediction, making~\eqref{eq:AA2} a conservative (pessimistic) upper bound on the noise level.

\section{Integral Expansion of Steady-State Baseline Variance}
\label{app:integral_variance}
The stationary thermal noise covariance integrates quadratically over the single-electron impulse response. Setting the substitution variable $u = t - t_j$ maps the differential domain $du = -dt_j$ and boundaries $(-\infty, t) \to (\infty, 0)$. Processing this analytically simplifies the sum over orthogonal poles:
\begin{align}
    \sigma_{base}^2 & = F_{ENF} \nu_{DCR} \left(I_{cell}^{macro}\right)^2 \nonumber                                                                                                                                         \\
                    & \quad \times \int_{\infty}^0 \sum_{x,y} A_x A_y e^{-u\left(\frac{1}{\tau_x} + \frac{1}{\tau_y}\right)} (-du) \nonumber                                                                                \\
                    & = F_{ENF} \nu_{DCR} \left(I_{cell}^{macro}\right)^2 \nonumber                                                                                                                                         \\
                    & \quad \times \sum_{x,y} A_x A_y \left[ \frac{-1}{\frac{1}{\tau_x} + \frac{1}{\tau_y}} e^{-u \left( \frac{1}{\tau_x} + \frac{1}{\tau_y} \right)} \right]_0^\infty \nonumber                            \\
                    & = F_{ENF} \nu_{DCR} \left(I_{cell}^{macro}\right)^2 \nonumber                                                                                                                                         \\
                    & \quad \times \sum_{x,y} \left( \frac{-A_x A_y}{\frac{1}{\tau_x} + \frac{1}{\tau_y}} \right) \nonumber \\
                    & \quad \times \left( \lim_{u \to \infty} e^{-u\left(\frac{1}{\tau_x} + \frac{1}{\tau_y}\right)} - e^0 \right) \nonumber \\
                    & = F_{ENF} \nu_{DCR} \left(I_{cell}^{macro}\right)^2 \nonumber \\
                    & \quad \times \sum_{x,y} \left( \frac{-A_x A_y}{\frac{1}{\tau_x} + \frac{1}{\tau_y}} \right) (0 - 1) \nonumber                                                  \\
                    & = F_{ENF} \nu_{DCR} \left(I_{cell}^{macro}\right)^2 \sum_{x,y} \frac{A_x A_y \tau_x \tau_y}{\tau_x + \tau_y}
\end{align}

\section{Explicit Expansion of the Baseline Variance Cross-Terms}
\label{app:baseline_crossterms}
Equation~\eqref{eq:sigma_base} expresses the baseline variance as a double sum $\sum_{x,y} A_x A_y \tau_x \tau_y / (\tau_x + \tau_y)$. For the three poles $x \in \{d, p1, p2\}$, this yields nine terms (three diagonal and six cross-terms). The diagonal terms contribute:
\begin{equation}
    \sigma_{diag}^2 = F_{ENF}\,\nu_{DCR}\,(I_{cell}^{macro})^2 \sum_{x} \frac{A_x^2 \tau_x}{2}
    \label{eq:DD1}
\end{equation}
while the cross-terms are:
\begin{align}
    \sigma_{cross}^2 &= F_{ENF}\,\nu_{DCR}\,(I_{cell}^{macro})^2 \nonumber \\
    &\quad \times 2\!\sum_{x<y}\! A_x A_y \frac{\tau_x \tau_y}{\tau_x + \tau_y}
    \label{eq:DD2}
\end{align}
Using the numerical values $A_d = 0.98$, $A_{p1} = 0.015$, $A_{p2} = 0.005$, $\tau_d = 0.5$~ns, $\tau_{p1} = 1.5$~ns, $\tau_{p2} = 4$~ns:
\begin{itemize}
    \item Dominant term: $A_d^2 \tau_d/2 = 0.9604 \times 0.25 = 0.2401$~ns
    \item Largest cross-term: $2A_d A_{p1}\,\tau_d\tau_{p1}/(\tau_d+\tau_{p1}) = 0.01103$~ns
    \item Next-largest remaining cross-term: $2A_d A_{p2}\,\tau_d\tau_{p2}/(\tau_d+\tau_{p2}) = 0.00436$~ns; the other diagonal and cross-terms are below about $2 \times 10^{-4}$~ns
\end{itemize}
The fast discharge pole therefore dominates the baseline variance by ${\approx}\,94\%$, confirming that the secondary current-response poles $p1$ and $p2$ contribute negligibly to the steady-state noise floor.

\section{Derivation of Fisher Information for Poisson Rate Estimation}
\label{app:fisher_info_derivation}
Consider, at fixed DOI $z$ and with all other waveform parameters treated as known, a non-homogeneous Poisson point process with rate $\lambda_{tot}(t; t_{int}) = \lambda_{dyn}(t - t_{int}; z) + \nu_{DCR}$, where $t_{int}$ is the unknown interaction time to be estimated. The log-likelihood of observing $N_{\mathrm{ev}}$ events at times $\{t_1, \ldots, t_{N_{\mathrm{ev}}}\}$ over the post-interaction observation window $[0, T]$ is~\cite{Snyder1991Random}:
\begin{equation}
    \ell(t_{int}) = \sum_{j=1}^{N_{\mathrm{ev}}} \ln \lambda_{tot}(t_j; t_{int}) - \int_0^T \lambda_{tot}(t; t_{int})\,dt \nonumber
\end{equation}
Taking the derivative with respect to $t_{int}$ and noting $\frac{\partial \lambda_{tot}}{\partial t_{int}} = -\dot{\lambda}_{dyn}(t - t_{int}; z)$ (since $\nu_{DCR}$ is independent of $t_{int}$), the score function is:
\begin{equation}
    S = -\sum_{j=1}^{N_{\mathrm{ev}}} \frac{\dot{\lambda}_{dyn}(t_j - t_{int})}{\lambda_{tot}(t_j; t_{int})} + \int_0^T \dot{\lambda}_{dyn}(t - t_{int})\,dt \nonumber
\end{equation}
The Fisher Information $\mathcal{I} = \mathbb{E}[S^2]$ requires evaluating the expectation of the squared score. Defining the local score weight $\chi(t) \equiv \dot{\lambda}_{dyn}(t-t_{int})/\lambda_{tot}(t; t_{int})$ and $B \equiv \int_0^T \dot{\lambda}_{dyn}(t-t_{int})\,dt$ for brevity:
\begin{equation}
    S^2 = \left(\sum_j \chi(t_j)\right)^2 - 2B \sum_j \chi(t_j) + B^2 \nonumber
\end{equation}

For a Poisson process, Campbell's theorem gives the first and second moment identities:
\begin{align}
    \mathbb{E}\!\left[\sum_j \chi(t_j)\right] &= \int_0^T \chi(t)\,\lambda_{tot}(t)\,dt \nonumber \\
    \mathbb{E}\!\left[\left(\sum_j \chi(t_j)\right)^{\!2}\right] &= \int_0^T \chi^2(t)\,\lambda_{tot}(t)\,dt \nonumber \\
    &\quad + \left(\int_0^T \chi(t)\,\lambda_{tot}(t)\,dt\right)^{\!2} \nonumber
\end{align}
The second identity follows from the independence of increments: the squared sum decomposes into diagonal terms ($j=j'$, yielding the integral of $\chi^2 \lambda_{tot}$) and off-diagonal terms ($j \neq j'$, yielding the squared mean).

Substituting $\chi(t) = \dot{\lambda}_{dyn}/\lambda_{tot}$:
\begin{align}
    \int_0^T \chi(t)\,\lambda_{tot}(t)\,dt &= \int_0^T \dot{\lambda}_{dyn}\,dt = B \nonumber \\
    \int_0^T \chi^2(t)\,\lambda_{tot}(t)\,dt &= \int_0^T \frac{\dot{\lambda}_{dyn}^2}{\lambda_{tot}}\,dt \nonumber
\end{align}
Assembling $\mathbb{E}[S^2]$:
\begin{align}
    \mathcal{I} &= \underbrace{\int_0^T \frac{\dot{\lambda}_{dyn}^2}{\lambda_{tot}}\,dt + B^2}_{\mathbb{E}[(\sum \chi)^2]} - \underbrace{2B \cdot B}_{2B\,\mathbb{E}[\sum \chi]} + B^2 \nonumber \\
    &= \int_0^T \frac{\dot{\lambda}_{dyn}^2}{\lambda_{tot}}\,dt + B^2 - 2B^2 + B^2 \nonumber \\
    &= \int_0^T \frac{\left[\dot{\lambda}_{dyn}(t; z)\right]^2}{\lambda_{dyn}(t; z) + \nu_{DCR}} \, dt \nonumber
\end{align}
Extending $T \to \infty$:
\begin{equation}
    \mathcal{I}(z) = \int_{0}^{\infty} \frac{\left[\dot{\lambda}_{dyn}(t; z)\right]^2}{\lambda_{dyn}(t; z) + \nu_{DCR}} \, dt
\end{equation}
This is exact for any Poisson point process and requires no Gaussian or white-noise approximation. The $B^2$ terms cancel identically, confirming that no separate dependence on the integrated derivative $B = \int \dot{\lambda}_{dyn}\,dt$ appears as an additive term. However, the Fisher Information retains an intrinsic dependence on both the \emph{shape} and the \emph{amplitude} of the rate function through the ratio $\dot{\lambda}_{dyn}^2/\lambda_{tot}$: scaling $\lambda_{dyn} \to A\lambda_{dyn}$ with $\nu_{DCR} = 0$ yields $\mathcal{I} \to A\mathcal{I}$, reflecting the fundamental statistical scaling with total photon count.

\section{Waveform-Observation Fisher Information and the Role of $i_{ser}(t)$}
\label{app:waveform_fisher}

The point-process Fisher Information derived in Appendix~\ref{app:fisher_info_derivation} assumes an idealized observer that has direct access to the avalanche triggering times. If instead the estimator operates on the analog SiPM current waveform or its sampled digitized version, the single-photoelectron response kernel $i_{ser}(t)$ must appear explicitly because every avalanche is filtered by that kernel before observation.

\textit{Step 1: Mean waveform under a time-shift parameter.} Let the unknown interaction time be denoted locally by $\vartheta \equiv t_{int}$. Under the waveform-observation model, the deterministic mean current is the shifted dynamic current:
\begin{equation}
    \mu_{\vartheta}(t; z) = I_{dyn}(t-\vartheta; z) = \int_{-\infty}^{t} \lambda_{dyn}(u-\vartheta; z) \, i_{ser}(t-u) \, du
    \label{eq:wf_mean}
\end{equation}
For the shift parameter $\vartheta$, the derivative of the mean waveform is therefore:
\begin{equation}
    \frac{\partial \mu_{\vartheta}(t; z)}{\partial \vartheta} = -\dot{I}_{dyn}(t-\vartheta; z)
    \label{eq:wf_mean_derivative}
\end{equation}
Equation~\eqref{eq:wf_mean} makes explicit that, unlike the point-process model, the waveform mean depends on the single-photoelectron response through the convolution kernel $i_{ser}(t)$.

\textit{Step 2: Waveform covariance kernel.} Under the filtered compound-Poisson approximation of Section~\ref{sec:poisson_noise_modeling}, plus additive electronics noise, the continuous-time covariance kernel may be written as:
\begin{equation}
    \begin{aligned}
        K_{\vartheta}(t,s; z)
            &\approx F_{ENF} \int_{-\infty}^{\infty} \lambda_{tot}(u-\vartheta; z) \\
            &\quad \times i_{ser}(t-u) \, i_{ser}(s-u) \, du + \sigma_{elec}^2 \, \delta(t-s)
    \end{aligned}
    \label{eq:wf_covariance}
\end{equation}
where $\lambda_{tot}(t; z) \approx \lambda_{dyn}(t; z) + \nu_{DCR}$ and $\delta(\cdot)$ is the Dirac delta. On the diagonal, this kernel reduces to the instantaneous variance envelope already used in Section~\ref{sec:poisson_noise_modeling}, namely $K_{\vartheta}(t,t; z) = \sigma_{total}^2(t-\vartheta; z)$ under the same approximation hierarchy. Equation~\eqref{eq:wf_covariance} shows that $i_{ser}(t)$ affects not only the mean waveform but also the temporal correlation structure of the shot noise.

\textit{Step 3: Fisher Information for sampled Gaussian waveforms.} Suppose the waveform is sampled at times $\{t_1, \ldots, t_N\}$, yielding an observation vector $\mathbf{y} \sim \mathcal{N}(\boldsymbol{\mu}_{\vartheta}, \mathbf{\Sigma}_{\vartheta})$ with entries $[\boldsymbol{\mu}_{\vartheta}]_k = \mu_{\vartheta}(t_k; z)$ and $[\mathbf{\Sigma}_{\vartheta}]_{kl} = K_{\vartheta}(t_k,t_l; z)$. The exact Fisher Information for the Gaussian waveform model is then:
\begin{equation}
    \begin{aligned}
        \mathcal{I}_{wf}(z)
            &= \left(\frac{\partial \boldsymbol{\mu}_{\vartheta}}{\partial \vartheta}\right)^{\!T} \mathbf{\Sigma}_{\vartheta}^{-1} \left(\frac{\partial \boldsymbol{\mu}_{\vartheta}}{\partial \vartheta}\right) \\
            &\quad + \frac{1}{2} \operatorname{Tr}\!\left( \mathbf{\Sigma}_{\vartheta}^{-1} \frac{\partial \mathbf{\Sigma}_{\vartheta}}{\partial \vartheta} \mathbf{\Sigma}_{\vartheta}^{-1} \frac{\partial \mathbf{\Sigma}_{\vartheta}}{\partial \vartheta} \right)
    \end{aligned}
    \label{eq:wf_fisher_general}
\end{equation}
The first term measures how strongly the mean waveform changes with the time shift, while the second term accounts for the time-shift dependence of the waveform covariance itself.

\textit{Step 4: Practical reduction.} If the covariance varies only weakly with the shift parameter across the narrow leading-edge window dominating the timing estimate, the second term in \eqref{eq:wf_fisher_general} may be neglected to first order. If one further adopts a local white-noise approximation,
\begin{equation}
    K_{\vartheta}(t,s; z) \approx \sigma_{total}^2(t-\vartheta; z) \, \delta(t-s)
    \label{eq:wf_white_noise}
\end{equation}
then the waveform Fisher Information reduces to:
\begin{equation}
    \mathcal{I}_{wf}^{(1)}(z) \approx \int_{0}^{\infty} \frac{\dot{I}_{dyn}^2(t; z)}{\sigma_{total}^2(t; z)} \, dt
    \label{eq:wf_fisher_reduced}
\end{equation}
This is the waveform-level analogue of the event-level point-process result \eqref{eq:fisher_info}. In contrast to \eqref{eq:fisher_info}, however, the single-photoelectron response enters explicitly through both $I_{dyn}(t; z) = (\lambda_{dyn} * i_{ser})(t; z)$ and the covariance kernel \eqref{eq:wf_covariance}. Consequently, any broadening of $i_{ser}(t)$ reduces the attainable timing information by both smoothing the leading-edge slope and redistributing the shot-noise power over time.

\section{From Fisher Information to Coincidence CTR FWHM}
\label{app:ctr_crlb_derivation}
This appendix derives the explicit relationship between the Poisson Fisher Information $\mathcal{I}(z)$ and the coincidence timing resolution CTR expressed as FWHM.

\textit{Step 1: Gaussian FWHM conversion.} For a Gaussian-distributed timing error with probability density $p(\Delta t) = \frac{1}{\sqrt{2\pi}\sigma}\exp(-\Delta t^2 / 2\sigma^2)$, the FWHM is defined by the condition $p(\Delta t_{1/2}) = \frac{1}{2}p(0)$. This yields:
\begin{align}
    \exp\!\left(-\frac{\Delta t_{1/2}^2}{2\sigma^2}\right) &= \frac{1}{2} \nonumber \\
    \Delta t_{1/2}^2 &= 2\sigma^2 \ln 2 \nonumber \\
    \mathrm{FWHM} &= 2\Delta t_{1/2} = 2\sqrt{2\ln 2}\,\sigma \approx 2.355\,\sigma \nonumber
\end{align}

\textit{Step 2: Coincidence variance.} For any unbiased single-detector estimator, the Cram\'er-Rao inequality gives $\operatorname{Var}(\hat t_{int}) \ge \mathcal{I}^{-1}(z)$. We denote the bound itself by $\sigma_{CRLB}^2(z) = \mathcal{I}^{-1}(z)$. In a coincidence measurement, two independent detectors each contribute an independent timing error. The measured time difference $\Delta t_{coinc} = t_1 - t_2$ has variance:
\begin{align}
    \sigma_{coinc}^2(z) &= \sigma_{CRLB,1}^2(z) + \sigma_{CRLB,2}^2(z) \nonumber \\
                        &= 2\sigma_{CRLB}^2(z) = \frac{2}{\mathcal{I}(z)} \nonumber
\end{align}
where the second equality holds for identical detector pairs.

\textit{Step 3: Combined result.} Applying the Gaussian FWHM conversion to the coincidence standard deviation:
\begin{align}
    \mathrm{CTR}_{CRLB}(z) &= 2\sqrt{2\ln 2}\,\sigma_{coinc}(z) \nonumber \\
    &= 2\sqrt{2\ln 2} \cdot \sqrt{\frac{2}{\mathcal{I}(z)}} \nonumber \\
    &= 2\sqrt{2\ln 2} \cdot \sqrt{2} \cdot \frac{1}{\sqrt{\mathcal{I}(z)}} \nonumber \\
    &= 4\sqrt{\ln 2} \cdot \frac{1}{\sqrt{\mathcal{I}(z)}} \approx \frac{3.330}{\sqrt{\mathcal{I}(z)}}
\end{align}
This confirms that the coincidence CTR CRLB depends solely on the Poisson Fisher Information integral, which in turn encapsulates all detector physics: scintillation kinetics ($\tau_{eff}$, $N_0$), optical transport ($\sigma_{TTS}$), dynamic saturation ($M$, $\tau_{rec}$), photon detection efficiency ($q$), and dark count rate ($\nu_{DCR}$).

\section{Energy-Resolution Fisher Bounds for an Arbitrary Deposited Energy}
\label{app:energy_resolution_fisher}

This appendix extends the point-process Cram\'er-Rao construction used for the interaction-time bound to deposited-energy estimation. The timing problem estimates a shift parameter $t_{int}$ and is therefore controlled by the time derivative $\dot{\lambda}_{dyn}$. The energy problem estimates the local deposited energy $E$ and is controlled by the energy derivative $\partial_E\lambda_{dyn}$. Throughout this appendix, the DOI $z$, detector parameters, and interaction time are treated as known unless explicitly promoted to nuisance parameters.

\textit{Step 1: Energy parameterization of the forward model.} In the main text, the initial excitation-center population is denoted by $N_0$ and is fixed for each numerical case. For an energy-resolution bound, this scalar is promoted locally to a differentiable mean function
\begin{equation}
    \bar N_0(E),
    \label{eq:energy_n0_function}
\end{equation}
where $E$ is the deposited energy. If scintillator nonproportionality is neglected over the energy interval of interest, the linear specialization is
\begin{equation}
    \bar N_0(E) = Y_0 E,
    \label{eq:energy_linear_yield}
\end{equation}
with $Y_0$ the mean number of initial excitation centers per unit energy. The 511~keV reference used in Section~\ref{sec:verification} corresponds to $N_0=16000$, giving $Y_0 \approx 16000/511 \approx 31.3~\mathrm{keV}^{-1}$ under this linear calibration.

The self-absorption cascade in Section~II-A gives the escaping-light fraction, denoted by $\rho_{esc}$,
\begin{equation}
    \rho_{esc} = \frac{1-\epsilon a}{1-\epsilon a\eta}.
    \label{eq:energy_rho_esc}
\end{equation}
For fixed DOI and optical transport efficiency, the expected total number of photons incident on the SiPM surface, denoted by $\bar n_{ph,\infty}(E,z)$, is therefore
\begin{equation}
    \bar n_{ph,\infty}(E,z) = \bar N_0(E)\rho_{esc}k_{trans}(z).
    \label{eq:energy_incident_photons}
\end{equation}
The instantaneous incident photon rate is the same transport-convolved rate as in \eqref{eq:rph_tts}, now with its dependence on $E$ made explicit:
\begin{equation}
    r_{ph}(t;E,z) = k_{trans}(z)\left[Y_{mod}(\cdot;E)*f_{TTS}(\cdot;z)\right](t).
    \label{eq:energy_rph}
\end{equation}
When the scintillation time constants and transport kernel are energy independent over the local interval, the energy dependence enters only through the population scale:
\begin{equation}
    r_{ph}(t;E,z) = \bar N_0(E)\psi_{ph}(t;z),
    \label{eq:energy_rph_shape}
\end{equation}
where $\psi_{ph}(t;z)$ is the incident photon-rate shape per initial excitation center. The dynamic SiPM recovery model is then
\begin{align}
    \frac{d}{dt}N_{busy}(t;E,z)
        &= \lambda_{dyn}(t;E,z)-\frac{N_{busy}(t;E,z)}{\tau_{rec}}, \label{eq:energy_dynamic_rec} \\
    \lambda_{dyn}(t;E,z)
        &= \alpha r_{ph}(t;E,z)\left[M-N_{busy}(t;E,z)\right], \label{eq:energy_lambda_dyn} \\
    \alpha &= -\ln\left(1-\frac{q}{M}\right). \nonumber
\end{align}
Thus $E$ changes not only the photon-rate amplitude but also the microcell-availability trajectory when finite-cell saturation becomes appreciable.

\textit{Step 2: Ideal primary-trigger Fisher Information.} Consider first an ideal observer that has direct access to the primary avalanche triggering times, rather than to the filtered analog current. Over a finite observation window $[0,T]$, the total point-process rate is
\begin{equation}
    \lambda_{tot}(t;E,z) = \lambda_{dyn}(t;E,z)+\nu_{DCR},
    \label{eq:energy_lambda_tot}
\end{equation}
where $\nu_{DCR}$ is independent of $E$. For observed event times $\{t_j\}_{j=1}^{N_{\mathrm{ev}}}$, the Poisson log-likelihood is
\begin{equation}
    \ell(E) = \sum_{j=1}^{N_{\mathrm{ev}}}\ln\lambda_{tot}(t_j;E,z)
        - \int_0^T\lambda_{tot}(t;E,z)\,dt.
    \label{eq:energy_poisson_likelihood}
\end{equation}
Taking the energy derivative gives the score
\begin{equation}
    S_E = \sum_{j=1}^{N_{\mathrm{ev}}}
        \frac{\partial_E\lambda_{dyn}(t_j;E,z)}{\lambda_{tot}(t_j;E,z)}
        - \int_0^T\partial_E\lambda_{dyn}(t;E,z)\,dt.
    \label{eq:energy_score}
\end{equation}
Define
\begin{align}
    h_E(t) &\equiv
        \frac{\partial_E\lambda_{dyn}(t;E,z)}{\lambda_{tot}(t;E,z)}, \nonumber \\
    B_E &\equiv \int_0^T\partial_E\lambda_{dyn}(t;E,z)\,dt.
    \label{eq:energy_score_weight}
\end{align}
Campbell's theorem for a non-homogeneous Poisson process gives
\begin{align}
    \mathbb{E}\!\left[\sum_j h_E(t_j)\right]
        &= \int_0^T h_E(t)\lambda_{tot}(t;E,z)\,dt = B_E, \nonumber \\
    \operatorname{Var}\!\left[\sum_j h_E(t_j)\right]
        &= \int_0^T h_E^2(t)\lambda_{tot}(t;E,z)\,dt. \nonumber
\end{align}
As in Appendix~\ref{app:fisher_info_derivation}, the integrated-derivative terms cancel in $\mathbb{E}[S_E^2]$, yielding
\begin{equation}
    \mathcal{I}_E(E,z;T) = \int_0^T
        \frac{\left[\partial_E\lambda_{dyn}(t;E,z)\right]^2}
             {\lambda_{dyn}(t;E,z)+\nu_{DCR}}\,dt.
    \label{eq:energy_fisher_finite}
\end{equation}
When the acquisition window covers the full scintillation pulse, the limiting expression is
\begin{equation}
    \boxed{
    \mathcal{I}_E(E,z) = \int_0^{\infty}
        \frac{\left[\partial_E\lambda_{dyn}(t;E,z)\right]^2}
             {\lambda_{dyn}(t;E,z)+\nu_{DCR}}\,dt .
    }
    \label{eq:energy_fisher}
\end{equation}
Equation~\eqref{eq:energy_fisher} is the deposited-energy analogue of \eqref{eq:fisher_info}. The replacement is
\begin{equation}
    \dot{\lambda}_{dyn}(t;z) \quad \longrightarrow \quad \partial_E\lambda_{dyn}(t;E,z).
    \label{eq:energy_derivative_replacement}
\end{equation}
The single-cell response $i_{ser}(t)$ does not appear in this ideal bound because the observation is the underlying primary-trigger point process itself.

For any unbiased single-detector estimator of $E$ at fixed $z$,
\begin{equation}
    \operatorname{Var}(\hat E) \ge \sigma_{E,CRLB}^2(E,z)
        = \frac{1}{\mathcal{I}_E(E,z)}.
    \label{eq:energy_crlb_variance}
\end{equation}
At a reference energy or photopeak center $E_0$, the relative FWHM energy-resolution bound is
\begin{equation}
    \boxed{
    \begin{aligned}
    R_{E_0,primary}(z)
        &\equiv \Delta E_{\mathrm{FWHM},CRLB}(E_0,z)/E_0 \\
        &= \frac{2\sqrt{2\ln2}}{E_0\sqrt{\mathcal{I}_E(E_0,z)}} \\
        &\approx \frac{2.355}{E_0\sqrt{\mathcal{I}_E(E_0,z)}} .
    \end{aligned}
    }
    \label{eq:energy_relative_primary}
\end{equation}
Unlike coincidence timing, no additional factor of $\sqrt{2}$ appears here because this is a single-detector energy-estimation problem.

\textit{Step 3: Energy sensitivity ODE.} The practical computation of \eqref{eq:energy_fisher} requires $\partial_E\lambda_{dyn}$. Differentiating \eqref{eq:energy_dynamic_rec} with respect to $E$ gives a stable sensitivity equation. Let
\begin{equation}
    S_E(t;E,z) \equiv \partial_E N_{busy}(t;E,z).
    \label{eq:energy_sensitivity_definition}
\end{equation}
For the separable rate model \eqref{eq:energy_rph_shape},
\begin{equation}
    \partial_E r_{ph}(t;E,z)
        = \frac{d\ln\bar N_0(E)}{dE}\,r_{ph}(t;E,z),
    \label{eq:energy_rph_derivative_general}
\end{equation}
which reduces to
\begin{equation}
    \partial_E r_{ph}(t;E,z) = \frac{r_{ph}(t;E,z)}{E}
    \label{eq:energy_rph_derivative_linear}
\end{equation}
when $\bar N_0(E)=Y_0E$. If nonproportionality or energy-dependent pulse-shape changes are important, their derivatives should be retained inside $\partial_E r_{ph}$; the sensitivity ODE below is unchanged.

Differentiating the dynamic recovery equation gives
\begin{equation}
    \boxed{
    \begin{aligned}
    \frac{dS_E}{dt}
        &= \alpha\,\partial_E r_{ph}(t;E,z) \\
        &\quad \times \left[M-N_{busy}(t;E,z)\right] \\
        &\quad - \left[\alpha r_{ph}(t;E,z)+\tau_{rec}^{-1}\right] \\
        &\quad \times S_E(t;E,z),
    \end{aligned}
    }
    \label{eq:energy_sensitivity_ode}
\end{equation}
with $S_E(-\infty;E,z)=0$. The corresponding energy derivative of the dynamic activation rate is
\begin{equation}
    \boxed{
    \begin{aligned}
    \partial_E\lambda_{dyn}(t;E,z)
        &= \alpha\,\partial_E r_{ph}(t;E,z) \\
        &\quad \times \left[M-N_{busy}(t;E,z)\right] \\
        &\quad - \alpha r_{ph}(t;E,z)S_E(t;E,z).
    \end{aligned}
    }
    \label{eq:energy_lambda_derivative}
\end{equation}
The first term is the direct increase in photon-driven trigger rate with deposited energy. The second term is the finite-microcell correction: increasing $E$ also increases the busy-cell population, thereby reducing the available-cell factor $M-N_{busy}$.

\textit{Step 4: Linear low-occupancy limit.} The full expression must reduce to the familiar photon-counting limit. In the linear, unsaturated, dark-count-free regime,
\begin{align}
    \lambda_{dyn}(t;E,z) &\approx \lambda_{lin}(t;E,z), \nonumber \\
    \lambda_{lin}(t;E,z) &= E\,s(t;z), \nonumber \\
    \partial_E\lambda_{lin}(t;E,z)&=s(t;z).
    \label{eq:energy_linear_lambda}
\end{align}
Substitution into \eqref{eq:energy_fisher} gives
\begin{equation}
    \mathcal{I}_E(E,z) = \int_0^{\infty}\frac{s^2(t;z)}{E s(t;z)}\,dt
        = \frac{1}{E}\int_0^{\infty}s(t;z)\,dt.
    \label{eq:energy_linear_fisher_step}
\end{equation}
Since the mean primary photoelectron count is
\begin{equation}
    \bar N_{pe}(E,z) = \int_0^{\infty}\lambda_{lin}(t;E,z)\,dt
        = E\int_0^{\infty}s(t;z)\,dt,
    \label{eq:energy_npe_linear}
\end{equation}
one obtains
\begin{equation}
    \mathcal{I}_E(E,z)=\frac{\bar N_{pe}(E,z)}{E^2}.
    \label{eq:energy_linear_fisher_npe}
\end{equation}
Therefore, the relative FWHM energy-resolution bound $R_{E,\mathrm{FWHM}}$ satisfies
\begin{equation}
    \boxed{
    \frac{\sigma_E}{E} \ge \frac{1}{\sqrt{\bar N_{pe}(E,z)}},
    \qquad
    R_{E,\mathrm{FWHM}} \ge \frac{2.355}{\sqrt{\bar N_{pe}(E,z)}}.
    }
    \label{eq:energy_poisson_limit}
\end{equation}
Equation~\eqref{eq:energy_fisher} is thus the extension of the Poisson photoelectron-counting limit to finite rise time, optical transport, DCR, recovery, and saturation.

When $Y_0$, $\rho_{esc}$, $k_{trans}$, and $q$ are locally energy independent, the low-occupancy mean primary photoelectron count $\bar N_{pe,lin}(E_0,z)$ is
\begin{equation}
    \bar N_{pe,lin}(E_0,z) \approx qY_0E_0\rho_{esc}k_{trans}(z),
    \label{eq:energy_npe_scaling}
\end{equation}
and the ideal low-occupancy scaling becomes
\begin{equation}
    \boxed{
    R_{E_0,\mathrm{Poisson}}(z)
        \approx \frac{2.355}{\sqrt{qY_0E_0\rho_{esc}k_{trans}(z)}} .
    }
    \label{eq:energy_poisson_scaling}
\end{equation}
This gives the familiar $E_0^{-1/2}$ relative-resolution improvement only within the linear regime. The full dynamic Fisher expression need not follow this simple scaling at high occupancy or in the presence of scintillator nonproportionality.

\textit{Step 5: Static saturation check.} A closed-form saturation check is obtained by ignoring intra-event recovery and observing only the final number of fired microcells. In the linear light-yield case, write
\begin{equation}
    \bar n_{ph,\infty}(E,z)=b(z)E,
    \qquad
    b(z)=Y_0\rho_{esc}k_{trans}(z).
    \label{eq:energy_b_definition}
\end{equation}
The probability that a given microcell has fired at least once is
\begin{equation}
    p(E,z)=1-\exp[-\alpha b(z)E].
    \label{eq:energy_static_p}
\end{equation}
Approximating the final occupancy by $K\sim\mathrm{Binomial}(M,p)$, the binomial Fisher Information for $E$ is
\begin{equation}
    \mathcal{I}_{E,stat}(E,z)
        = M\frac{[\partial_Ep(E,z)]^2}{p(E,z)[1-p(E,z)]}.
    \label{eq:energy_static_fisher_general}
\end{equation}
Since $\partial_Ep=\alpha b(z)\exp[-\alpha b(z)E]$,
\begin{equation}
    \boxed{
    \mathcal{I}_{E,stat}(E,z)
        = M[\alpha b(z)]^2
          \frac{\exp[-\alpha b(z)E]}{1-\exp[-\alpha b(z)E]} .
    }
    \label{eq:energy_static_fisher}
\end{equation}
The corresponding relative FWHM bound is
\begin{equation}
    \boxed{
    \begin{aligned}
    R_{E,stat}(E,z) \ge \frac{2.355}{E}
        \left[
        M[\alpha b(z)]^2
        \right. \\
        \left.
        \times\frac{\exp[-\alpha b(z)E]}{1-\exp[-\alpha b(z)E]}
        \right]^{-1/2} .
    \end{aligned}
    }
    \label{eq:energy_static_resolution}
\end{equation}
For $\alpha bE\ll1$, \eqref{eq:energy_static_fisher} reduces to $\mathcal{I}_{E,stat}\approx M\alpha b/E\approx qb/E=\bar N_{pe,lin}(E,z)/E^2$, recovering \eqref{eq:energy_linear_fisher_npe}. For large $\alpha bE$, the exponential factor suppresses information because additional photons increasingly encounter already fired microcells.

\textit{Step 6: Charge and waveform observation bounds.} The primary-trigger bound is an ideal benchmark. Energy spectra are more commonly estimated from integrated charge, pulse area, pulse height, or a full sampled waveform. In those observation models, the single-cell current kernel and compound avalanche statistics enter explicitly.

Let $G$ be the total avalanche multiplicity initiated by one primary trigger. As in Section~\ref{sec:poisson_noise_modeling}, crosstalk is represented by the mean gain $\langle G\rangle$ and excess noise factor
\begin{align}
    F_{ENF} &= \frac{\mathbb{E}[G^2]}{\langle G\rangle^2}, \nonumber \\
    F_{ENF} &= \frac{1}{1-P_{ct}}
    \quad\text{(unbounded Borel approx.).}
    \label{eq:energy_enf}
\end{align}
Let $Q_{cell}^{macro}$ denote the mean-gain-mapped integrated charge of one fired microcell. An ideal linear integrated-charge model can then be written for the charge mean $\mu_Q(E,z)$ and charge variance $\sigma_Q^2(E,z)$ as
\begin{align}
    \mu_Q(E,z) &= Q_{cell}^{macro}\bar N_{pe}(E,z), \label{eq:energy_charge_mean} \\
    \sigma_Q^2(E,z) &\approx (Q_{cell}^{macro})^2F_{ENF} \nonumber \\
        &\quad \times [\bar N_{pe}(E,z)+\nu_{DCR}T]
        + \sigma_{Q,elec}^2. \label{eq:energy_charge_variance}
\end{align}
For a Gaussian charge observation,
\begin{equation}
    \boxed{
    \mathcal{I}_{E,Q}
        = \frac{[\partial_E\mu_Q(E,z)]^2}{\sigma_Q^2(E,z)}
        + \frac{[\partial_E\sigma_Q^2(E,z)]^2}{2\sigma_Q^4(E,z)} .
    }
    \label{eq:energy_charge_fisher}
\end{equation}
When dark counts, electronics noise, and the covariance-derivative term are negligible, this reduces to
\begin{equation}
    \boxed{
    \frac{\sigma_E^2}{E^2} \gtrsim \frac{F_{ENF}}{\bar N_{pe}(E,z)},
    \qquad
    R_{E,\mathrm{charge}} \gtrsim 2.355\sqrt{\frac{F_{ENF}}{\bar N_{pe}(E,z)}} .
    }
    \label{eq:energy_charge_limit}
\end{equation}

At the full waveform level,
\begin{equation}
    I_{dyn}(t;E,z) = \int_{-\infty}^{t}\lambda_{dyn}(u;E,z)i_{ser}(t-u)\,du,
    \label{eq:energy_waveform_mean_continuous}
\end{equation}
and
\begin{equation}
    \partial_E I_{dyn}(t;E,z)
        = \int_{-\infty}^{t}\partial_E\lambda_{dyn}(u;E,z)i_{ser}(t-u)\,du.
    \label{eq:energy_waveform_derivative_continuous}
\end{equation}
For sampled waveform observations $\mathbf{y}\sim\mathcal{N}(\boldsymbol{\mu}_E,\mathbf{\Sigma}_E)$ with $[\boldsymbol{\mu}_E]_n=I_{dyn}(t_n;E,z)$, the Gaussian waveform Fisher Information is
\begin{equation}
    \boxed{
    \begin{aligned}
    \mathcal{I}_{E,wf}
        &= (\partial_E\boldsymbol{\mu}_E)^T\mathbf{\Sigma}_E^{-1}(\partial_E\boldsymbol{\mu}_E) \\
        &\quad + \frac{1}{2}\operatorname{Tr}\!\left(
        \mathbf{\Sigma}_E^{-1}\partial_E\mathbf{\Sigma}_E
        \mathbf{\Sigma}_E^{-1}\partial_E\mathbf{\Sigma}_E
        \right).
    \end{aligned}
    }
    \label{eq:energy_waveform_fisher}
\end{equation}
With a locally white-noise reduction analogous to \eqref{eq:wf_white_noise},
\begin{equation}
    \boxed{
    \mathcal{I}_{E,wf}^{(1)}(E,z)
        \approx \int_0^{\infty}
        \frac{[\partial_E I_{dyn}(t;E,z)]^2}{\sigma_{total}^2(t;E,z)}\,dt .
    }
    \label{eq:energy_waveform_fisher_reduced}
\end{equation}
The resulting waveform-observation bound is
\begin{equation}
    \boxed{
    R_{E_0,wf}(z) \gtrsim \frac{2.355}{E_0\sqrt{\mathcal{I}_{E,wf}(E_0,z)}} .
    }
    \label{eq:energy_waveform_resolution}
\end{equation}
This level is closer to measured spectra than the ideal primary-trigger bound because $i_{ser}(t)$, filtered shot-noise covariance, crosstalk variance, electronics noise, gain calibration, and baseline uncertainty can all reduce the available information.

\textit{Step 7: Scintillator fluctuations, DOI variation, and nuisance parameters.} The detector Fisher bounds above condition on a deterministic mean relation $\bar N_0(E)$. Real scintillators also exhibit intrinsic light-yield fluctuations and nonproportional response. A compact variance model is
\begin{equation}
    \mathbb{E}[N_0|E]=Y_0E,
    \qquad
    \operatorname{Var}(N_0|E)=F_{sc}Y_0E+\sigma_{np}^2(E),
    \label{eq:energy_scint_variance_n0}
\end{equation}
where $F_{sc}$ represents scintillator excess fluctuation and $\sigma_{np}^2(E)$ represents nonproportionality-related spread. Propagating through $E=N_0/Y_0$ gives
\begin{equation}
    \sigma_{E,scint}^2(E)
        = \frac{F_{sc}E}{Y_0}+\frac{\sigma_{np}^2(E)}{Y_0^2}.
    \label{eq:energy_scint_variance_e}
\end{equation}
If scintillator fluctuations, detector readout fluctuations, and residual DOI collection nonuniformity are approximately independent, and if $\mathcal{I}_{E,det}$ denotes the detector-level energy Fisher Information for the selected observation model while $\sigma_{E,coll}^2(E,z)$ denotes residual light-collection nonuniformity not removed by DOI correction, a practical total floor can be written as
\begin{equation}
    \boxed{
    \begin{aligned}
    \Delta E_{\mathrm{FWHM},total}(E,z)
        &\gtrsim 2.355
        \left\{
        [\mathcal{I}_{E,det}(E,z)]^{-1}
        \right. \\
        &\quad \left.
        +\sigma_{E,scint}^2(E)
        +\sigma_{E,coll}^2(E,z)
        \right\}^{1/2} .
    \end{aligned}
    }
    \label{eq:energy_total_fwhm}
\end{equation}
Here $\mathcal{I}_{E,det}$ may be chosen as the primary-trigger, charge, or waveform Fisher Information.

If unknown interaction time, DOI, gain, baseline, or integration-window parameters are estimated jointly with $E$, the scalar inverse-Fisher bound is replaced by a block of a joint inverse matrix. For the point-process model with
\begin{equation}
    \boldsymbol{\theta}=(E,t_{int},z),
    \qquad
    \lambda_{tot}(t;\boldsymbol{\theta})
        = \lambda_{dyn}(t-t_{int};E,z)+\nu_{DCR},
    \label{eq:energy_joint_theta}
\end{equation}
the Fisher matrix is
\begin{equation}
    \boxed{
    [\boldsymbol{\mathcal{I}}(\boldsymbol{\theta})]_{ij}
        = \int_0^{\infty}
        \frac{
        \partial_{\theta_i}\lambda_{dyn}(t;\boldsymbol{\theta})
        \partial_{\theta_j}\lambda_{dyn}(t;\boldsymbol{\theta})
        }{
        \lambda_{dyn}(t;\boldsymbol{\theta})+\nu_{DCR}
        }\,dt .
    }
    \label{eq:energy_joint_fisher}
\end{equation}
The energy variance bound is then
\begin{equation}
    \boxed{
    \sigma_E^2 \ge [\boldsymbol{\mathcal{I}}^{-1}(\boldsymbol{\theta})]_{EE}.
    }
    \label{eq:energy_joint_bound}
\end{equation}
Adding nuisance parameters cannot improve the bound relative to the fixed-parameter scalar case; it can only preserve or degrade the attainable energy resolution.

\textit{Step 8: Reference 511~keV numerical scale.} The preceding equations apply to an arbitrary reference energy $E_0$. For comparison with the numerical parameter set in Section~\ref{sec:verification}, set $E_0=511~\mathrm{keV}$, $N_0=16000$, $\epsilon a=0.12$, $\eta=0.83$, $k_{trans}=0.40$, $q=0.50$, $M=7100$, and $P_{ct}=0.20$. The escape fraction is
\begin{equation}
    \rho_{esc}=\frac{1-0.12}{1-0.12\times0.83}\approx0.977.
    \label{eq:energy_511_rho}
\end{equation}
The expected incident photon count is
\begin{equation}
    \bar n_{ph,\infty}(511~\mathrm{keV})
        \approx 16000\times0.977\times0.40
        \approx 6.26\times10^3.
    \label{eq:energy_511_nph}
\end{equation}
The corresponding low-occupancy primary photoelectron count is
\begin{equation}
    \bar N_{pe,lin}(511~\mathrm{keV})
        \approx q\bar n_{ph,\infty}
        \approx 3.13\times10^3.
    \label{eq:energy_511_npe}
\end{equation}
Thus the ideal primary-counting FWHM limit is
\begin{equation}
    R_{511,\mathrm{Poisson}}
        \approx \frac{2.355}{\sqrt{3128}}
        \approx 4.2\%.
    \label{eq:energy_511_poisson}
\end{equation}
Under the unbounded Borel crosstalk approximation, $F_{ENF}=1/(1-0.20)=1.25$, giving the ideal charge-statistics scale
\begin{equation}
    R_{511,\mathrm{charge}}
        \gtrsim 2.355\sqrt{\frac{1.25}{3128}}
        \approx 4.7\%.
    \label{eq:energy_511_charge}
\end{equation}
Finally,
\begin{equation}
    \alpha = -\ln\left(1-\frac{0.50}{7100}\right)\approx7.04\times10^{-5},
    \qquad
    \alpha\bar n_{ph,\infty}\approx0.44,
    \label{eq:energy_511_saturation_scale}
\end{equation}
showing that the 511~keV case is not in deep saturation, but finite microcell occupancy is already large enough that the dynamic sensitivity calculation \eqref{eq:energy_sensitivity_ode}--\eqref{eq:energy_lambda_derivative} is the appropriate general procedure. For other energies, the same equations should be re-evaluated at $E_0$ using $\bar N_0(E_0)$, $d\bar N_0/dE|_{E_0}$, the corresponding dynamic occupancy trajectory, and the selected observation model.

Combining the ideal and practical levels, and using the integral definition in \eqref{eq:energy_fisher}, the compact result is
\begin{equation}
    \boxed{
    \begin{aligned}
    R_{E_0,primary}(z)
        &= \frac{2.355}{E_0}
        [\mathcal{I}_E(E_0,z)]^{-1/2},
    \end{aligned}
    }
    \label{eq:energy_final_primary}
\end{equation}
while a waveform or charge spectrum should use
\begin{equation}
    \boxed{
    \begin{aligned}
    R_{E_0,det}(z)
        &\gtrsim \frac{2.355}{E_0}
        [\mathcal{I}_{E,det}(E_0,z)]^{-1/2} .
    \end{aligned}
    }
    \label{eq:energy_final_detector}
\end{equation}
with $\mathcal{I}_{E,det}=\mathcal{I}_{E,Q}$ or $\mathcal{I}_{E,wf}$, and with additional scintillator, electronics, calibration, and DOI terms added through \eqref{eq:energy_total_fwhm} or through the joint Fisher matrix \eqref{eq:energy_joint_fisher}.

\section{Derivation of the Experimental Fitting Models from the Linear Pulse Framework}
\label{app:fitting_biexp_derivation}

This appendix traces the derivation of the three experimental fitting models~\eqref{eq:fit_biexp}--\eqref{eq:fit_ringing} (Section~IV-A) from the first-principles linear pulse model developed in Section~II.

\textit{Step 1: The full analytical pulse model.}
In the unsaturated (linear) regime, the macroscopic SiPM output current at depth of interaction $z$ is given by \eqref{eq:ilin_emg}:
\begin{align}
    I_{lin}(t; z) &= C \sum_{x \in \{d, p1, p2\}} A_x \Big[ \tilde{\tau}_{x,eff} \big( \operatorname{EMG}(t; \mu_z, \sigma_z, \tau_{eff}) \nonumber \\
    & \quad - \operatorname{EMG}(t; \mu_z, \sigma_z, \tau_x) \big) \nonumber \\
    & \quad - \tilde{\tau}_{x,r} \big( \operatorname{EMG}(t; \mu_z, \sigma_z, \tau_r) \nonumber \\
    & \quad - \operatorname{EMG}(t; \mu_z, \sigma_z, \tau_x) \big) \Big]
    \label{eq:S1}
\end{align}
where $C = q k_{trans} C_{gen} I_{cell}^{macro}$ is the global scale factor, $\operatorname{EMG}$ is defined in \eqref{eq:emg_def}, and the coupling constants $\tilde{\tau}_{x,eff}$ and $\tilde{\tau}_{x,r}$ encode the pole interactions between the scintillation kinetics and each SiPM response branch $x$.

\textit{Step 2: Dominant-pole reduction.}
In practical SiPMs, the primary fast avalanche discharge pole $x = d$ carries the dominant fraction of the total cell current ($A_d \gg A_{p1}, A_{p2}$). In the representative parameter set used in Section~\ref{sec:verification}, this dominance is explicit: $A_d = 0.98$, $A_{p1} = 0.015$, and $A_{p2} = 0.005$. Additionally, the intrinsic thermalization rise time $\tau_r \sim 70$~ps is much shorter than the tens-of-nanoseconds apparent rise and decay constants extracted by the phenomenological waveform fits, so it is not separately identifiable in those fitted macroscopic parameters. Taking the limit $\tau_r \to 0$ within this experimental reduction, the coupled $\tilde{\tau}_{x,r}$ branch vanishes in the full analytical expression (Appendix~\ref{app:ideal_kernel}); this is a statement about the complete product term, not about the isolated EMG factor alone. Similarly, isolating only the dominant pole $x = d$ and absorbing the coupling constant $\tilde{\tau}_{d,eff}$ into a redefined amplitude, the full linear model reduces to:
\begin{align}
    I_{lin}(t; z) &\approx A_{eff} \Big[ \operatorname{EMG}(t; \mu_z, \sigma_z, \tau_{eff}) \nonumber \\
                  & \quad\; - \operatorname{EMG}(t; \mu_z, \sigma_z, \tau_d) \Big]
    \label{eq:S2}
\end{align}
where $A_{eff} = C A_d \tilde{\tau}_{d,eff}$ is the macroscopic pulse amplitude. This two-EMG difference is the dominant-pole analytical pulse shape.

\textit{Step 3: From EMG to bi-exponential.}
In the regime where the Gaussian broadening width $\sigma_z$ is small compared to both exponential time constants ($\sigma_z \ll \tau_{eff}, \tau_d$), the EMG kernel reduces to its exponential limit. This follows directly from \eqref{eq:emg_def}: for fixed $t > \mu_z$, the complementary-error-function argument tends to $-\infty$ as $\sigma_z/\tau \to 0$, so $\operatorname{erfc}(\cdot) \to 2$, whereas for $t < \mu_z$ it tends to $+\infty$, so $\operatorname{erfc}(\cdot) \to 0$. The resulting pointwise limit is:
\begin{equation}
    \operatorname{EMG}(t; \mu_z, \sigma_z, \tau) \;\xrightarrow{\;\sigma_z/\tau \to 0\;}\; e^{-(t - \mu_z)/\tau} \, u(t - \mu_z)
    \label{eq:S3}
\end{equation}
Substituting this limit into \eqref{eq:S2} and identifying $t_0 \equiv \mu_z$ as the observable pulse onset time, $\tau_{eff} \to \tau_d^{\mathrm{fit}}$ (the fitted decay constant, which absorbs the convolved scintillation--SiPM pole), and $\tau_d \to \tau_r^{\mathrm{fit}}$ (the fitted rise constant, absorbing the fast avalanche discharge), we recover the bi-exponential fit model~\eqref{eq:fit_biexp}:
\begin{equation}
    I_{biexp}(t) = A \left( e^{-(t-t_0)/\tau_d^{\mathrm{fit}}} - e^{-(t-t_0)/\tau_r^{\mathrm{fit}}} \right) u(t - t_0) + b
    \label{eq:S4}
\end{equation}
where the constant baseline $b$ accounts for the DC pedestal of the digitizer, which is not part of the scintillation physics but is always present in real waveforms. The parameter correspondence is summarized as:
\begin{align}
    A &\;\leftarrow\; A_{eff} = C A_d \tilde{\tau}_{d,eff} \nonumber \\
    \tau_d^{\mathrm{fit}} &\;\leftarrow\; \tau_{eff} = \frac{\tau}{1 - \epsilon a \eta} \quad \text{(modified by SiPM convolution)} \nonumber \\
    \tau_r^{\mathrm{fit}} &\;\leftarrow\; \tau_d^{\mathrm{SiPM}} \quad \text{(primary avalanche discharge pole)}
    \label{eq:S5}
\end{align}
Note that both fitted time constants are \emph{macroscopic effective} parameters; $\tau_d^{\mathrm{fit}} \approx 118$~ns reflects the intrinsic $\tau_{eff} = 44.4$~ns broadened by SiPM recovery and readout bandwidth, while $\tau_r^{\mathrm{fit}} \approx 36$~ns absorbs TTS, electronic bandwidth, and the sub-ns thermalization into a single resolvable rise pole.

\textit{Step 4: Linear-baseline extension.}
In DC-coupled readout chains, slow thermal drift and charge accumulation on coupling capacitors produce a linear baseline wander $kt$ superimposed on the constant pedestal $b_0$. This is a purely electronic artifact with no counterpart in the scintillation physics. Replacing the constant $b \to kt + b_0$ yields the 6-parameter model~\eqref{eq:fit_linbase}.

\textit{Step 5: Restoring the Gaussian convolution.}
The sharp-edge approximation \eqref{eq:S3} neglects the finite TTS and electronic bandwidth broadening, which manifests as a smoothed rising edge. Retaining the full EMG convolution from \eqref{eq:S2} rather than taking the $\sigma \to 0$ limit, the pulse core becomes $\mathcal{E}(t; t_0, \tau_d^{\mathrm{fit}}, \sigma) - \mathcal{E}(t; t_0, \tau_r^{\mathrm{fit}}, \sigma)$, where $\mathcal{E}$ is the EMG kernel \eqref{eq:ecg_kernel}. This re-introduces the Gaussian width $\sigma$ as an explicit fit parameter, encoding the combined DOI-averaged TTS and finite electronic bandwidth:
\begin{equation}
    \sigma^{\mathrm{fit}} \;\leftarrow\; \sqrt{\sigma_{TTS}^2(\bar{z}) + \sigma_{elec}^2}
    \label{eq:S6}
\end{equation}
where $\bar{z}$ denotes the DOI-averaged TTS contribution and $\sigma_{elec}$ is the electronic bandwidth contribution.

\textit{Step 6: Damped-sinusoidal ringing correction.}
The SiPM die is connected to the readout through bond wires (effective series inductance $L_{\mathrm{RLC}}$) and parasitic junction capacitance $C_j$, forming an under-damped RLC circuit. The transient response of such a circuit to an impulsive current injection is a causal damped sinusoid:
\begin{equation}
    h_{RLC}(t) = A_{ring} \, e^{-t/\tau_{ring}} \sin(\omega t + \phi) \, u(t)
    \label{eq:S7}
\end{equation}
where $\tau_{ring} = 2L_{\mathrm{RLC}}/R_{\mathrm{RLC}}$ is the damping time, $\omega = \sqrt{1/(L_{\mathrm{RLC}}C_j) - 1/(2L_{\mathrm{RLC}}/R_{\mathrm{RLC}})^2}$ is the damped natural frequency, $R_{\mathrm{RLC}}$ is the effective damping resistance, $A_{ring}$ is the oscillation amplitude (scaling with the bond-wire/PCB impedance mismatch), and $\phi$ is the initial phase. Superimposing this parasitic artifact onto the Gaussian-convolved bi-exponential yields the 11-parameter ringing-corrected model~\eqref{eq:fit_ringing}. The four ringing parameters ($A_{ring}, \tau_{ring}, \omega, \phi$) have no counterpart in the scintillation--SiPM physics; they are purely electronic interface artifacts.

\textit{Summary.}
The progression from first-principles theory to experimental fit models follows a systematic chain of physically justified approximations:
\begin{align}
    I_{lin}(t; z) &\;\xrightarrow[\text{dominant pole}]{\tau_r \to 0}\; \text{Eq.~\eqref{eq:S2}} \;\xrightarrow[\sigma \to 0]{\text{sharp edge}}\; I_{biexp} \nonumber \\
    &\;\xrightarrow[\text{drift}]{+kt}\; I_{linbase} \;\xrightarrow[\sigma,\;\text{RLC}]{\text{restore}}\; I_{ring}
    \label{eq:S8}
\end{align}
Each step either relaxes a mathematical limit ($\tau_r \to 0$, $\sigma \to 0$) or accounts for a specific physical artifact (DC drift, parasitic ringing). The 11-parameter ringing-corrected model is therefore the most general experimentally accessible reduction of the full analytical framework in the unsaturated regime.


\section{Residual Diagnostic Metrics}
\label{app:residual_diagnostics}
The residual diagnostics in Fig.~\ref{fig:residual_diag} employ the following standard signal-processing tools, applied to the fit residual sequence $\{r_i\}_{i=1}^{n}$ sampled at interval $\Delta t$:

\textit{Power Spectral Density (PSD).} Estimated via the Welch periodogram~\cite{Welch1967Use}: the residual sequence is partitioned into overlapping segments of length $L$ with 50\% overlap, each segment is windowed (Hanning), and the squared magnitude of the discrete Fourier transform is averaged across segments:
\begin{equation}
    \hat{S}(f) = \frac{1}{K U} \sum_{j=1}^{K} \left| \sum_{m=0}^{L-1} w_m \, r_{m+s_j} \, e^{-i2\pi f m \Delta t} \right|^2
    \label{eq:psd}
\end{equation}
where $K$ is the number of segments, $s_j$ the starting index of segment $j$, $w_m$ the window function, and $U = \sum w_m^2 / L$ the window normalization.

\textit{Autocorrelation Function (ACF).} The sample autocorrelation at lag $\ell$ is:
\begin{equation}
    \hat{\rho}(\ell) = \frac{\sum_{i=1}^{n-\ell} r_i \, r_{i+\ell}}{\sum_{i=1}^{n} r_i^2}, \qquad \ell = 0, 1, \ldots
    \label{eq:acf}
\end{equation}
with the 95\% confidence interval for white noise given by $\pm 1.96/\sqrt{n}$.

\textit{Rolling Root-Mean-Square (RMS).} The local RMS envelope over a sliding window of half-width $w$ samples centered at index $i$ is:
\begin{equation}
    \mathrm{RMS}(i) = \sqrt{\frac{1}{2w+1} \sum_{j=i-w}^{i+w} r_j^2}
    \label{eq:rolling_rms}
\end{equation}

\section{Derivation of the Experimental Fitting ODE from the Theoretical Framework}
\label{app:fitting_ode_derivation}

This appendix traces the derivation of the experimental fitting ODE~\eqref{eq:ode_fit} from the first-principles dynamic recovery model~\eqref{eq:dynamic_ode}, and correspondingly justifies the structure of the fitting current model~\eqref{eq:fit_dynamic}.

\textit{Step 1: The first-principles ODE.}
In Section~II, the dynamic microcell recovery is governed by the ODE~\eqref{eq:dynamic_rec}:
\begin{equation}
    \frac{d}{dt} N_{busy}(t; z) = \alpha \, r_{ph}(t; z) \bigl[M - N_{busy}(t; z)\bigr] - \frac{N_{busy}(t; z)}{\tau_{rec}}
    \label{eq:R1}
\end{equation}
where $\alpha = -\ln(1 - q/M)$ is the effective per-photon triggering parameter, $r_{ph}(t; z) = k_{trans}(Y_{mod} * f_{TTS}(\cdot\,; z))(t)$ is the DOI-dependent photon arrival rate at the sensor surface (incorporating scintillation kinetics, transport efficiency, and TTS convolution), $M$ is the total microcell count, and $\tau_{rec}$ is the microcell recovery time constant.

\textit{Step 2: Identifying the experimentally inaccessible quantities.}
In the theoretical framework, the photon rate $r_{ph}(t; z)$ is constructed from the intrinsic scintillation profile $Y_{mod}(t)$ convolved with the geometry-dependent transit time spread $f_{TTS}(t; z)$. In the experimental validation context, we face two fundamental limitations:
\begin{enumerate}
    \item The DOI coordinate $z$ is unknown on a per-pulse basis; hence $f_{TTS}(t; z)$ cannot be evaluated explicitly.
    \item The absolute photon rate $r_{ph}(t; z)$ is not directly measurable from the voltage waveform; only the macroscopic current $I(t)$, which is proportional to the instantaneous firing rate convolved with the single-cell response, is observable.
\end{enumerate}

\textit{Step 3: Effective photon rate proxy.}
In the linear regime (negligible saturation, $N_{busy} \ll M$), the macroscopic voltage waveform is proportional to the photon arrival rate:
\begin{equation}
    V_{lin}(t) \propto \lambda_{lin}(t) \otimes i_{ser}(t) = \alpha \, r_{ph}(t) \, M \otimes i_{ser}(t)
    \label{eq:R2}
\end{equation}
The experimentally fitted bi-exponential template $g(t - t_0) = (e^{-(t-t_0)/\tau_d^{\mathrm{fit}}} - e^{-(t-t_0)/\tau_r^{\mathrm{fit}}})\,u(t - t_0)$, when convolved with a Gaussian kernel $G_\sigma$, provides a phenomenological representation of $V_{lin}(t)$ that implicitly absorbs:
\begin{itemize}
    \item the intrinsic bi-exponential scintillation profile $Y_{mod}(t)$,
    \item the DOI-averaged TTS broadening (via the Gaussian $G_\sigma$, which effectively plays the role of $f_{TTS}$),
    \item the multi-exponential single-cell response $i_{ser}(t)$ (whose dominant pole is absorbed into the fitted $\tau_d^{\mathrm{fit}}$).
\end{itemize}
We therefore identify $g(t - t_0)$ as an effective proxy for the photon rate shape, up to a multiplicative scale factor.

\textit{Step 4: Substitution and parameter absorption.}
Replacing the theoretical product $\alpha \, r_{ph}(t; z)$ in \eqref{eq:R1} with a single effective coupling parameter $\beta$ multiplying the experimentally determined template $g(t - t_0)$:
\begin{equation}
    \alpha \, r_{ph}(t; z) \;\longrightarrow\; \beta \, g(t - t_0)
    \label{eq:R3}
\end{equation}
where $\beta$ is an effective saturation coupling strength. Since $g(t-t_0)$ is dimensionless in \eqref{eq:fit_dynamic}, $\beta$ carries units of inverse time (ns$^{-1}$ in the fits reported here) and absorbs:
\begin{equation}
    \beta \;\sim\; \alpha \cdot k_{trans} \cdot C_{gen} \cdot \kappa_{EMG}
    \label{eq:R4}
\end{equation}
Here $\alpha$ is the effective per-photon triggering parameter, $k_{trans}$ is the crystal-to-sensor transport efficiency, $C_{gen}$ is the generation rate constant from the scintillation model, and $\kappa_{EMG}$ is a shape-matching correction factor that accounts for the difference between the true photon rate profile (multi-exponential with TTS convolution) and the simplified bi-exponential template $g(t)$. Because the bi-exponential template has already been phenomenologically fitted to the observed pulse shape, the residual shape mismatch is small, and $\kappa_{EMG} \approx 1$; the dominant contribution to $\beta$ is the product $\alpha k_{trans} C_{gen}$, which has units of inverse time and scales with the absolute photon flux intensity.

Substituting \eqref{eq:R3} into \eqref{eq:R1} immediately yields the fitting ODE~\eqref{eq:ode_fit}:
\begin{equation}
    \frac{dN_{busy}}{dt} = \beta \, g(t - t_0) \bigl(M - N_{busy}\bigr) - \frac{N_{busy}}{\tau_{rec}}
    \label{eq:R5}
\end{equation}
This ODE is structurally identical to the first-principles model \eqref{eq:R1}, preserving both the state-dependent depletion nonlinearity $(M - N_{busy})$ and the exponential recovery $N_{busy}/\tau_{rec}$. Only the input driving function has been replaced by its experimentally observable proxy.

\textit{Step 5: Current model construction.}
From \eqref{eq:i_dyn}, the theoretical macroscopic current is $I_{dyn}(t) = \int \lambda_{dyn}(t') \, i_{ser}(t - t') \, dt'$, where $\lambda_{dyn} = \alpha \, r_{ph} (M - N_{busy})$. In the fitting framework, we approximate this convolution integral by a product-form model:
\begin{equation}
    I_{dyn}(t) \approx A \left[ g(t - t_0) \cdot \frac{M - N_{busy}(t)}{M} \right] \otimes G_\sigma
    \label{eq:R6}
\end{equation}
This approximation is justified as follows. In the linear regime, $N_{busy}/M \ll 1$, so the saturation factor $(M - N_{busy})/M \to 1$ and the model reduces to $A \, [g \otimes G_\sigma]$---the standard ringing-corrected bi-exponential fit. When saturation becomes appreciable, the factor $(M - N_{busy}(t))/M < 1$ multiplicatively suppresses the pulse amplitude, predominantly near the peak where $N_{busy}$ is maximal. This captures the leading-order saturation effect: the instantaneous reduction of the available microcell pool. The product-form approximation neglects the temporal smearing of the saturation correction by $i_{ser}$---i.e., it assumes the saturation modulation varies slowly relative to the single-cell current pulse width. For the representative SiPM parameter set, the relevant single-cell current poles ($0.5$--$4$~ns) are short compared with the fitted macroscopic decay envelope ($\tau_d^{\mathrm{fit}} \approx 120$~ns), so the quasi-static approximation is expected to introduce only a small error in the fitted envelope.

\textit{Summary.}
Equations~\eqref{eq:fit_dynamic}--\eqref{eq:ode_fit} constitute a physically motivated, experimentally tractable reduction of the first-principles framework (Section~II). The structural form of the ODE is preserved exactly; the only simplification is the replacement of the microscopically computed photon rate $\alpha \, r_{ph}(t; z)$ by the phenomenological proxy $\beta \, g(t - t_0)$, with $\beta$ serving as the sole free parameter that quantifies the effective saturation coupling strength. This approach retains the essential nonlinear physics (microcell depletion and exponential recovery) while remaining compatible with nonlinear least-squares fitting to digitized oscilloscope waveforms.

\section{From Diagonal Gaussian Likelihood to IRLS DOI Estimation}
\label{app:doi_irls_derivation}

This appendix derives the practical DOI objectives \eqref{eq:doi_wls_linear} and \eqref{eq:doi_wls_dyn} from the diagonal Gaussian approximation used in Section~V, and then shows how that approximation yields a workable Iteratively Reweighted Least-Squares (IRLS) solver.

\textit{Step 1: Diagonal Gaussian waveform likelihood.}
Let $\boldsymbol{\varphi}$ denote a generic DOI-estimation parameter vector, let $m_n(\boldsymbol{\varphi})$ denote the predicted mean of sample $y_n$ at time $t_n$, and let $s_n(\boldsymbol{\varphi}) > 0$ denote the corresponding model variance. Under the diagonal approximation adopted in Section~V, the sampled waveform likelihood factorizes as
\begin{equation}
    p(\mathbf{y}\mid\boldsymbol{\varphi}) = \prod_{n=1}^{N}
    \frac{1}{\sqrt{2\pi s_n(\boldsymbol{\varphi})}}
    \exp\!\left[
        -\frac{\left(y_n - m_n(\boldsymbol{\varphi})\right)^2}{2 s_n(\boldsymbol{\varphi})}
    \right].
    \label{eq:doi_diag_pdf}
\end{equation}
Discarding the additive constant $\frac{N}{2}\log(2\pi)$ gives the negative log-likelihood
\begin{equation}
    \mathcal{L}(\boldsymbol{\varphi}) = \frac{1}{2}\sum_{n=1}^{N}
    \left[
        \log s_n(\boldsymbol{\varphi})
        + \frac{\left(y_n - m_n(\boldsymbol{\varphi})\right)^2}{s_n(\boldsymbol{\varphi})}
    \right].
    \label{eq:doi_diag_nll_generic}
\end{equation}
For the linear DOI fit, $m_n = g_{ro} I_{lin}(t_n - t_{int}; z) + k t_n + b_0$ and $s_n = \sigma_{total}^2(t_n; z)$, so \eqref{eq:doi_diag_nll_generic} reduces directly to \eqref{eq:doi_wls_linear}. For the dynamic DOI fit, $m_n = g_{ro} I_{dyn}(t_n - t_{int}; z, N_0) + k t_n + b_0$ and $s_n = \sigma_{total}^2(t_n; z, N_0)$, which gives \eqref{eq:doi_wls_dyn}.

\textit{Step 2: Exact score of the diagonal likelihood.}
Define the residual $r_n(\boldsymbol{\varphi}) = y_n - m_n(\boldsymbol{\varphi})$. Differentiating \eqref{eq:doi_diag_nll_generic} with respect to an arbitrary component $\varphi_\ell$ yields
\begin{equation}
    \frac{\partial \mathcal{L}}{\partial \varphi_\ell}
    = -\sum_{n=1}^{N} \frac{r_n}{s_n}\,\frac{\partial m_n}{\partial \varphi_\ell}
    + \frac{1}{2}\sum_{n=1}^{N}
    \left(
        \frac{1}{s_n} - \frac{r_n^2}{s_n^2}
    \right)
    \frac{\partial s_n}{\partial \varphi_\ell}.
    \label{eq:doi_diag_grad}
\end{equation}
Equation~\eqref{eq:doi_diag_grad} shows why the exact estimator is not ordinary weighted least squares: the second summation accounts for the parameter dependence of the heteroscedastic variance model.

\textit{Step 3: Frozen-variance surrogate and IRLS weights.}
At iteration $m$, let $\boldsymbol{\varphi}^{(m)}$ be the current estimate and freeze the variances to
\begin{equation}
    s_n^{(m)} = s_n\!\left(\boldsymbol{\varphi}^{(m)}\right).
    \label{eq:doi_frozen_var}
\end{equation}
The resulting surrogate objective is
\begin{equation}
    Q^{(m)}(\boldsymbol{\varphi}) = \frac{1}{2}\sum_{n=1}^{N}
    \left[
        \log s_n^{(m)}
        + \frac{\left(y_n - m_n(\boldsymbol{\varphi})\right)^2}{s_n^{(m)}}
    \right].
    \label{eq:doi_irls_surrogate}
\end{equation}
Because $\log s_n^{(m)}$ is constant within the $m$th subproblem, minimizing \eqref{eq:doi_irls_surrogate} is equivalent to minimizing a weighted residual norm with weights
\begin{equation}
    w_n^{(m)} = \frac{1}{s_n^{(m)}}.
    \label{eq:doi_irls_weights}
\end{equation}
Repeatedly recomputing \eqref{eq:doi_irls_weights} from the updated forward model yields the IRLS procedure referenced in Section~V.

\textit{Step 4: Exact weighted update for the linear nuisance block.}
Partition the parameter vector as $\boldsymbol{\varphi} = (\boldsymbol{\eta}, \mathbf{a})$, where $\boldsymbol{\eta} = (z, t_{int})$ in the linear model or $\boldsymbol{\eta} = (z, t_{int}, N_0)$ in the dynamic model, while $\mathbf{a} = (g_{ro}, k, b_0)^T$ collects the coefficients entering the mean waveform linearly. For fixed $\boldsymbol{\eta}$, define the design row
\begin{equation}
    \mathbf{x}_n^T(\boldsymbol{\eta}) =
    \begin{cases}
        \bigl[I_{lin}(t_n - t_{int}; z),\; t_n,\; 1\bigr], & \text{linear model} \\
        \bigl[I_{dyn}(t_n - t_{int}; z, N_0),\; t_n,\; 1\bigr], & \text{dynamic model}
    \end{cases}
    \label{eq:doi_design_row}
\end{equation}
so that $m_n(\boldsymbol{\varphi}) = \mathbf{x}_n^T(\boldsymbol{\eta})\mathbf{a}$. Stacking these rows into $\mathbf{X}(\boldsymbol{\eta})$, stacking the waveform samples into $\mathbf{y} = [y_1, \ldots, y_N]^T$, and defining $\mathbf{W}^{(m)} = \operatorname{diag}(w_1^{(m)}, \ldots, w_N^{(m)})$, the weighted least-squares subproblem for $\mathbf{a}$ is
\begin{equation}
    \hat{\mathbf{a}}^{(m+1)} = \arg\min_{\mathbf{a}}
    \left(\mathbf{y} - \mathbf{X}\mathbf{a}\right)^T
    \mathbf{W}^{(m)}
    \left(\mathbf{y} - \mathbf{X}\mathbf{a}\right).
    \label{eq:doi_irls_linear_block}
\end{equation}
The normal equations are
\begin{equation}
    \mathbf{X}^T \mathbf{W}^{(m)} \mathbf{X}\,\hat{\mathbf{a}}^{(m+1)}
    = \mathbf{X}^T \mathbf{W}^{(m)} \mathbf{y},
    \label{eq:doi_irls_normal_eq}
\end{equation}
which gives the closed-form update
\begin{equation}
    \hat{\mathbf{a}}^{(m+1)}
    = \left(\mathbf{X}^T \mathbf{W}^{(m)} \mathbf{X}\right)^{-1}
    \mathbf{X}^T \mathbf{W}^{(m)} \mathbf{y}
    \label{eq:doi_irls_linear_solution}
\end{equation}
whenever the weighted design matrix is nonsingular.

\textit{Step 5: Nonlinear DOI-sensitive update.}
After substituting \eqref{eq:doi_irls_linear_solution} back into \eqref{eq:doi_irls_surrogate}, the remaining subproblem depends only on $\boldsymbol{\eta}$. Define the stacked model waveform as $\mathbf{m}(\boldsymbol{\eta}, \mathbf{a}) = [m_1(\boldsymbol{\varphi}), \ldots, m_N(\boldsymbol{\varphi})]^T$. A practical step is then obtained by weighted Gauss--Newton or a weighted trust-region method. Let $\mathbf{r}^{(m)} = \mathbf{y} - \mathbf{m}(\boldsymbol{\eta}^{(m)}, \hat{\mathbf{a}}^{(m+1)})$ and let $\mathbf{J}^{(m)} = \partial \mathbf{m}/\partial \boldsymbol{\eta}$ denote the residual vector and Jacobian under frozen weights. The Gauss--Newton increment then satisfies
\begin{equation}
    \left(\mathbf{J}^{T} \mathbf{W}^{(m)} \mathbf{J}\right)
    \Delta\boldsymbol{\eta}^{(m)}
    = \mathbf{J}^{T} \mathbf{W}^{(m)} \mathbf{r}^{(m)}.
    \label{eq:doi_irls_gn}
\end{equation}
The nonlinear parameter block is updated as $\boldsymbol{\eta}^{(m+1)} = \boldsymbol{\eta}^{(m)} + \Delta\boldsymbol{\eta}^{(m)}$, followed, when needed, by projection of the DOI estimate onto $[0, L_{cry}]$.

\textit{Step 6: Practical solver sequence.}
The DOI estimator implied by \eqref{eq:doi_wls_linear} and \eqref{eq:doi_wls_dyn} can therefore be implemented as follows:
\begin{enumerate}
    \item Initialize $z$ from \eqref{eq:doi_init_sigma}, optionally refine it with \eqref{eq:doi_init_mu}, and initialize $t_{int}$, $g_{ro}$, $k$, and $b_0$ from a preliminary EMG fit; for the dynamic model, initialize $N_0$ from pulse area or peak amplitude.
    \item Evaluate the current variance profile $s_n^{(m)}$ and the weights \eqref{eq:doi_irls_weights}.
    \item Update the linear nuisance block exactly through \eqref{eq:doi_irls_linear_solution}.
    \item Update $(z, t_{int})$ or $(z, t_{int}, N_0)$ through \eqref{eq:doi_irls_gn} or a weighted trust-region step.
    \item Recompute the variance profile and repeat until the relative change in $\mathcal{L}_{lin}$ or $\mathcal{L}_{dyn}$ falls below a prescribed tolerance.
\end{enumerate}
This derivation makes the computational role of the diagonal approximation explicit: it replaces the exact sampled-waveform likelihood by a sequence of weighted least-squares subproblems while preserving the DOI dependence of both the mean waveform and the heteroscedastic variance model.

\bibliographystyle{ieeetr}
\bibliography{references}

\end{document}